\documentclass[aip,jcp,amsmath,amssymb,reprint,superscriptaddress,floatfix]{revtex4-2}
\usepackage{graphicx}
\usepackage{dcolumn}
\usepackage{bm}
\usepackage{gensymb}
\usepackage{tikz}
\usepackage{hyperref}
\usepackage[caption=false]{subfig}
\usepackage{xcolor}

\graphicspath{{../../figures/}}

\begin{document}

\title{Dispersion and Orientation patterns in nanorod-infused polymer melts}

\author{Navid Afrasiabian}
\email{nafrasia@uwo.ca}
\affiliation{Department of Physics and Astronomy, University of Western Ontario, 1151 Richmond Street, London, Ontario, Canada, N6A 3K7}
\author{Venkat Balasubramanian}
\affiliation{Department of Applied Mathematics, University of Western Ontario, 1151 Richmond Street, London, Ontario, Canada, N6A 5B7}
\author{Colin Denniston}
\email{cdennist@uwo.ca}
\affiliation{Department of Physics and Astronomy, University of Western Ontario, 1151 Richmond Street, London, Ontario, Canada, N6A 3K7}

\date{12 December 2022}

\begin{abstract}
	Introducing nanorods into a polymeric matrix can enhance the physical and mechanical properties of the resulting material. In this paper, we focus on understanding the dispersion and orientation patterns of nanorods in an unentangled polymer melt, particularly as a function of nanorod concentration, using Molecular Dynamics (MD) simulations. The system is comprised of flexible polymer chains and multi-thread nanorods that are equilibrated in the NPT ensemble. All interactions are purely repulsive except for those between polymers and rods.  Results with attractive versus repulsive polymer-rod interactions are compared and contrasted. The concentration of rods has a direct impact on the phase behaviour of the system. At lower concentrations rods phase separate into nematic clusters, while at higher concentrations more isotropic and less structured rod configurations are observed. A detailed examination of the conformation of the polymer chains near the rod surface shows extension of the chains along the director of the rods (especially within clusters).  The dispersion and orientation of the nanorods is a result of the competition between depletion entropic forces responsible for the formation of rod clusters, the enthalpic effects that improve mixing of rods and polymer, and entropic losses of polymers interpenetrating rod clusters. 
\end{abstract}

\maketitle

\section{\label{sec:level1}Introduction}
Mixtures of anisotropic fillers, especially nanofibres (NFs) and nanotubes (NTs) in polymer matrices have shown a great potential to produce high performance materials and therefore have received a lot of attention from scientific and engineering communities \cite{Iijima1991, Ajayan1994, Arash2014, Lau2006, Zhu2007, Pal2016, Sahoo2010, Moniruzzaman2006, Lozano2001, Mordkovich2003}. On top of the intrinsic properties of the nanorods, their distribution and orientation in the polymer matrix, interaction with the matrix, and aspect ratio play a crucial role in the overall performance of the material \cite{Al-Saleh2011, Kumar2002, Coleman2006a, Paul2018}. Larger aspect ratio of the nanofillers is known to increase the efficiency of the polymer nanocomposite \cite{Coleman2006a, Paul2018} and previous computational works have studied the effect of nanorod length \cite{Toepperwein2011, Sapkota2017}. However, in this study, we have polymer chains and nanorods with fixed lengths in all simulations and instead turn our focus to the effect of concentration on dispersion patterns of nanorods in a system with polymer-rod attractions.

One of the main barriers in enhancing properties of polymeric materials through adding NTs or NFs is the formation of aggregates which leads to problems such as non-uniform stress distribution and slippage \cite{Xie2005, Coleman2006}. In spite of the development of preparation and processing techniques such as in situ polymerization and surface modification that have been successful in promoting better dispersion of nanorods in a polymer matrix \cite{Hirsch2002, Severini2002, Ramasubramaniam2003, Zhu2003, Xie2005, Grossiord2006, Vaisman2006, Sahoo2010, Ma2010, Kalra2010, Al-Saleh2011}, there is a need for a deeper understanding of the underlying physics that leads to the observed phase behaviour in nanorod-polymer systems.  As a result, it has been under an extensive examination both theoretically \cite{Lekkerkerker1994,Bolhuis1997, Surve2007, Hall2010} and computationally \cite{Starr2002, Savenko2006, Toepperwein2011, Hu2013a, Hore2013, Gao2014, Sankar2015, Milchev2020,Lu2021, Erigi2021}. Savenko and Dijkstra conducted Monte Carlo simulations of a polymer-nanorod system using an effective Hamiltonian that accounted for the effect of the polymer matrix implicitly \cite{Savenko2006}. In this study, the polymers were assumed to be noninteracting but excluded from the surface of the rods by a distance equal to their radius of gyration. Their results showed rod packing, and formation of nematic aggregates. Although understanding nonabsorbing systems provide a great insight into depletion effects, they are not common in practical applications due to a high number of aggregates and poorer performance.

Surface treatments like functionality or adding sizing agents not only results in more uniform dispersion of nanorods but also improves the interfacial interactions between the rods and the polymer matrix which is critical for achieving good mechanical properties \cite{Tang1997, Zhang2013,Zhang2013a, Karger-Kocsis2015, Paul2018}. Stronger interfacial adhesion facilitates the stress transfer from the matrix to the nanorods thus improving the interfacial shear stress (IFSS) strength and performance of the nanocomposite \cite{Park2014, Aztatzi-Pluma2016, Liao2001, Xu2002, Wong2003}. Therefore, it is interesting and useful to investigate polymer nanocomposite (PNCs) systems with attractive polymer-rod interactions. By means of Monte Carlo (MC) and Molecular Dynamics (MD) simulations, Toepperwein et al. studied a system of nanorods in an entangled polymer matrix where all interactions(i.e. rod-rod, polymer-polymer, and rod-polymer) were attractive \cite{Toepperwein2011}. They used stronger polymer-nanorod interactions to mimic a more realistic system and examined the effects of particle size, aspect ratio and volume fraction. They observed a well-dispersed mixture for shorter rods while the 16-mer rods phase separated to aligned aggregates. Our results, for an unentangled melt and 16-mer rods, manifest similar patterns of highly-ordered rod clusters. Another interesting observation by Toepperwein et al., which is observed in our simulation as well, was the presence of polymer chains between the rods within the clusters despite their packed structure. 

In a molecular dynamics study, Gao et al. investigated the effect of inter-component interaction strength, temperature, filler concentration, cross-linking density, external shear, aspect ratio, and nanorod grafting on the dispersion patterns and kinetics\cite{Gao2014}. For a system in which the polymers were attracted to both rods and polymers, and rods repelled rods, they found that there exists an optimum moderate polymer-rod attraction strength that promotes good dispersion. They categorized the driver of aggregate formation to be polymer-bridging (single polymers attached to two or more rods) or direct agregation (entropic depletion-like effect). Their simulations showed that lower nanorod loading provides better filler distribution. However, as will be shown later, our simulations predict higher isotropy and lower order in the system as more nanorods are packed into the system. 

In a recent article, Lu, Wu, and Jayaraman conducted MD simulations on polymer-rod nanocomposites with homogeneous and patchy surface to understand the effect of nanorod design on final PNC morphology \cite{Lu2021}. In this study, the polymer-polymer, and polymer-nanorod interactions were purely repulsive while the nanorods interacted with an attractive potential. For short nanorods, they observed percolated nanorod structure for the system with patchy rods whereas the simple nanorods phase-separated to a cluster. In the case of long nanorods, both designs exhibited formation of ordered aggregates, either finite-sized or percolating. They also looked into the conformation of polymers at the nanorod interface where they discovered that although the average radius of gyration ($R_g$) of the polymers remained the same as that of a pure melt, the interfacial chains stretched out and expanded. We will also examine the interfacial behaviour of polymer chains for a system with polymer-nanorod attractions.

High- and ultrahigh-density ($>$50\%) polymer nanocomposites have been less studied due to the practical difficulties in their production. However, recently several methods have been implemented to overcome some of these barriers \cite{Huang2015,Manohar2017,Venkatesh2019}. It has also been shown that high- and ultrahigh-density polymer nanocomposites exhibit exceptional properties such as improved toughness \cite{Jiang2018}and have potential to be used in energy storage and conversion devices \cite{Swain2022}. Therefore, they are of high interest. Since both lower and higher concentrations are practically relevant, it is important to understand the impact of concentration on the properties of a polymer-nanorod composite system and discover any possible behaviour changes from lower to higher concentrations. Hence, in this work we study the effect of concentration on the phase behaviour of a polymer-nanorod melt over a wide range of concentrations up to 0.44 particle fractions.

The dispersion patterns of nanorods in a pool of attracting polymer chains is a less explored field and is the focus of this paper. To get a better idea of what is driving the phase behavior in the system we contrast our results with those of a system with purely repulsive nanorod-polymer interactions but otherwise identical.  Using molecular dynamics, we simulated a polymer-nanorod melt where all interactions were repulsive except for polymer-rod and looked at the dispersion and orientation of rods as well as the conformation of polymer chains at the rod interface. In the next section, we go over the simulation setup and details. Section \ref{sec:level3} is dedicated to results and discussion where we first describe the dispersion patterns of the rods by means of auto-correlation of a number density, and rod-rod distances and then delve into orientational behaviour of the nanocomposite melt. We sum up the paper by pointing out the main findings in section \ref{sec:level4}.

\begin{figure}[htbp]
	\subfloat[\label{fig:chainpolymer}]{        	
		\includegraphics[width=0.25\textwidth ]{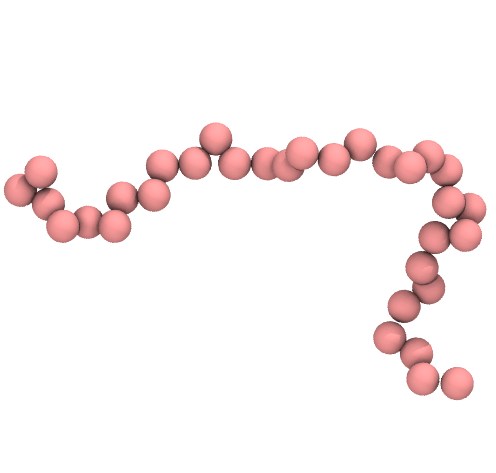}
	}
	\subfloat[\label{fig:chainrod1}]{        	
		\includegraphics[width=0.25\textwidth ]{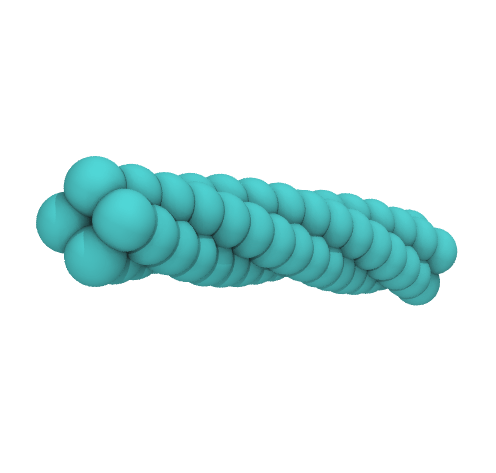}
	}
	
	\subfloat[\label{fig:iniconf}]{        	
		\includegraphics[width=0.3\textwidth ]{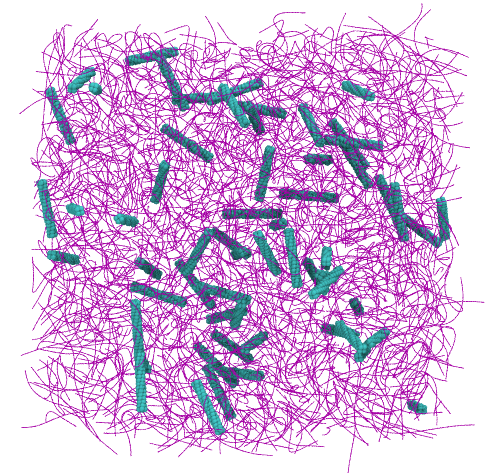}
	}

	\caption{\footnotesize{An example of polymer chain is shown in (a) while (b) shows a lateral view of a nanorod. A random initial configuration shown in (c) (before energy minimization), is generated using moltemplate package for each realization. The nanorods are shown in cyan(green) and the polymer chains are shown as purple lines. The VMD software was used for this visualizations \cite{Humphrey1996}.}}
	\label{fig:setup}
\end{figure}

\section{\label{sec:level2} Model and Simulation Method}

In this work, we adopt a coarse-grained approach to model the polymer-nanorod mixture. The melt is comprised of a mix of polymer chains and rigid nanorods. Each polymer molecule is composed of $n_{p} = 32$ consecutive beads (monomers) connected via Finitely Extensible Non-linear Elastic (FENE) bonds (cf. Fig.~\ref{fig:chainpolymer}). The Kremer-Grest FENE potential
\begin{multline}
	\scriptsize
	\label{eqn:fenebond}
	\small
	U_{b}= -\frac{1}{2}K R_{b}^2 \ln\left[1-\left(\frac{r}{R_{b}}\right)^2\right]\\
		+ 4\epsilon_b\left[\left(\frac{\sigma}{r}\right)^{12}-\left(\frac{\sigma}{r}\right)^6+\frac{1}{4}\right]H\left(2^{\frac{1}{6}}\sigma\right)
\end{multline}
is used to implement all the bonded interactions necessary for the monomer inter-connectivity\cite{Kremer1990}. The first term on the RHS is attractive in nature with $K = 30 \epsilon_b \sigma^{-2}$ being the effective elastic constant where $\epsilon_b = 1.0 \epsilon$ and $\epsilon = \frac{m\sigma^2}{\tau^2}$ is the Lennard-Jones energy constant. Here, $m$ is the Lennard-Jones(LJ) unit mass, $\sigma$ is the LJ length scale, and $\tau$ is the LJ time scale.  $R_{b} = 1.5\sigma$ is the maximum bond extension in any direction. On the other hand, the second term on the RHS represents the repulsive portion of the potential with the cut-off length $r_{c} = 2^{1/6}\sigma$, at the minimum if the LJ potential, enforced through the Heaviside function $H(x)$. This form of potential also eliminates nonphysical bond crossings \cite{Kremer1990}. The excluded volume of the polymer chains is implemented through a repulsive $12-6$ LJ potential similar to the one used for the FENE bonds between the polymer beads. The interaction cutoff of the polymer-polymer interactions is also set to $r_{pp} = 2^{1/6}\sigma$ and the strength is $\epsilon_{pp} = 1.0 \epsilon$. Interactions between the polymers and rods are also of a $12-6$ LJ potential form with strength $\epsilon_{rp} = 1.0\epsilon$. However, the cutoff length for these interactions is set to $r_{rp} = 2.5\sigma$ creating attraction between the polymer chains and the nanorods.  To help understand the effect of attractive forces, we compare our results to systems that are identical except for lacking the attractive part of the potential by setting $r_{rp}=2^{1/6}\sigma$ (and adding a constant shift so that they are zero at the cutoff).

The rigid rods in our system consist of four individual threads (sub-rods) which are assembled in a helical pattern as shown in Fig.\ref{fig:chainrod1}. Each thread is comprised of $n_{r} = 16$ point particles (monomers) of mass $m$ and diameter $\sigma$ where $m$ and $\sigma$ are the LJ unit mass and length, respectively. The monomers are interconnected along the backbone via the FENE bonds described in Eq. \ref{eqn:fenebond} and rigidity of the rods is ensured by a harmonic angle potential $U_{h}$ for every monomer \emph{triad}
\begin{equation}
	\label{eqn:harmonic}
	U_{h} = k \left(\theta - \theta_{0}\right)^2
\end{equation}
where $k = 1000$ (LJ units) is the spring constant, $\theta$ is the angle formed by a triad at any given time during the simulation and $\theta_{0}$ is its equilibrium value. A rigid conformation is obtained by penalizing any bending of the rods by setting $\theta_{0} = 180\degree$ during the energy minimization step. However, at equilibration and production stages, each rod is treated as a rigid body to reduce computational cost of the simulation without compromising the physics. The diameter and length of the rods are respectively $D \approx 2.35 \sigma$ and $L \approx 13.35\sigma$ giving them an aspect ratio of about $5.5$. Atoms are spaced approximately $1.15\sigma$ along the rod and $0.85 \sigma$ between atoms in neighboring threads.  The multi-thread design of the nanorods is different from most previous studies where single thread nanorods \cite{Toepperwein2011, Toepperwein2012, Gao2014, Park2014, Park2014a, Sankar2015, Gao2015, Shen2018, Paiva2019, Milchev2020, Lu2021}, hollow nanotubes \cite{Karatrantos2011, Karatrantos2016}, or smooth (sphero)cylinders were used \cite{Bolhuis1997,Tuinier2007, Wang2021}. This gives the nanorods a new surface roughness, which has been shown to play a role in the phase behaviour of polymer-nanorod composites \cite{Lu2021a}. Moreover, the multi-threaded nanorods are incommensurate with the polymers (where the typical atomic spacing is around $1.35 \sigma$) eliminating possible artificial adhesion at the surface \cite{Robbins2000, Berman2018}. This makes the multi-threaded design a good candidate for further studies of fibre pullout and interfacial slippage. All the simulations are done using the open source package LAMMPS \cite{Thompson2022}. The details of the simulation and equilibration procedures are discussed in the next subsection.

One of the main objectives of the present work is to investigate the effect of nanorod inclusion on the conformation of the polymer chains in the melt, especially the chains at the interface of the nanorods. As a measure of the shape and size of the polymers, we calculated the radius of gyration tensor $\bm{R}_g$ of a chain from the particle coordinates as
\begin{equation}
	R_{g_{\alpha\beta}}^2 = \frac{1}{M^2}\left[\sum_{i=1}^{n_p} m_i (r_{i,\alpha} - r_{com,\alpha})(r_{i,\beta} - r_{com,\beta})\right]
\end{equation}
where $R_{g_{\alpha\beta}}^2$ is the element of the tensor $\bm{R}_g^2$ on the $\alpha$th row and $\beta$th column, M is the total mass of the chain, $n_p$ is the number of beads in the chain, and $m_i$ is the mass of the $i$th bead. The $r_{i,\alpha}$ represents the position of the $i$th bead in $\alpha = x,y,z$ direction and similarly, the $r_{i,\beta}$ is the position of the $i$th bead in $\beta = x,y,z$ direction. The $r_{com}$ is the position of the centre of mass of the polymer chain. Then the $|R_g|$ was found by
\begin{equation}
	|R_g| = \sqrt{\lambda_1^2 + \lambda_2^2 + \lambda_3^2}
\end{equation}
where $\lambda_i$ is the $i$th eigenvalue of the gyration tensor. A set of 10 realizations of a pure melt, at the same temperature and pressure as the production runs examined in the results section, are run and the average radius of gyration of the polymer is measured to be $R_0 \approx 3\sigma$. This value is used as a reference throughout the paper.

A fixed number of polymer chains $N_p = 1000$ is used  across realizations whereas the total number of rigid rods $N_{r}$ in the melt is varied in order to achieve different concentrations of nanofillers. We quantify the concentration of the rods $\phi_{c}$ by simply taking the ratio of the total number of rod monomers to the total number of particles $N$
\begin{equation}
	\centering
	\phi_{c} = \frac{\text{number of rod beads}}{\text{total number of beads}}= \frac{(4 \cdot N_{r} )n_{r}}{N} 
\end{equation}
where $n_{p} = 32$, $n_{r} = 16$, $N_p = 1000$, $N=N_{p} n_{p}+(4 \cdot N_{r} )n_{r}$, $60\le N_r \le500$, and $0.1\le \phi_c \le0.5$. 

Lastly, to improve results statistically, for each value of $\phi_{c}$, 10 independent realizations were carried out and the results were averaged over the realizations. Throughout the paper, all quantities are presented in dimensionless LJ units.

\begin{figure*}[!htbp]
	\subfloat{
		\includegraphics[width=\textwidth]{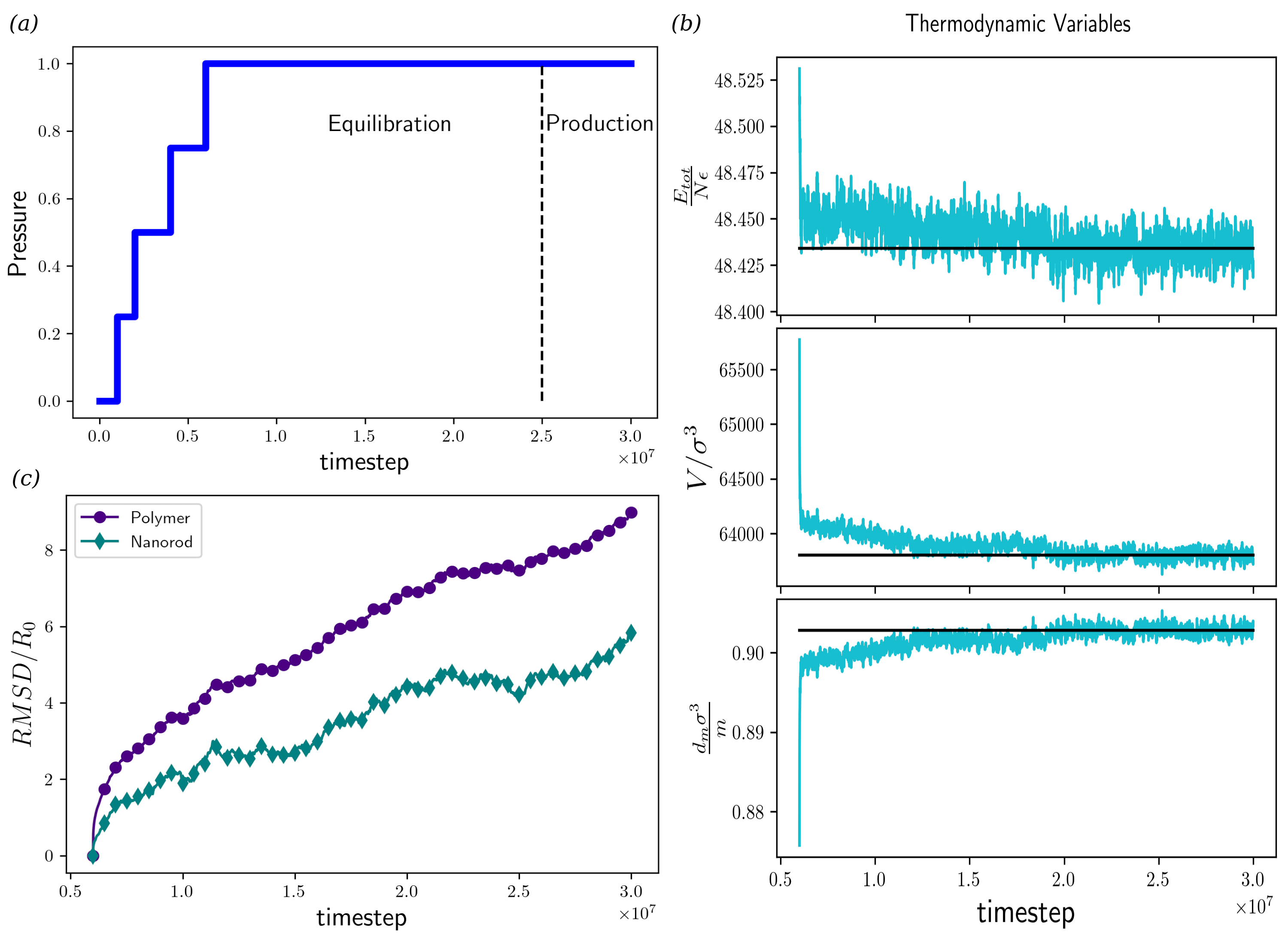}
	}

	\caption{\footnotesize{The equilibration scheme is shown in (a). (b) shows the total energy, volume, and mass density during the last stage of equilibration and the production for a system of $\phi_c = 0.44$. In (c), the RMSD for the nanorods and polymers as a function of time are demonstrated. The value of the RMSD is normalized by the pure melt average radius of gyration $R_0$.}}
	
	\label{fig:thermo_var}
\end{figure*}

\subsection{Equilibration procedure}

The simulation is started from a random initial configuration (see Fig.\ref{fig:iniconf}) generated using the \emph{moltemplate} package \cite{Jewett2021}. Following the procedure in \cite{Auhl2003}, a soft interaction of the form
\begin{equation}
	U = A\left[1 + \text{cos}\left(\frac{\pi r}{r_c}\right)\right] \text{ for } r<r_c
\end{equation}
is then applied between all the components where $r_c$ is the cut-off length. The potential amplitude $A$ is linearly increased from 0 to $100\epsilon$ over $1000\tau$. This allows for the overlapping particles to rearrange themselves without making the simulation unstable. At this point, the molecules are at a reasonable separations to be able to turn on the LJ interactions (and the soft potential $U$ is turned off). As the last step before equilibration, the box (system) is relaxed.

The equilibration procedure consists of 4 stages as shown in Fig. \ref{fig:thermo_var}a. We start off the equilibration with a short NVT run where a Langevin thermostat is applied to keep the temperature at $T = 1$ in all subsequent steps. In the next step, we turn on a Berendsen barostat at the target pressure of $P = 0.25$ and let the system evolve under the NPT ensemble for 1 million steps. Then the system is compressed by increasing the pressure stepwise in three more steps up to the final $P = 1$. The packing density of the mixture is defined as $d_p=\frac{\text{Total Volume of particles}}{\text{Volume of simulation box}}$ and the final pressure is chosen to achieve a melt-like packing density of $0.3 \leq d_p \leq 0.5$ \cite{Hall2008, Hall2010, Lu2021}. The system is run at $P = 1$ and $T = 1$ for 9 million steps (or 19 million steps for the highest concentration) before the production. This careful stepwise protocol ensures that the mixture does not get trapped in a kinetically favourable state. To confirm that the system is in true equilibrium, we monitored the thermodynamic parameters as well as the root-mean-squared-displacement (RMSD). In Fig. \ref{fig:thermo_var} (b), the total energy, volume, and mass density of the system at the last stage of equilibration and the production for a typical system at $\phi_c = 0.44$ are illustrated. The values of these thermodynamic variables are levelled off and fluctuating around a mean value by the end of equilibration and during production. As the equilibration time increases with concentration, especially when nanorod-polymer attractions are at play, here, we only present the data for the highest concentration.

The structural equilibration of polymeric systems is a slower process compared to relaxation of thermodynamic variables \cite{Mozafar2022}. Therefore, we also measured the root-mean-squared-displacement of the particles as a function of time to make sure that the polymers and the rods move reasonable distances during equilibration \cite{Auhl2003, Gartner2019}. Fig. \ref{fig:thermo_var} (c) shows the evolution of the RMSD for the nanorods and polymers. As can be seen, both the rods and the polymers move multiple $R_0$, the average radius of gyration of a polymer in the pure melt, during the process validating the fact that the system is not stuck in a glassy state. We should mention that the particle RMSD is different from the RMSD of the centre of mass (COM). This takes into account the rotational motion of the molecules as well as the COM displacement which are both crucial to the equilibration process. However, we tracked the RMSD of the centre of mass of the constituents to make sure that the rotational motion is not taking over in the above graphs. The RMSD of COM for the rods and the polymers are plotted in Fig. \ref{fig:com_rmsd} (a) in the Appendix demonstrating they also move several $R_0$.  

To investigate the effect of attractive interactions on the phase behaviour of the nanorod-polymer melt, we compare some of the results for our system to a system with all repulsive interactions. The all-repulsive system is equilibrated following the same protocol as the attractive system. Similar to Fig. \ref{fig:thermo_var}, in Fig. \ref{fig:thermo_var_rep} of the Appendix, the evolution of the thermodynamic variables and particle RMSD at the highest concentration $\phi_c = 0.44$ are shown. The RMSD of centre of mass for the all-repulsive case is presented in Fig. \ref{fig:com_rmsd} (b).

In addition to the equilibration procedure described above, we also tested equilibration procedures involving parallel tempering (with up to 8 replicas) and just letting systems evolve for much longer time periods.  The final states found were the same to those described here but did not find those states any faster than the procedure described above.

\section{\label{sec:level3}Results and discussion}
\subsection{Dispersion and phase Separation}
\begin{figure}[t]
	\subfloat[\label{fig:system1_1}]{        	
		\includegraphics[width=0.22\textwidth ]{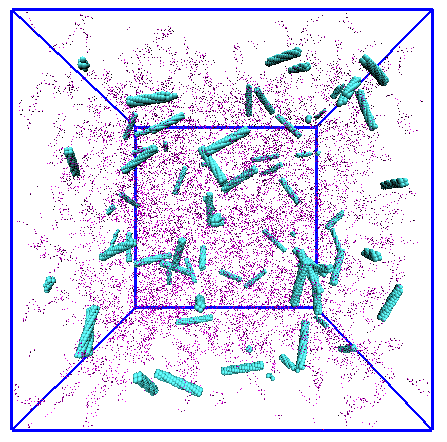}
	}
	\subfloat[\label{fig:system1_2}]{        	
		\includegraphics[width=0.22\textwidth ]{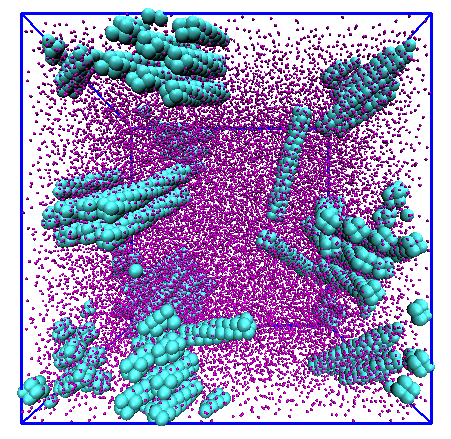}
	}

	\subfloat[\label{fig:system1_3}]{        	
	\includegraphics[width=0.22\textwidth ]{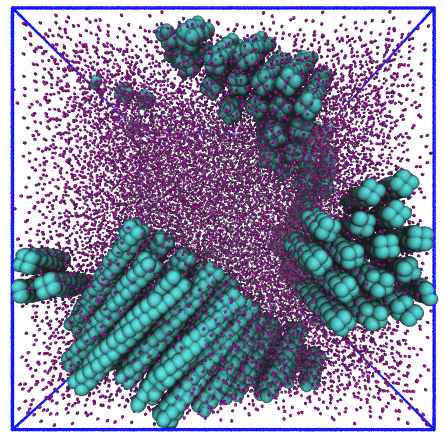}
	}
	\subfloat[\label{fig:system1_4}]{        	
		\includegraphics[width=0.22\textwidth ]{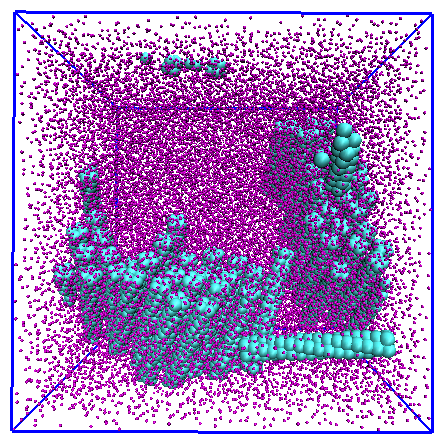}
	}
	\caption{\footnotesize{A nanorod-polymer melt with nanorod concentration $\phi=0.1$ and nanorod-polymer interaction strength $\epsilon_{rp} = 1$. As the system evolves, the (a) initial random configuration at $P=0$ progress to phase separate and form (d) distinct rod aggregates after equilibration at $P=1$. (b) and (c) show intermediate stages at $P=0.5$ and $P=0.75$ respectively. The considerable change in the system configuration is partial evidence of full equilibration of the system. (d) shows the final configuration of the system after $9\times 10^6$ equilibration steps and $10^6$ production steps. The matrix polymer chains are shown as purple dots for illustration purposes.}}
	\label{fig:system_prog}
\end{figure}

Rigid rods and nano particles have long been known to have poor dispersion in polymer melts. However achieving optimal dispersion of the rods throughout the melt is extremely important when considering the mechanical and structural properties of the resulting material. As mentioned earlier, chemically treating the surfaces of the rigid rods has shown to improve dispersion as it boosts their interactions with the polymer matrix \cite{Du2003, Zhao2016, Ma2010, Hirsch2002}. Therefore, due to their practical relevance, we mainly focus on a system of nanorod-polymer composite in which the polymer-rod interactions are attractive while all other interactions are hard-core repulsive. To understand the effect of the attractive polymer-rod interaction, the results are compared and contrasted with systems that have purely repulsive polymer-rod interactions but are otherwise identical.
\begin{figure*}[t]
	\subfloat[]{        	
		\includegraphics[width=0.225\textwidth ]{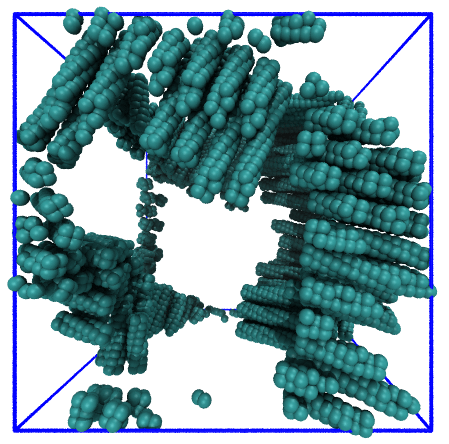}
	}
	\subfloat[]{        	
		\includegraphics[width=0.225\textwidth ]{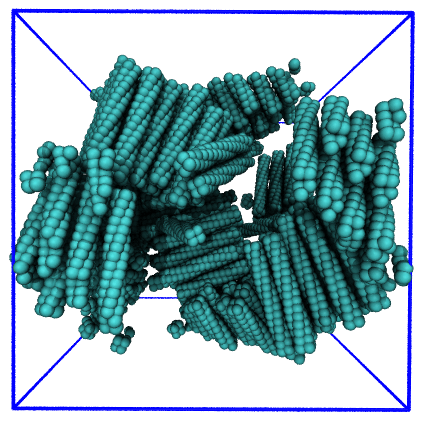}
	}
	\subfloat[]{        	
		\includegraphics[width=0.225\textwidth ]{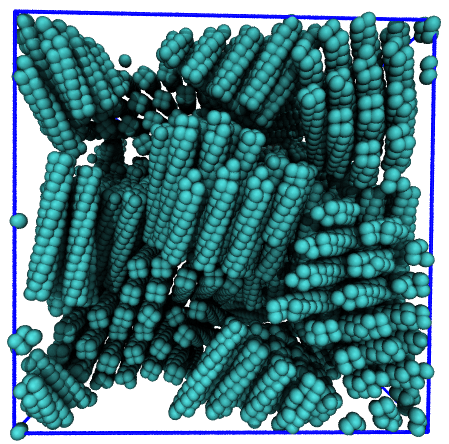}
	}
	\subfloat[]{        	
		\includegraphics[width=0.225\textwidth ]{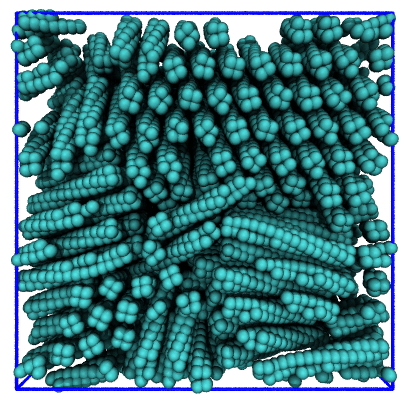}
	}
	
	\caption{\footnotesize{Snapshots of the rod-polymer system at concentrations (a) $\phi_c = 0.1$, (b) $\phi_c = 0.2$, (c) $\phi_c = 0.33$, and (d)$\phi_c = 0.44$ are shown. Initially, increasing the concentration of the rods results in the growth of the size of the clusters, but further increase breaks the clusters up and makes the system more uniformly mixed. This is attributed to the interplay of entropic and enthalpic effects.}}
	\label{fig:phi_snapshot}
\end{figure*}

In Fig. \ref{fig:system_prog}, we present snapshots of our system at the lowest concentration, i.e. $\phi_c = 0.1$, as it equilibrates. Fig. \ref{fig:system1_1} (a) shows the system at an initial stage of the equilibration. As can be seen, the system starts out in a fairly random configuration with a lot of empty space between the components (lower packing density and larger box size).  Fig. \ref{fig:system_prog} (b) shows the system after $2.5$ million timesteps at $P=0.5$. We can see the early stages of the agglomeration of rods with a few clusters of rods appearing. Fig. \ref{fig:system_prog} (c) depicts a later stage when $5$ million timesteps elapsed and at $P=0.75$. As can be seen, the evolution of clusters continues and the shape of the clusters changes. Fig. \ref{fig:system_prog} (d) shows the system at $P=1.0$ and packing density $d_p=0.33$ after $20$ million steps. We can see that distinct aggregates are formed and some regions are filled with only polymers (at lower concentrations). This phase separation has been observed in experiments \cite{Vaisman2006, Hore2013} as well. In most previous computational studies where formation of such clusters were studied, an attractive interaction between rods were at play \cite{Lu2021, Toepperwein2012}. However, the formation of such clusters in a system with rod-rod repulsive forces suggests the significance of entropic effects in this phenomenon.

\begin{figure}[!htbp]
	\centering
	\subfloat[\label{fig:crhorho}]{\includegraphics[width=0.4\textwidth]{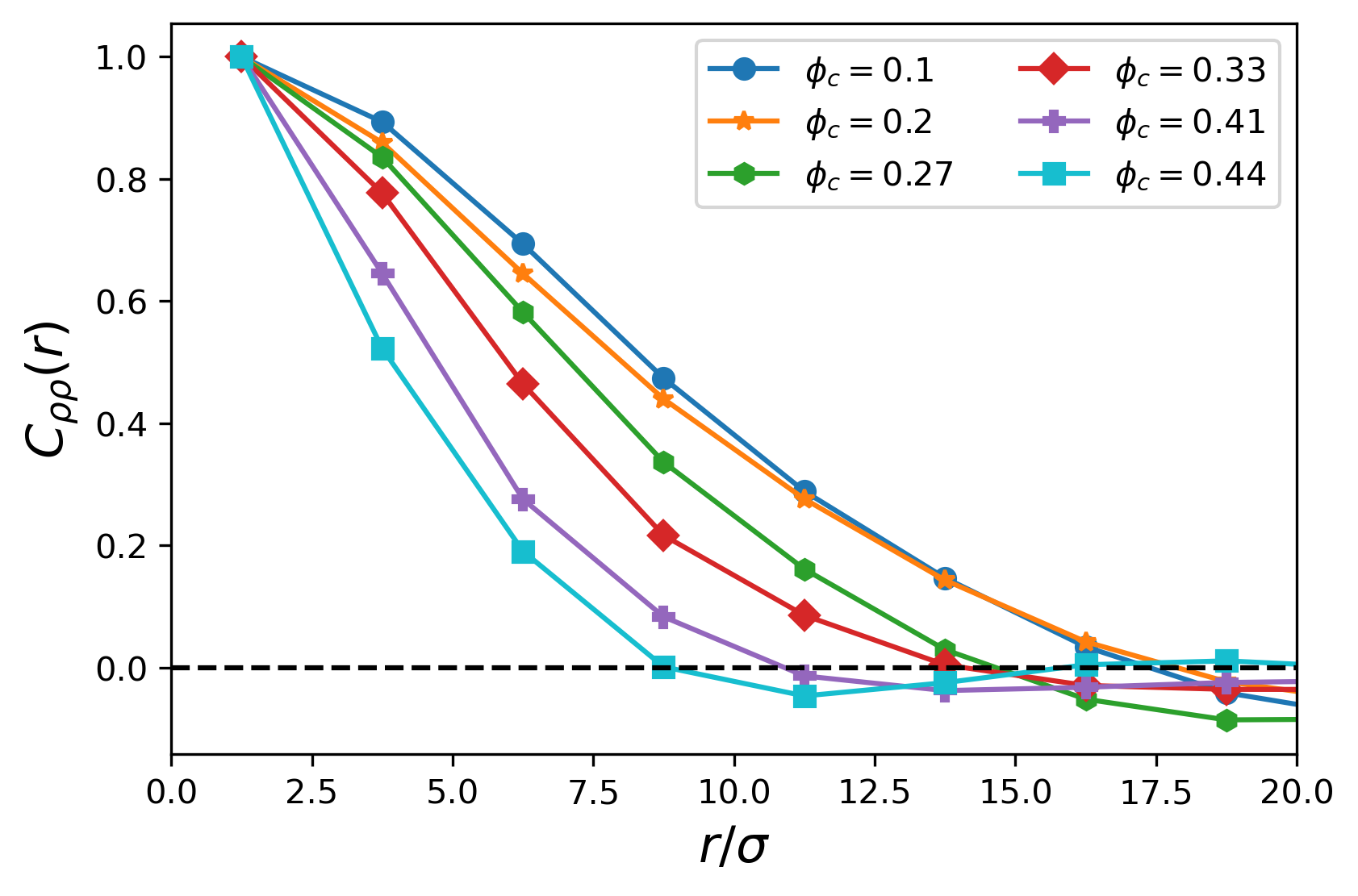}
	}
	
	\subfloat[\label{fig:acfphi}]{\includegraphics[width=0.4\textwidth]{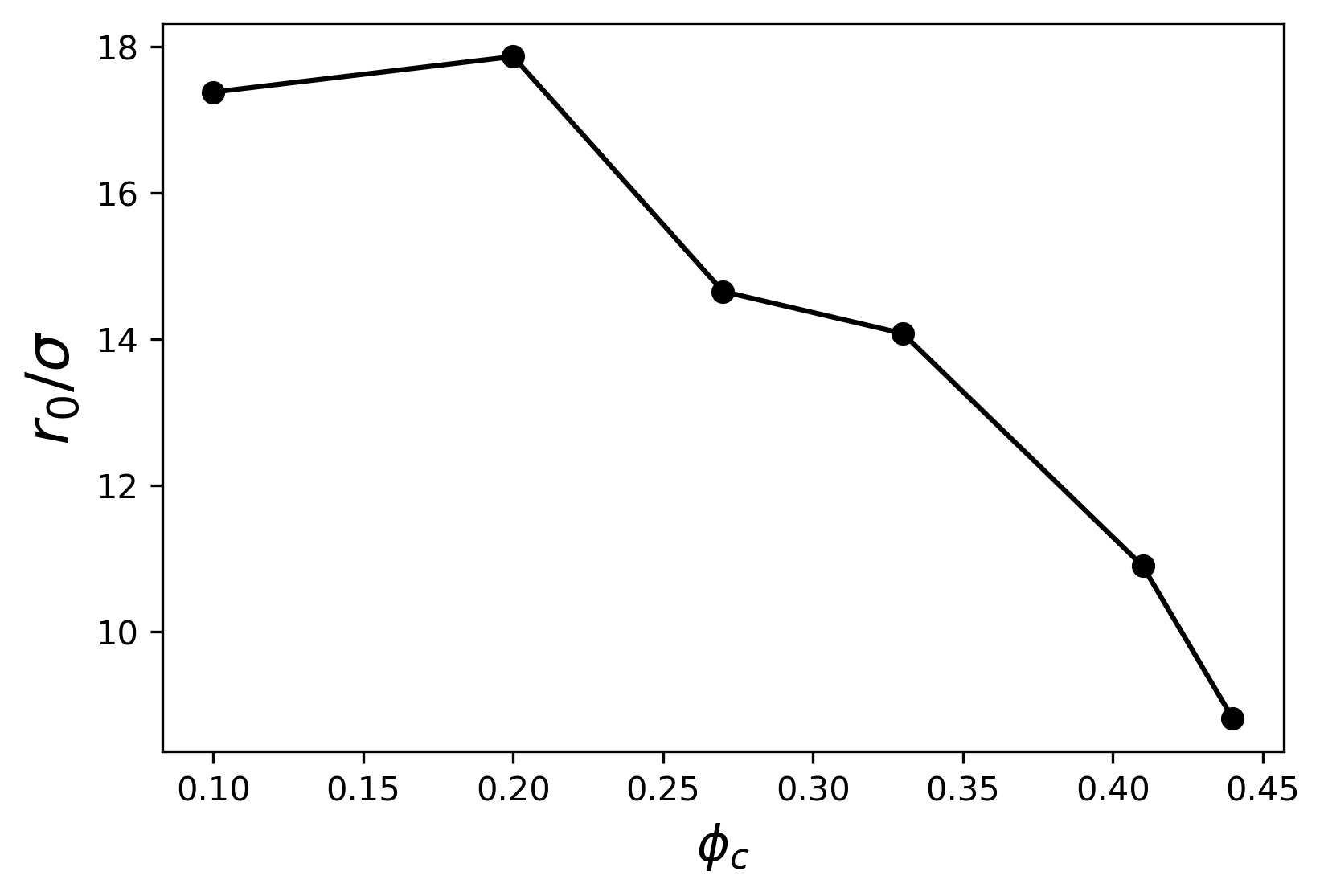}
	}
	\caption{\footnotesize{(a) shows the auto-correlation function of the number density difference $\rho_n$ as a function of radial distance from the origin for selected concentrations while (b) shows the intercept of the auto-correlation function with $C_{\rho \rho}(r) = 0$ axis as a function of concentration.}}
	
	\label{fig:number_density_autocorr}
\end{figure}

To quantitatively investigate the phase separation visually observed in the simulations, we divide the system into voxels and in each voxel we compute an order parameter related to the number density difference defined as
\begin{equation}
	\label{eqn:number_density}
	\rho_{n} = \frac{p_n - r_n}{p_n+r_n}
\end{equation}
where $p_n$ is the number of polymer monomers and $r_n$ the number of rod monomers in the voxel. 

To examine the ordering in the system it is useful to look at various spatial correlation functions.  Most studies examine the radial distribution function $g(r)$ (a mass-density correlation function). Some of the correlation functions we examine are similar, namely the density autocorrelation and rod-rod centre of mass correlation functions, but we will also analyze orientational correlation functions to provide a complete picture of the spatial correlation of the system. First, the auto-correlation function of the density-difference characterizes the distribution of particles inside the simulation box. The auto-correlation function for $\rho_n$ is
\begin{equation}
	C_{\rho \rho}(r) = \frac{\left<\rho_n(0).\rho_n(r)\right> - \left<\rho_n\right>^2}{\left<\rho_n^2\right>}
\end{equation}
and is found using Fast Fourier Transforms, and the Wiener-Khinchin theorem \cite{Mahnke2009}. Fig. \ref{fig:number_density_autocorr}(a) shows the $C_{\rho \rho}(r)$ as a function of radial distance from the reference point ($r=0$). Since we have a periodic boundary condition, we only plot the function for one octant of the simulation box. The 3D distribution obtained from the calculations is mapped onto the radial distance by averaging all the discrete values of the $C_{\rho \rho}(r)$ within distance $r$ and $r + \delta r$,  where $\delta r = 2.5\sigma$, and assigning the mean value to the point at $r$.

At lower concentrations, $C_{\rho\rho}$ drops from 1 at low $r$ and becomes negative beyond a characteristic length $r_0$.  In Fig. \ref{fig:number_density_autocorr}(b), the characteristic length $r_0$ corresponding to the zero-crossing of $C_{\rho\rho}$ as a function of the concentration is illustrated. We observe an overall decrease in $C_{\rho\rho}$ as we increase the concentration of rods which is reflected in $r_0$ going down as concentration goes up. From the definition of $\rho_n$, the decrease at higher concentrations shows that the correlation between the composition of voxels decreases as function of concentration which implies that the polymers and the rods are becoming better mixed in systems with higher rod concentrations. 

\begin{figure}[!htbp]
	\subfloat[\label{fig:dist_between_rods_a}]{	
		\includegraphics[width=0.4\textwidth]{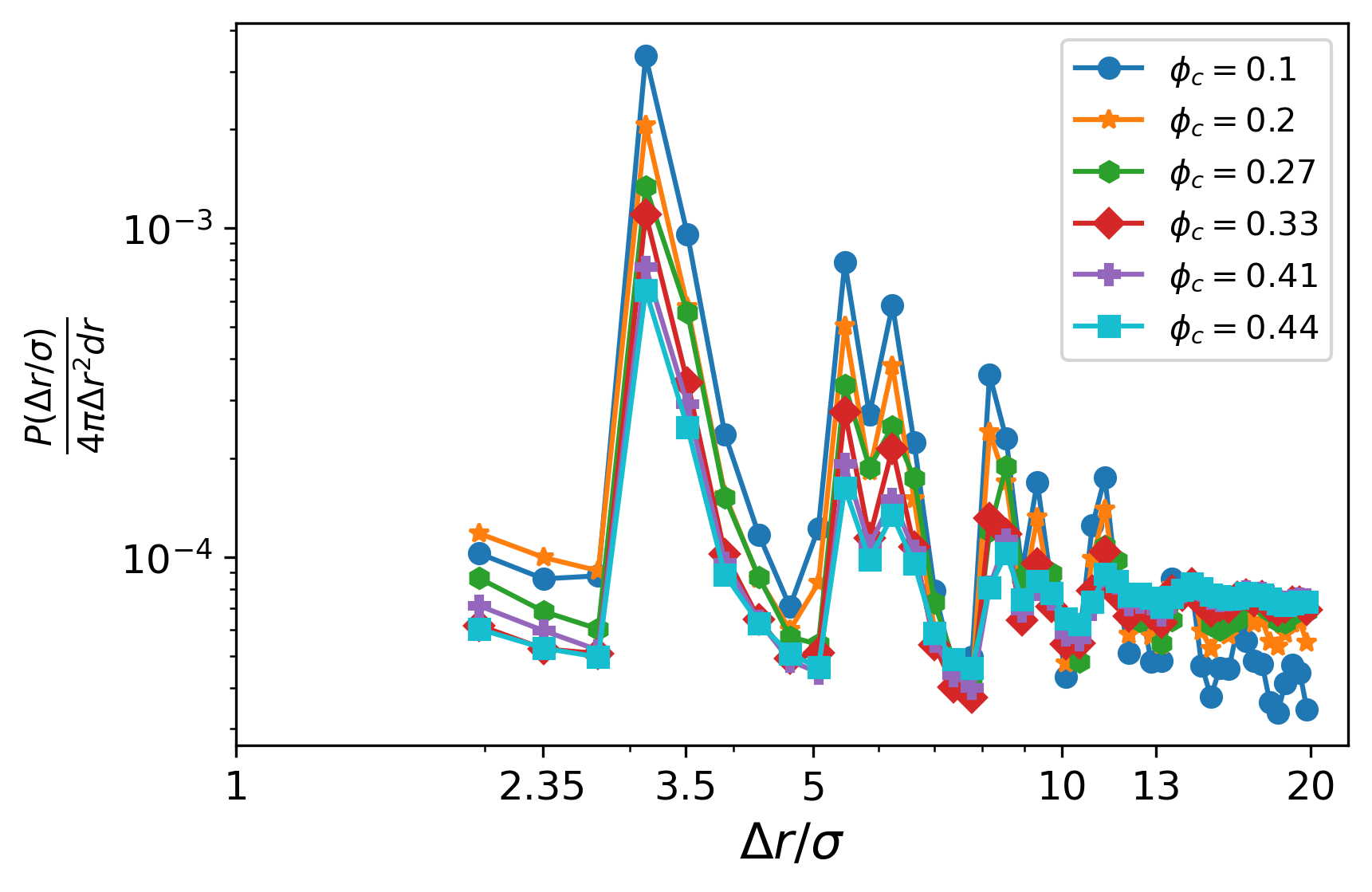}
	}
	
	\subfloat[\label{fig:dist_between_rods_b}]{
		\includegraphics[width = 0.4\textwidth]{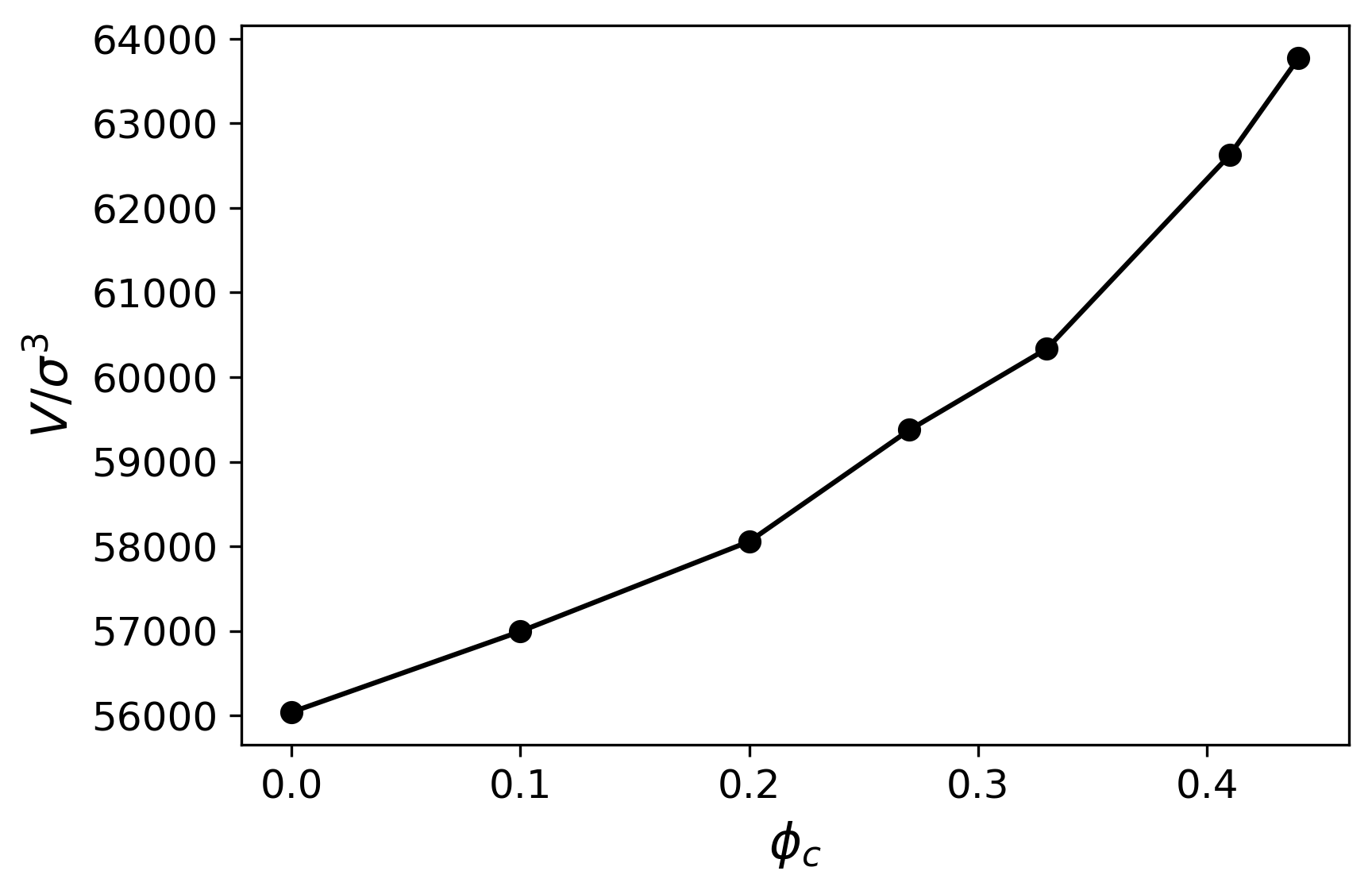}
	}
	\caption{\footnotesize{In (a), probability density function is shown for the
			pairwise distance between the centre of mass of the rigid rods in the melt. (b) shows the average volume of the system as a function of the concentration. The patterns in (a), and (b) suggest that the melt becomes more mixed and less ordered at concentrations higher than $\phi_c \approx 0.25$.}}
	
	\label{fig:dist_between_rods}
\end{figure}

Another important indicator of the structure of the system is the distance of nanorods from each other in the melt. In Fig. \ref{fig:dist_between_rods}(a), we show the probability of finding the centre of mass of the nanorods at a distance $\Delta r$ from each other. By increasing this shell radius to include larger distances, the number of rods within the shell increases just due to the larger volume. Therefore, we normalize the probability by the volume of the shell. Similar to $g(r)$ graphs, peaks in the rod-rod distance plot show spatial order in the system.  As can be seen, the graphs show a first peak around $3.15\sigma$ at all concentrations which corresponds to the distance between neighbouring rods. Since this value is larger than the direct contact distance of two rods, this shows that polymers interpenetrate the space between the rods.  Polymers between the rods were also observed directly in snapshots of the system configurations, such as the one shown in Fig. ~\ref{fig:phi_snapshot}d.

Moreover, in Fig. \ref{fig:dist_between_rods}(a), the peaks slowly diminish as the concentration goes up which implies there is less order at higher concentrations. The system becoming more mixed is also manifested in the average volume of the system. In Fig. \ref{fig:dist_between_rods}(b), we see the average volume versus concentration. The volume initially grows linearly, but beyond $\phi_c \approx 0.25$, as the system becomes less ordered (particularly the rods), the rate of growth increases. This is likely related to the well-known fact that a system of orientationally disordered rods require more volume than orientationally ordered rods at equilibrium.  The orientation of rods and the corresponding order of the system will be discussed in greater depth in the next section.

\begin{figure}[!htbp]
	\subfloat[\label{fig:repulsive_a}]{
		\includegraphics[width=0.45\textwidth]{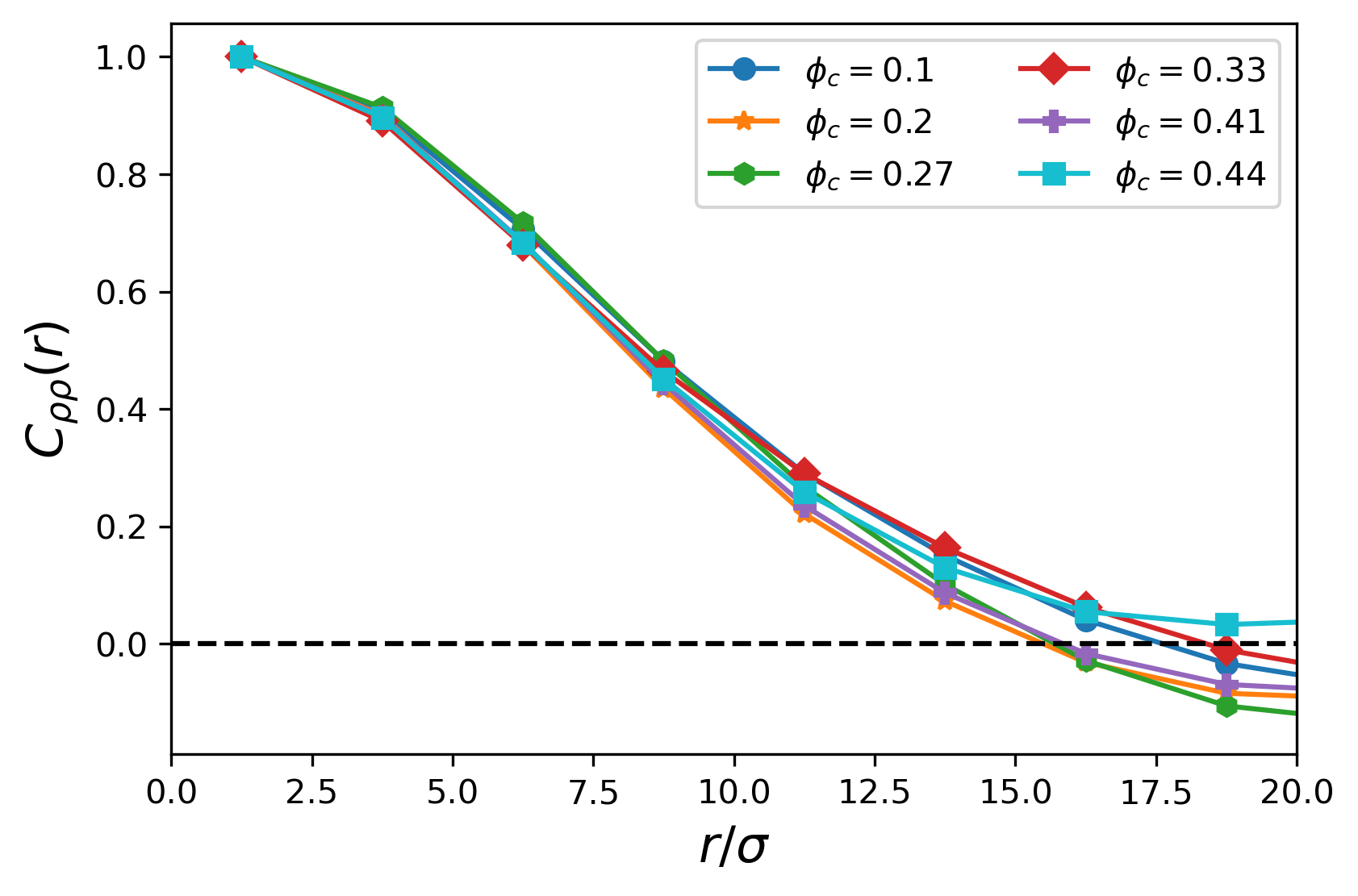}
	}
	
	\subfloat[\label{fig:repulsive_b}]{
		\includegraphics[width=0.45\textwidth]{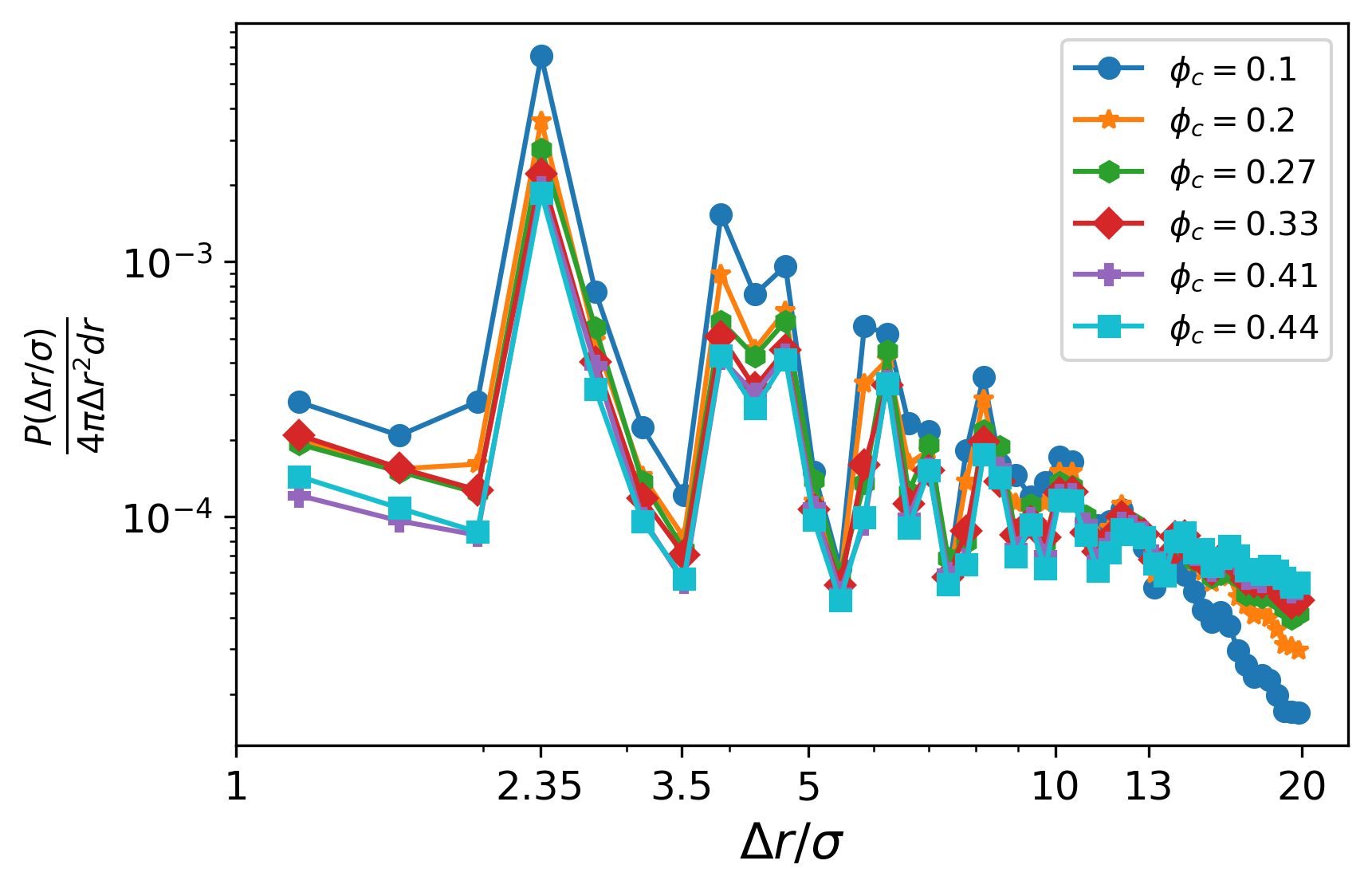}
	}
	
	\subfloat[\label{fig:repulsive_c}]{
		\includegraphics[width=0.38\textwidth]{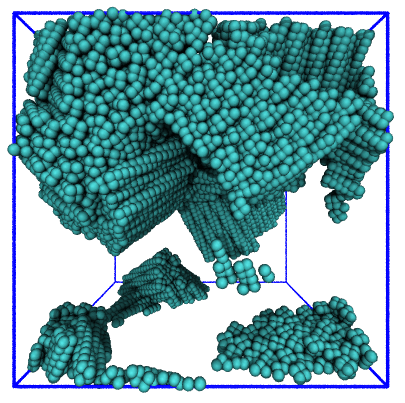}
	}
	\caption{\footnotesize{The auto-correlation function of the number density (a), and the rod-rod centre of mass distance (b) for a system with all-repulsive interactions are illustrated. In (c), a system at $\phi_c = 0.44$ with repulsive interactions between all components is visualized. (a)-(c) show both quantitatively and visually that rods aggregate via direct contact in the fully repulsive system. Regardless of the rod concentration, all the rods phase separate into a single cluster in this case.}}
	\label{fig:repulsive}
\end{figure}

To provide a better picture of the processes responsible for the above results, we compare the result of our system (with rod-polymer attractive interactions) to a system with all repulsive interactions.
Fig. \ref{fig:repulsive}(a) shows the $C_{\rho \rho}(r)$ for a system with repulsive forces between all components. Compared to Fig. \ref{fig:number_density_autocorr}(a), the graphs cross $C_{\rho \rho}(r) = 0$ at larger distances and we do not observe as significant a decrease in the zero-crossing as concentration increases. As a matter of fact, while the correlation curves are first decrease with concentration, they go up again beyond $\phi_c = 0.27$. The slower drop of $C_{\rho \rho}$ suggests that neighbouring voxels contain similar type of atoms. In other words, the rods and polymers are more fully phase separated. This is also observed from direct visualizations. Fig. \ref{fig:repulsive} (c) shows snapshots of an all-repulsive system at a concentration of $\phi_c=0.44$. A full phase separation is observed for the all-repulsive system. 

The lack of any attractive forces implies the phase separation for the system with all purely repulsive forces is entirely driven by entropic effects (excluded volume) similar to depletion forces seen in systems of spherical colloids of two different sizes \cite{Asakura1954, Vrij1977, Biben1996, Striolo2004}. This type of depletion-induced phase separation is also observed in mixtures of nanorods in non- or weakly-adsorbing polymer solutions \cite{Bolhuis1997, Savenko2006, Hall2010, Hu2013a, Sankar2015}.

In contrast, the rods in the corresponding system with attractive rod-polymer interactions are better dispersed in the polymer melt as shown in Fig. \ref{fig:phi_snapshot}(d).  In order to increase their contact surface with the rods, the polymers break up the large clusters of rods into smaller ones as well as penetrate into the space between rods within the clusters. This becomes more evident if we compare the rod-rod centre of mass results for the repulsive and attractive cases. Similar to Fig. \ref{fig:dist_between_rods} (a), Fig. \ref{fig:repulsive}(b) shows the probability density of rod-rod centre of mass distance. The maximum probability happens at a distance close to the diameter of the rods $\Delta r \approx 2.35\sigma$ which means the rods directly touch within a cluster for the repulsive system while for the system with rod-polymer attractions, the first peak happens at a larger distance which suggests that polymers are present between the rods within a cluster.  One might interpret this as polymers gluing the rods together and therefore determine formation of cluster to be energetically driven \cite{Gao2014, Erigi2021}. However, as shown by the $C_{\rho \rho}(r)$ graphs, introducing the rod-polymer attraction splits the larger clusters into smaller ones and does not promote formation of clusters. Although the polymers are present within clusters (polymer-bridged), polymer-bridging does not seem to be the source of aggregation at the specific interaction strengths studied in this work. The details of the effect of the rod-polymer attraction strength on the phase behaviour will be studied in a future work.

The difference in the behaviour of the fully repulsive system and attractive system can be interpreted as follows. The phase separation of the polymers and the rods in the repulsive system is an entropic process and since there is no other processes to compete with, increasing the number of rods does not alter the behaviour of the system significantly. However, in the presence of the rod-polymer attractive interactions, the enthalpic effect that tries to increase the contact surface of rods and polymers competes with the entropic effect pushing the system away from phase separation, and formation of clusters. As a result, increasing the number of rods steers the attractive system towards a more mixed configuration as it boosts the energetic interactions.

Another interesting result illustrated in Fig. \ref{fig:system1_3} is the orientation of rods within the aggregates. As can be seen, the rods within a cluster align laterally and in parallel. In other words, they form a nematic phase within each cluster. The formation of a nematic phase of nanorods in solutions of polymers has been mentioned in the literature \cite{Lekkerkerker1994, Vroege1992} and is the topic of the next section.
\subsection{Orientation and Order}
\subsubsection{Rod clusters}
In the previous section, we observed that the nanorods tend to phase separate into clusters that visually seem to have nematic order. In this section, we further investigate this possible ordering and phase behaviour.  We start with an examination of the orientational order of the rods as a function of their position. An orientational correlation $C_{rr}$ can be defined as
\begin{equation}\label{eqn:crr}
	C_{rr}\left(|\Delta \bm{r}|\right) = \left<|\bm{\hat{e}}^r_{i}(\bm{r}) \cdot \bm{\hat{e}}^r_{j}(\bm{r}+\Delta \bm{r})|\right>
\end{equation}
where $\bm{\hat{e}}^r_{i}$ and $\bm{\hat{e}}^r_{i}$ are the end-to-end vectors of the $i$th and $j$th nanorods and $\Delta r = |\Delta \bm{r}|$ is the distance between the centre of mass of the nanorods. The value of $C_{rr}$ is 1 for a fully orientationally ordered state and 0.5 for an isotropic state.
 
\begin{figure}[!htbp]
	\subfloat[\label{fig:crr_corr_att}]{
		\includegraphics[width=0.45\textwidth]{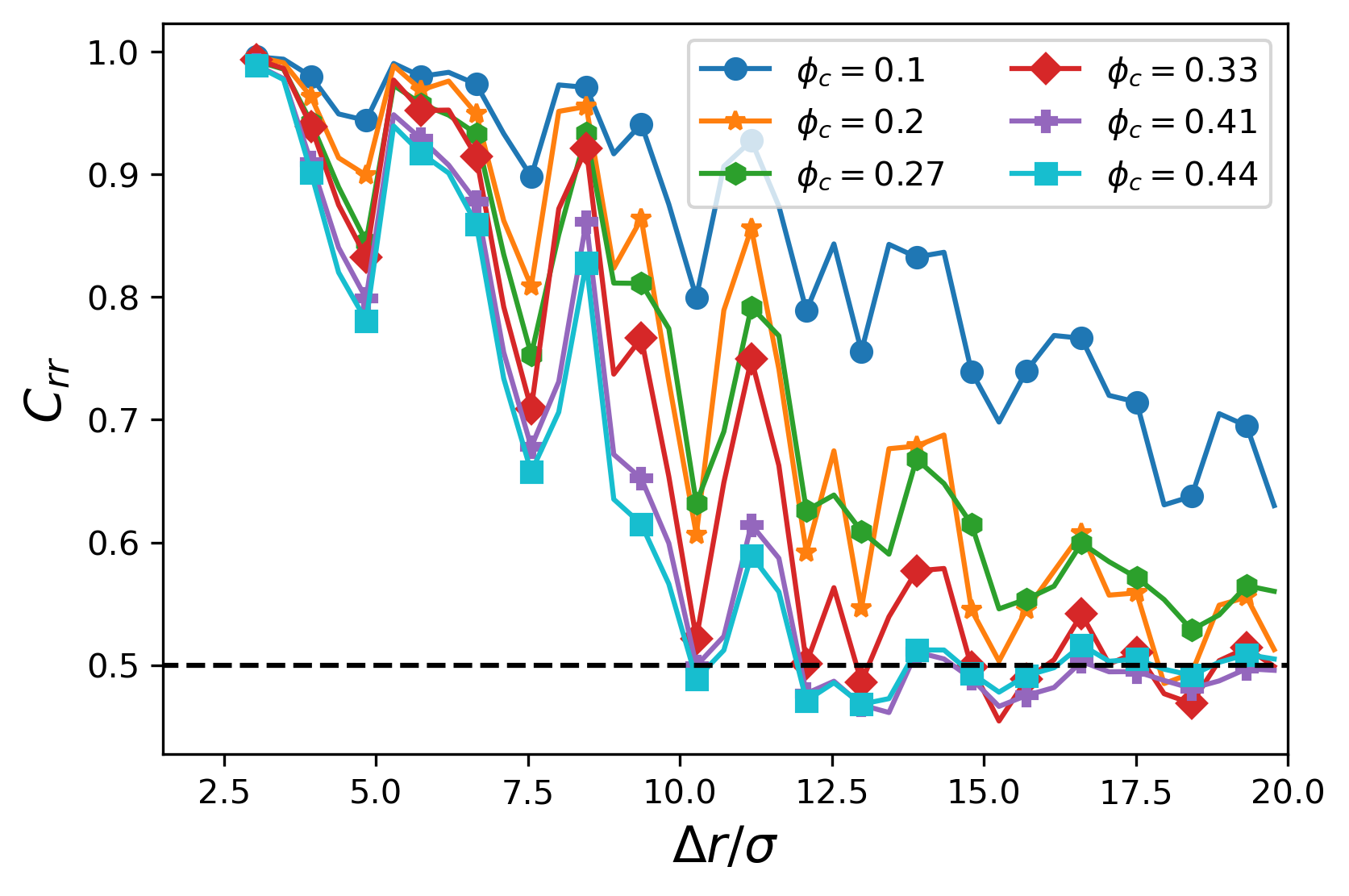}
	}

	\subfloat[\label{fig:crr_corr_rep}]{
		\includegraphics[width=0.45\textwidth]{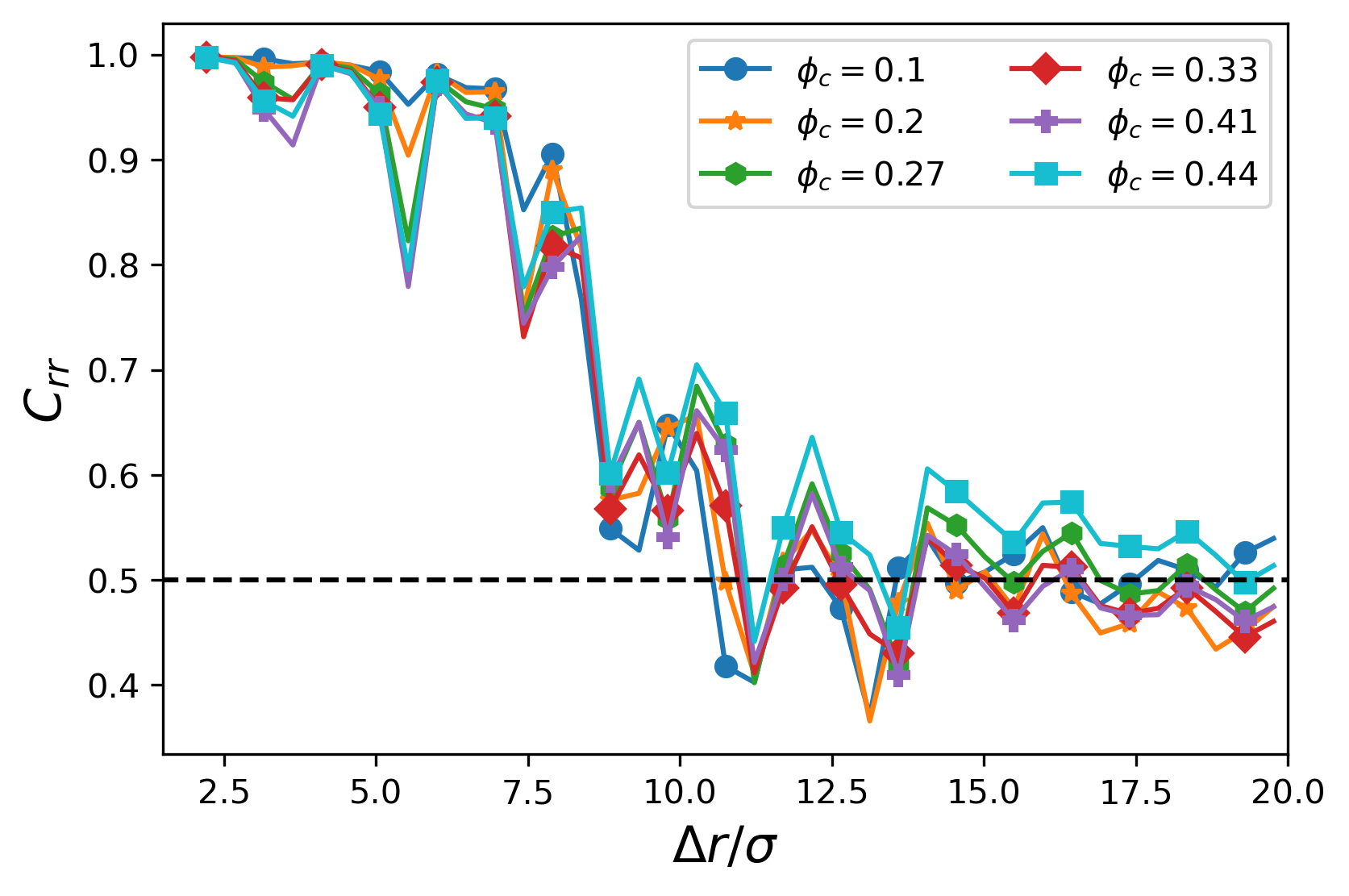}
	}
	\caption{\footnotesize{Orientational correlation between different rigid rod directors. (a) and (b) are the attractive and repulsive cases, respectively. The nearby rods align in the same direction which results in values close to 1 in low $\Delta r$ for both cases, but this value decreases as $\Delta r$ increases more gradually in the attractive case. The concentration seems to have small to nothing impact on the $C_{rr}$ pattern in the all repulsive system.}}
	\label{fig:crr_corr}
\end{figure}

Fig. \ref{fig:crr_corr} shows $C_{rr}$ as a function of $\Delta r$ for a range of concentrations for the attractive system (a) and repulsive system (b). In both systems, the neighbouring rods at short distances are very correlated and $C_{rr}$ takes values near one. However, moving away from a reference rod, the orientational correlation between the rod and other rods fades away and $C_{rr}$ decreases. At lower concentrations, this decay is slower in the presence of the attractive forces while in the all-repulsive system, $C_{rr}$ stays near 1 before a sharp drop at $\Delta r \approx 8\sigma$.  Moreover, we can see a shift towards smaller $C_{rr}$ values as the concentration of rods increases in the attractive system. This is on par with what we have already seen in Fig. \ref{fig:dist_between_rods}: the order of the nanocomposite melt diminishes somewhat as the concentration of the rods increases. In contrast, the concentration doesn't seem to have much effect on the orientational correlation of the all-repulsive system.

\begin{figure}[!htbp]
	\centering
	\includegraphics[width=0.45\textwidth]{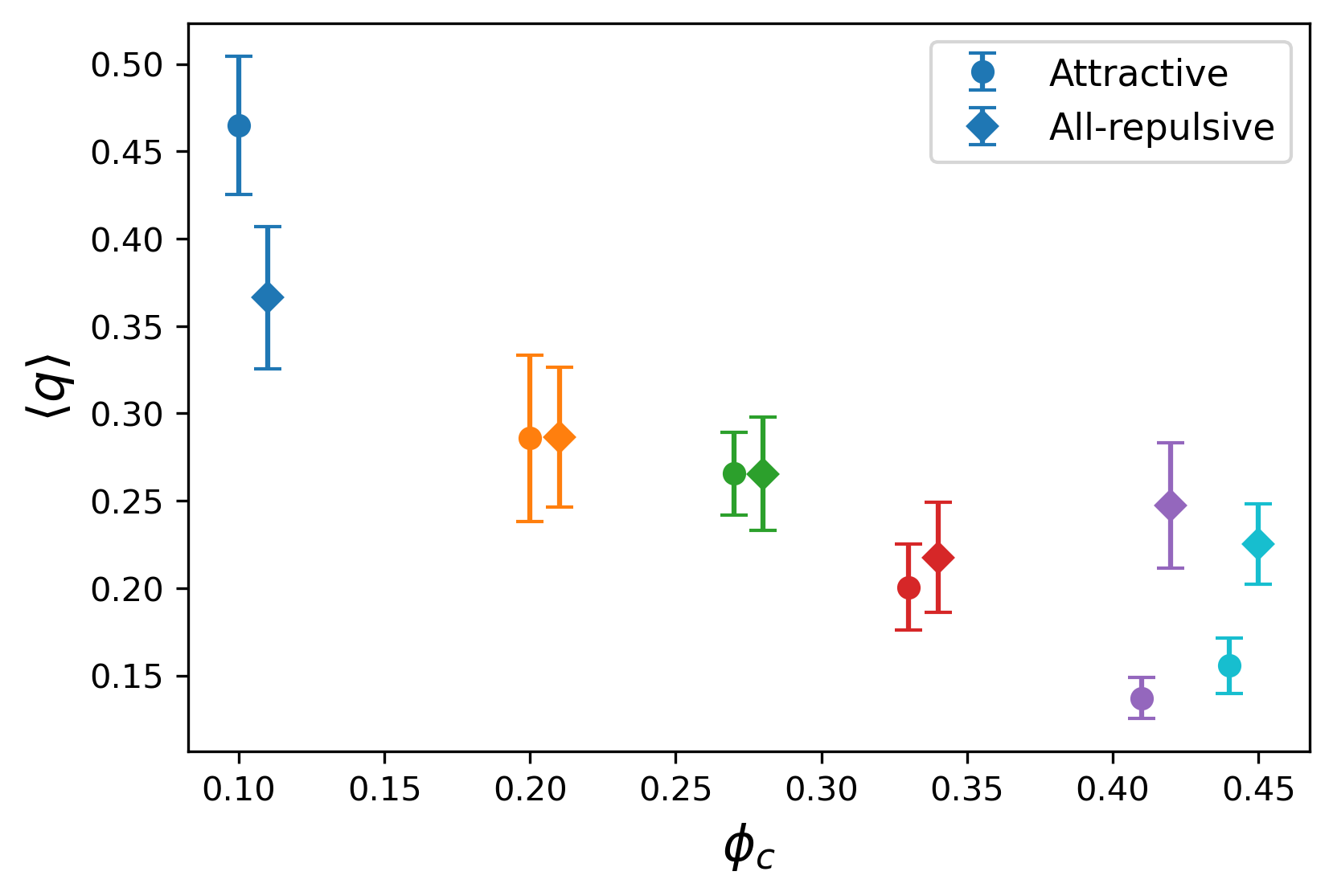}

	\caption{\footnotesize{The order paramter $\left<q\right>$ as a function of rod concentration is depicted. For the attractive system (solid cirles) we see a distinct decreasing pattern in the value of $\left<q\right>$. For the all-repulsive system (diamonds), the decreasing pattern is similarly observed, but it is not as monotonic and distinct.  Values for the all-repulsive system have been shifted right by 0.01 to make them easier to distinguish from the values for attractive system.}}
	\label{fig:order_parameter}
\end{figure}

To further investigate the order of the system, we look at the order parameter tensor $\bm{Q}$ and its eigenvalues. The $\bm{Q}$ is a traceless tensor defined as
\begin{equation}
	Q_{\alpha \beta}  = \left<\hat{e}_{\alpha} \hat{e}_{\beta} - \frac{1}{3}\delta_{\alpha \beta}\right>
\end{equation}
where $\hat{e}$ is the unit vector (along the length) of a rod, $\alpha, \beta = x, y, z$, and the angle brackets denote the expectation value over all rods \cite{Changizrezaei2017}. The eigenvalues of this matrix shows the order along the corresponding eigenvector. The average value of the largest eigenvalue $\left<q\right>$ over several realizations is plotted as a function of concentration in Fig. \ref{fig:order_parameter} where the attractive and repulsive cases are represented by circles and diamonds, respectively. One more time, we see that the overall order of the nanocomposite decreases as a function of concentration. Although this decrease is observed in both (a) and (b), it is more pronounced in the attractive case (a change from lowest to highest concentrations of around 0.3 for the attractive case versus 0.15 for the all-replusive). 

\begin{figure}[!htbp]
	\subfloat[\label{fig:s_r_att}]{
		\includegraphics[width=0.45\textwidth]{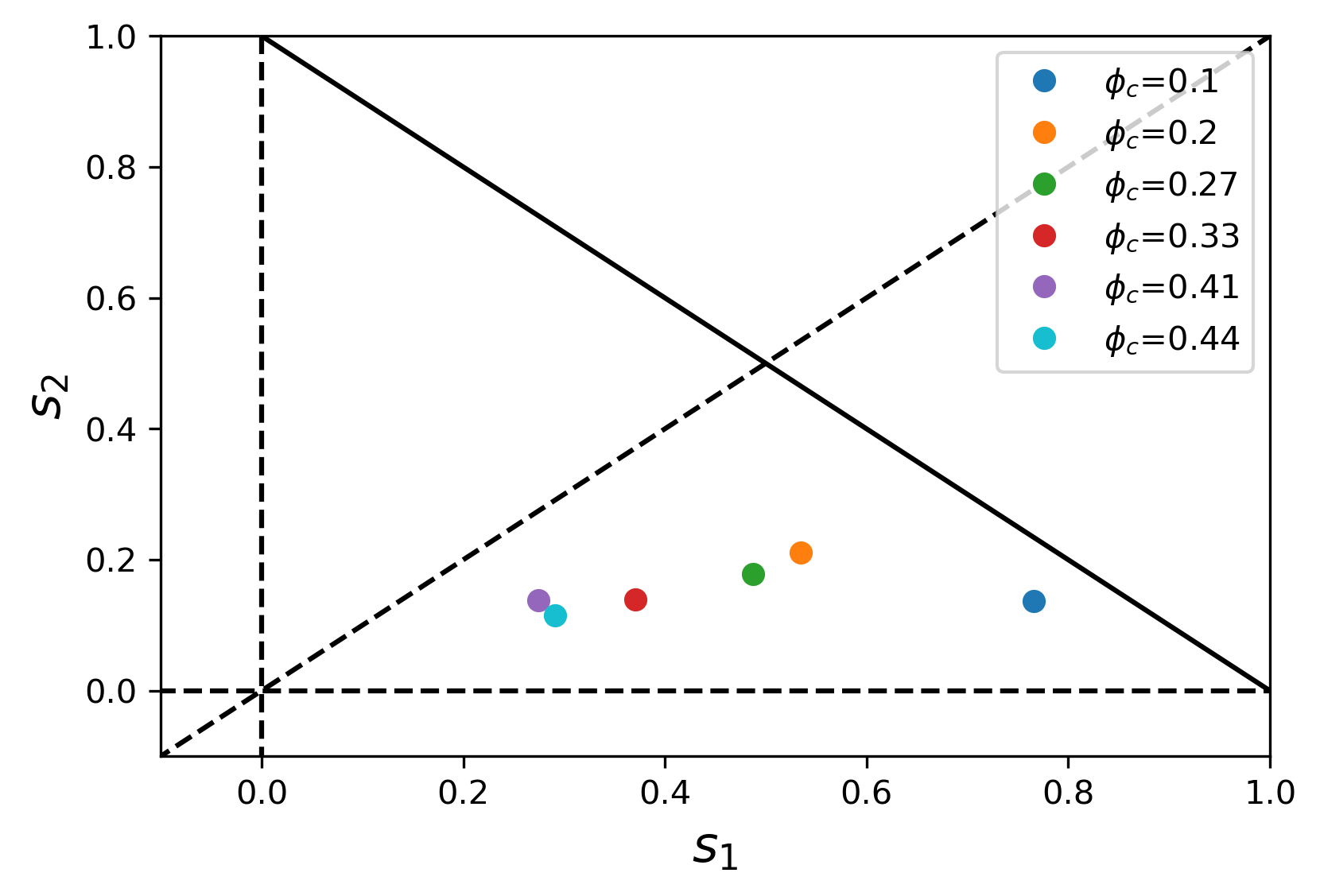}
	}

	\subfloat[\label{fig:s_r_rep}]{
		\includegraphics[width=0.45\textwidth]{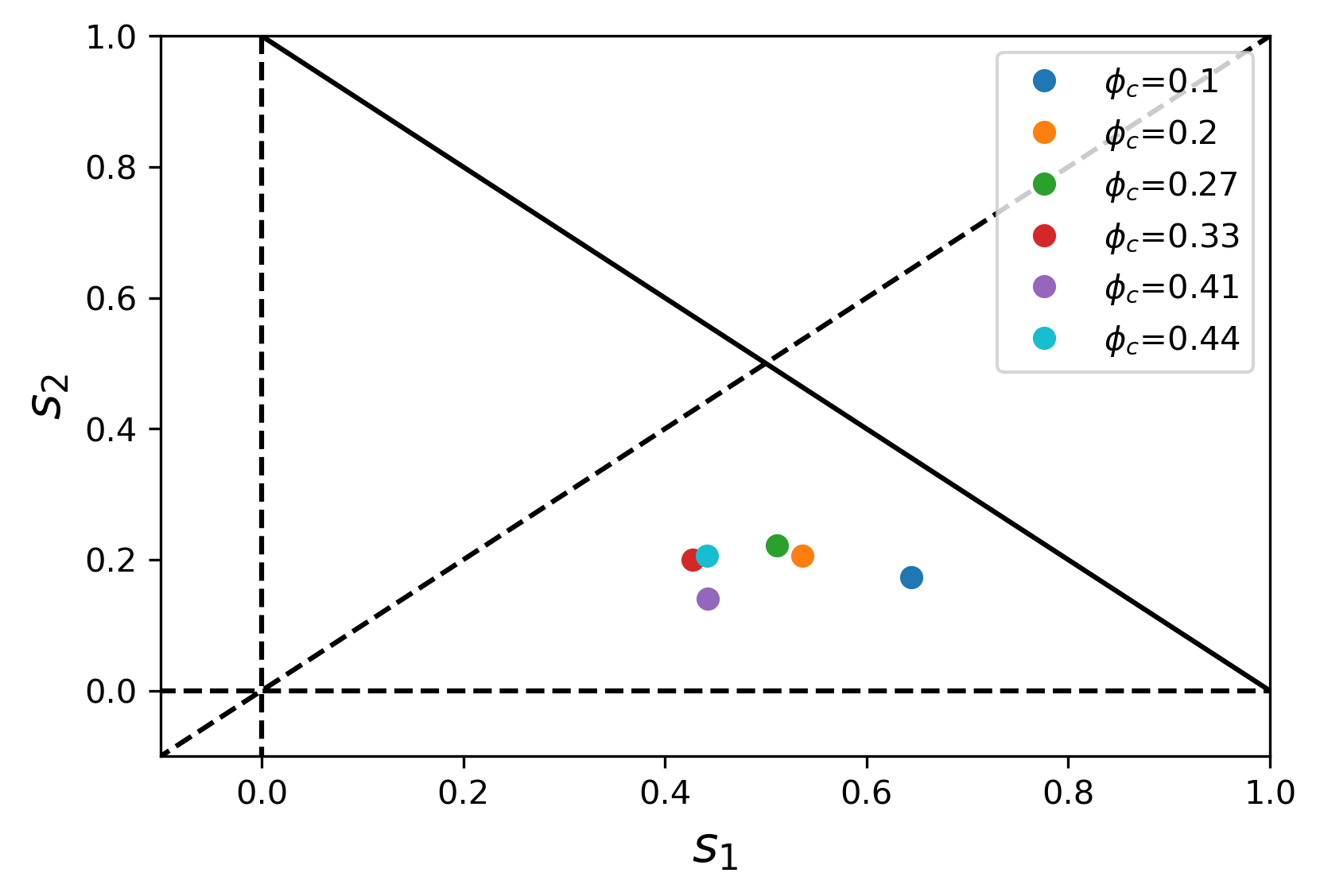}
	}
	\caption{\footnotesize{The eigenvalues of the order parameter $\bm{Q}$ are plotted in $s_1-s_2$ space. (a) shows the system with rod-polymer attrations and (b) shows the all-repulsive system. The isotropic state is at the origin $(0,0)$, the dashed-lines represent uniaxial states, and the rest of the region inside the triangle corresponds to biaxial states.}}
	\label{fig:s_r}
\end{figure}

Although $\left<q\right>$ shows the order along the global director (corresponding eigenvector), it does not provide a lot of information on the uni- or bi-axiality of the system and a look at the other eigenvalues of tensor $Q_{\alpha \beta}$ is necessary. We use a pair of order parameters $(s_1,s_2)$ as defined in \cite{Mottram2014, Majumdar2010}
\begin{align}
	s_1 &= q_1 - q_3\\
	s_2 &= q_2 - q_3
\end{align}
where $q_3\leq q_2 \leq q_1 = q$ are the eigenvalues of the $\bm{Q}$-tensor. Based on definition, the eigenvalues take values on the interval $\left[\frac{-1}{3}, \frac{2}{3}\right]$ which in turn translates to $s_1,s_2\in \left[-1, 1\right]$. However, since we have the condition $q_3\leq q_2 \leq q_1 = q$, all the points lie in the region $s_1,s_2 \in \left[0,1\right]$. Therefore, we only show this region of the $s_1-s_2$ triangle. In Fig. \ref{fig:s_r}, the order parameter pairs are plotted for different concentrations. The origin $(0,0)$ corresponds to the isotropic state, the dashed-lines represent uniaxial states, the rest show biaxial states. The boundaries of the triangle (black solid lines) are physically impossible states. In Fig. \ref{fig:s_r}(a), for the attractive system, we see a monotonic decrease in the value of $s_1$ as the function of concentration and the points move towards the origin as the concentration is increased.  Similar to  $\left<q\right>$, the decrease in $s_1$ as a function of concentration is again about half as much over the range of concentrations studied for the all-repulsive system compared to the system with attractive rod-polymer interactions.  

Comparing Fig. \ref{fig:s_r} (a) and (b), we can also see that overall, the attractive system has a more uniaxial order (the points are closer to the $x$-axis or the dashed-line). This agrees with what we have seen in Fig. \ref{fig:crr_corr} where $C_{rr}$ of the attractive system in Fig. \ref{fig:crr_corr} (a) shows deeper valleys compared to $C_{rr}$ of the all-repulsive case in Fig. \ref{fig:crr_corr} (b).

To summarize, at low concentrations, the system with attractive interactions has more orientational order and correlations over a longer range than the all-repulsive system.  This appears to be due to the presence of ribbon-like configurations of orientationally aligned rods stretching across the system (cf. Fig.~\ref{fig:phi_snapshot}a) in the attractive case whereas the rods are in more of a compact clump in the repulsive case.  At intermediate concentrations, the ribbons break up into clusters which are internally orientationally aligned and with sizes similar to those seen in the repulsive case (cf. Fig.~\ref{fig:crr_corr}). The size of these clusters seems related to that of the rod length (the correlations in Fig.~\ref{fig:crr_corr}b drop at a length scale that is almost exactly half of the rod length). This seems likely to be due to the fact, apparent in the snapshots (cf. Fig.~\ref{fig:phi_snapshot} and Fig.~\ref{fig:repulsive}c), that the rods do not just orient with their neighbour but also form layer-like clusters where the rod centers-of-mass line up in a plane with their long axis oriented normal to the plane.  Within the layer, the rods form close-packed structures. However, the orientation and layering of neighboring clusters are not correlated.  As such, as we add slightly more rods we primarily just add new clusters so the decay of the order parameter at intermediate concentrations is primarily due to averaging over more clusters.  At high concentrations, the orientational order for the all-repulsive system starts to level off, and perhaps even start to increase (cf. Fig.~\ref{fig:order_parameter}).  At the same time, the all-repulsive system starts to pick up some small longer range correlations in orientational order ($C_{rr} > 0.5$ in Fig.\ref{fig:crr_corr}b).  i.e. as we pack in more rods the all-repulsive system starts to gain some true long range liquid crystal-like ordering.  In contrast, the system with attractive interactions becomes more orientationally isotropic as we increase the number of rods.  As we saw in the previous subsection, the attractive system also becomes more mixed as we increase the number of rods whereas the all-repulsive system fully phase separated into a single clump of rods separate from the polymer melt.  This suggests that the interfacial polymers play a role in this difference in behaviour between the attractive and all-repulsive rod-polymer systems.

\subsubsection{\label{interface}Interfacial polymers}
The phase separation of rods, and nematic ordering within the aggregates, has important ramifications for the nanocomposite's mechanical properties. However, another factor that plays a determining role in understanding the mechanics of fracture of polymer-nanorod composites is the behaviour of polymer chains at the polymer-nanorod interface. Unlike nanorod dispersion patterns, the interfacial behaviour has not been studied significantly in the literature which is the motivation for the work in this section. In their recent article, Lu et. al found that the polymer chains near nanorod surfaces take on more extended conformations while the chains far away behave like chains in a pure melt \cite{Lu2021}. Our simulations tell a similar story. 

\begin{figure}[htbp!]
	\subfloat[\label{fig:dist_rg_att}]{
		\includegraphics[width=0.45\textwidth]{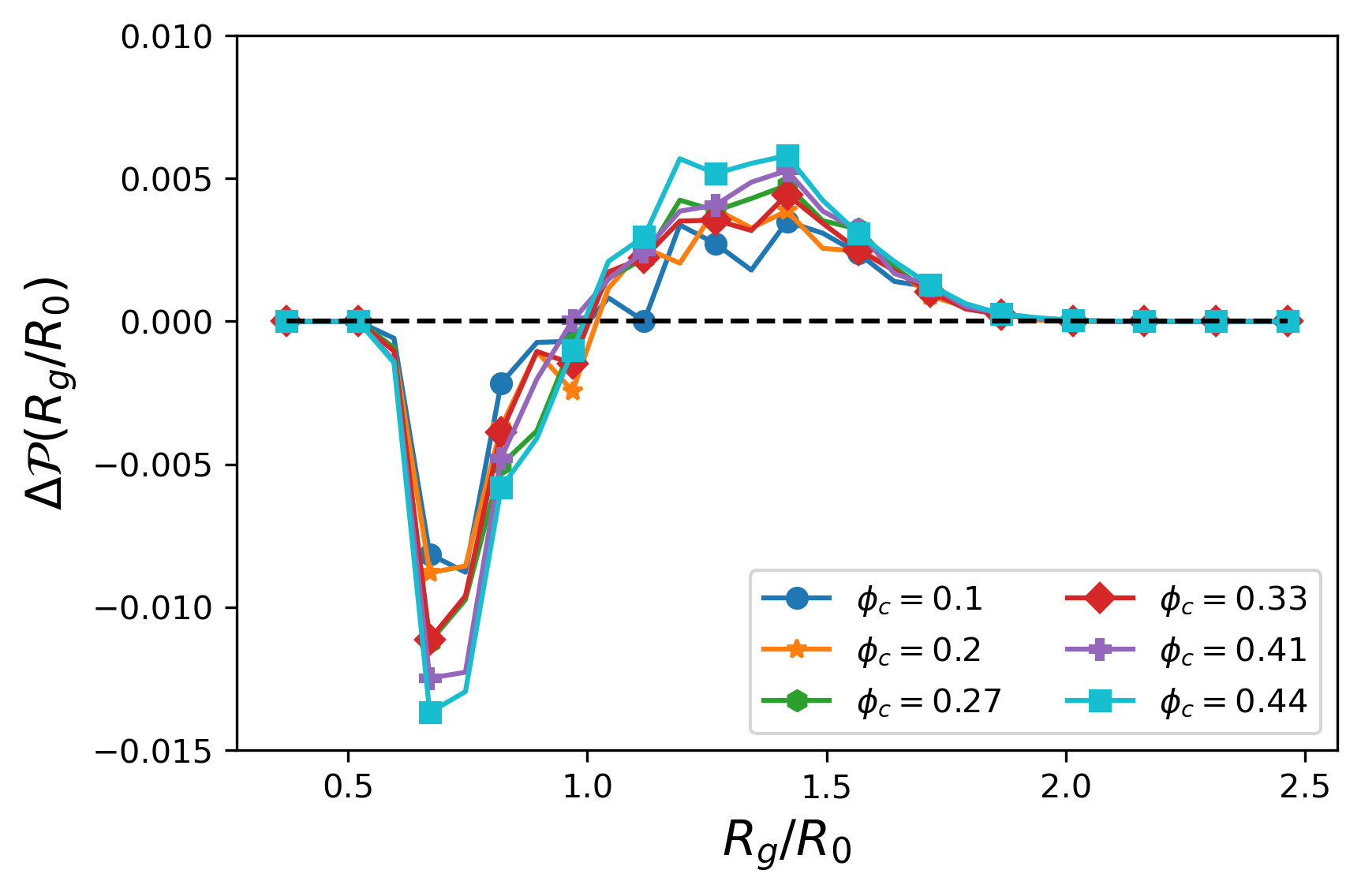}
	}

	\subfloat[\label{fig:dist_rg_rep}]{
		\includegraphics[width=0.45\textwidth]{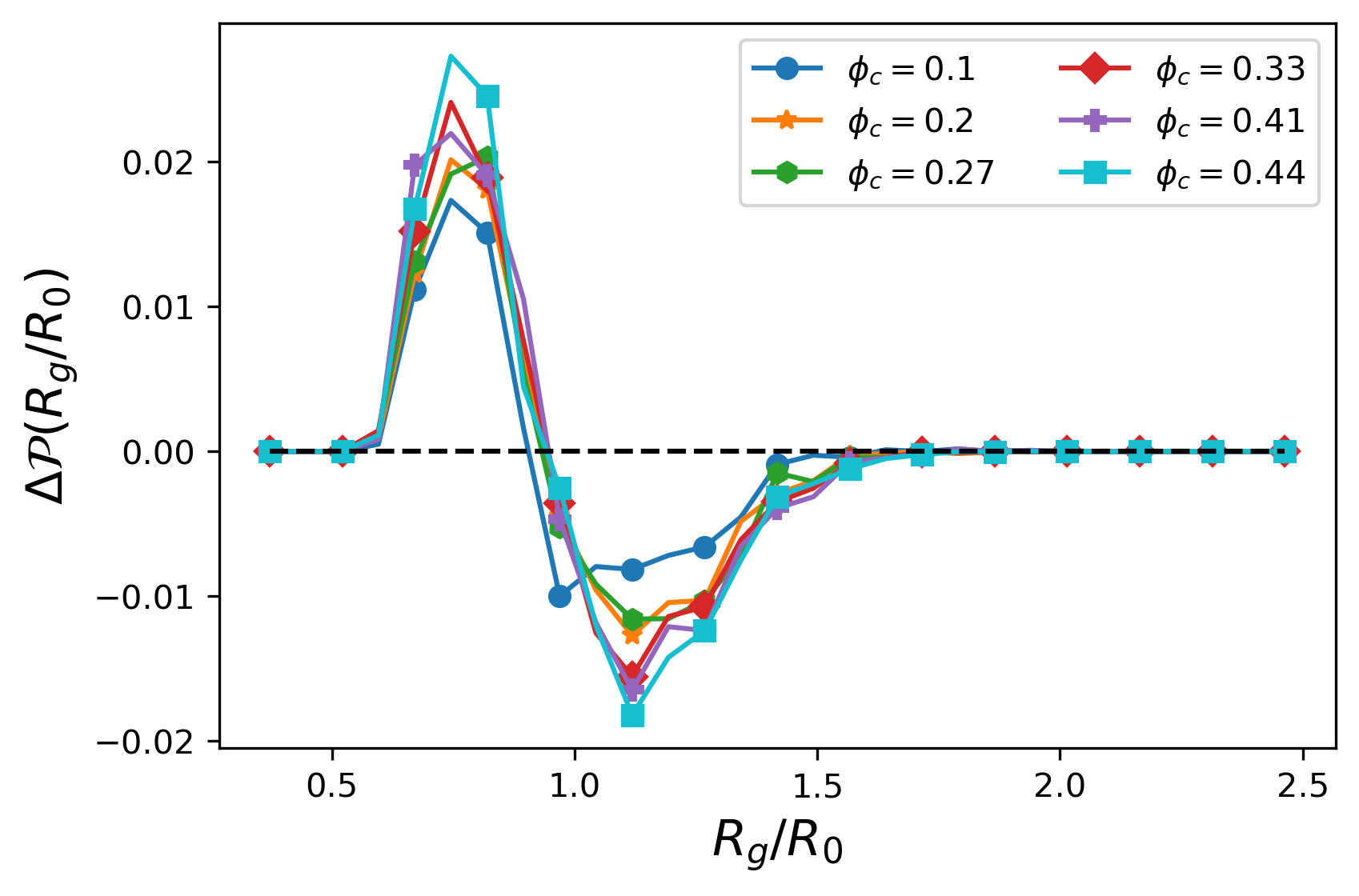}
	}
	\caption{\footnotesize{(a) depicts the difference of probability densities of the normalized radius of gyration of the near polymer chains and the pure melt in the original system while (b) shows the same quantity in the all-repulsive one.}}
	
	\label{fig:dist_rg}
\end{figure}

In a polymer melt, the radius of gyration of individual chains are distributed in a Gaussian distribution with mean $R_0$.  We measure the radius of gyration of the polymer chains within a distance of $5\sigma$ from the surface of rods ("nearby") and compare them to those in a melt (polymers far from the rods are still in the polymer melt phase so are distributed nearly identical to those in a pure melt). The difference in probability density of the radius of gyration for nearby chains from pure melt is shown for all concentrations in Fig. \ref{fig:dist_rg}. If we look at the difference between the probability densities for polymers near rods and the pure melt for the attractive polymer-rod interactions, as shown in Fig. \ref{fig:dist_rg}(a), we see a clear pattern where $\Delta \mathcal{P}$ is negative for $R_G/R_0 < 1$ and positive for $R_G/R_0 > 1$. This can be interpreted as near-rod polymers have fewer compact ($R_G < R_0$) and more expanded ($R_G > R_0$) chains. We also observe a longer tail in Fig. \ref{fig:dist_rg}(b) which means there are polymers that are extended up to twice their pure melt conformation. The system with purely repulsive interactions shows completely opposite behavior (cf. Fig. \ref{fig:dist_rg}b). In this case, near-rod polymers have more compact ($R_G < R_0$) and fewer expanded ($R_G > R_0$) chains.  

Note that while both of these case involve polymers nearby rods, we have seen earlier that in the attractive case these polymers completely interpenetrate the rod clusters whereas in the repulsive case these polymers are at the interfacial surface between a melt-like region of polymers and a single big cluster of rods (with no polymer interpenetration). In other words, in the repulsive case the near-rod polymers are effectively experiencing the effect of a hard wall that they cannot penetrate and so if we think of the polymer configuration as a random walk, when this walk "hits" the wall (formed by the cluster of rods) it is just reflected back into the melt. These reflections result in the observed more compact configurations seen here and in most other polymer melts near hard walls \cite{Aoyagi2001}.  

\begin{figure}[htpb]
	\subfloat[\label{fig:e2e_crp_corr_att}]{
		\includegraphics[width=0.5\textwidth]{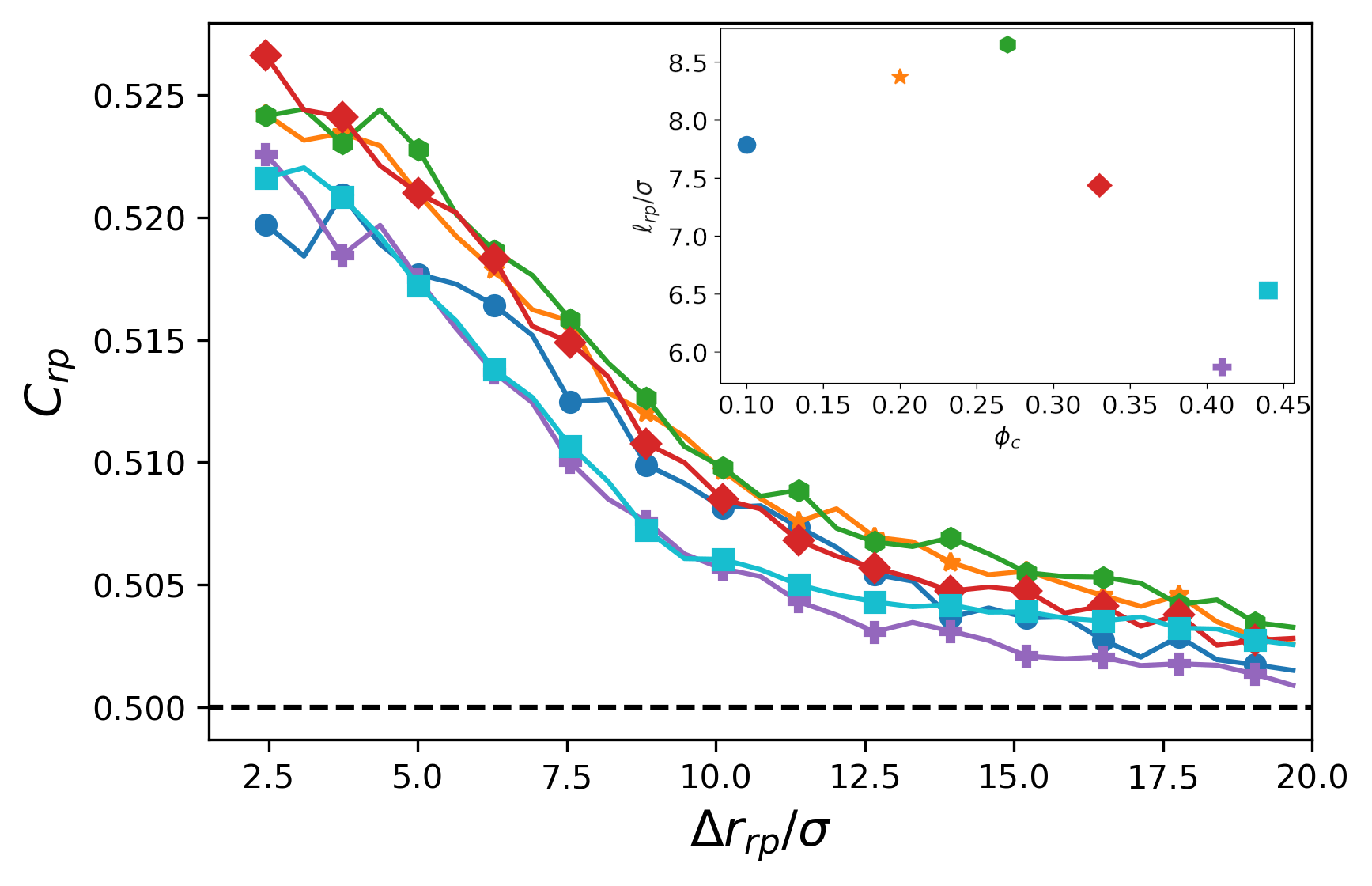}
	}
	
	\subfloat[\label{fig:e2e_crp_corr_rep}]{
		\includegraphics[width=0.5\textwidth]{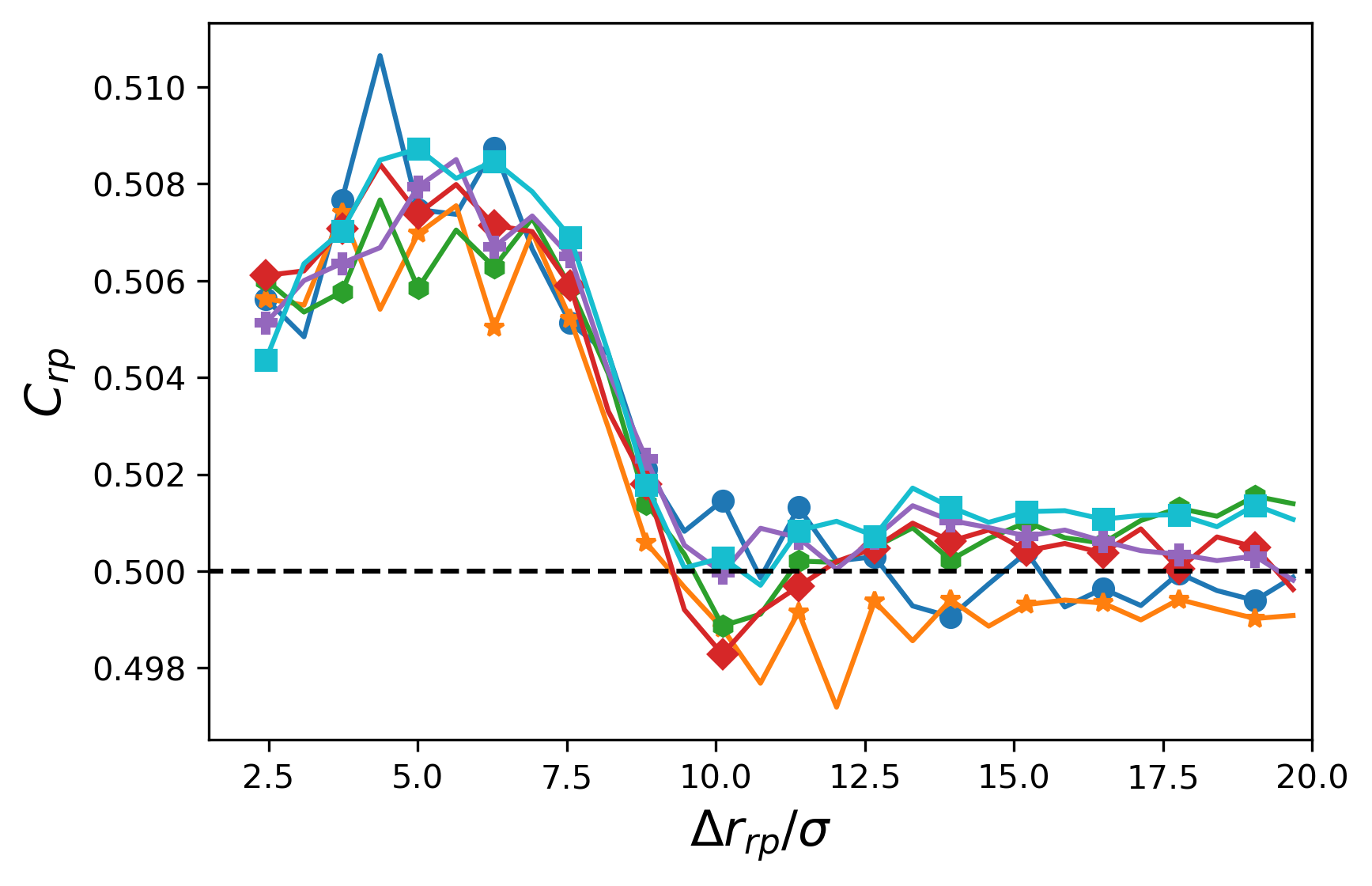}
	}
	\caption{\footnotesize{Orientational correlation $C_{rp}$ between director of rods and polymers' end-to-end vector are shown for (a) the original system and (b) the all-repulsive system. The inset in (a) shows the length scale $\ell_{rp}$ as a function of concentration.}}
	\label{fig:e2e_crp_corr}
\end{figure}

By contrast, the presence of the rods in the melt results in stretching of polymer chains for the attractive system.  However, we have not yet addressed the direction in which the polymers stretch. The chains can extend along the length of the rods or perpendicular to the direction of clusters. Therefore it is interesting to see if there is also some kind of orientational correlation between the rods and polymers. To measure this, we defined a rod-polymer correlation function such that
\begin{equation}\label{eqn:crp}
	C_{rp}\left(|\Delta \bm{r}|\right) = \left<|\bm{\hat{e}}^r_{i}(\bm{r}) \cdot \bm{\hat{e}}^p_{j}(\bm{r}+\Delta \bm{r})|\right>
\end{equation}
where $\bm{\hat{e}}^r_{i}$ and $\bm{\hat{e}}^p_{j}$ are the director of the $i$th rod and the end-to-end vector of the $j$th polymer and $\Delta r = |\Delta \bm{r}|$ is the distance between the centre of mass of the rod and polymer. Fig. \ref{fig:e2e_crp_corr} shows the values of this correlation parameter at different concentrations. 

As can be seen, for all concentrations in the attractive case shown in Fig. \ref{fig:e2e_crp_corr}a, polymers near rods are most elongated along the rod and the correlation decays as the distance increases. An interesting feature of this plot is the behaviour of $C_{rp}$ as a function of concentration. The overall correlation goes up as the concentration increases up to $\phi_c \approx 0.3$ and then declines. By fitting an exponential, we have defined a length scale $\ell_{rp}$ which is plotted in the inset as a function of concentration. The  $\ell_{rp}$ inclines at the beginning which corresponds to growth of cluster size and reaches a peak around $\phi_c = 0.3$. This agrees with what we have already seen in the previous sections. The initial rise in the number of rods results in larger ordered clusters where polymers sneak in between the rods and stretch along the director of the cluster. However, further increase in the number of rods leads to a less ordered system, particularly in the rod orientations, with an abundance of rod surfaces for the polymers to interact with. Therefore, the polymer chains show less preference to align with any specific rod.  

The correlations for the repulsive case, Fig. \ref{fig:e2e_crp_corr}b, are much lower in magnitude and are shorter range, dropping sharply to $0.5$ at fixed distance similar to what is seen for the rod-rod correlations in Fig.~\ref{fig:crr_corr_rep}. As noted above, these polymers are at the surface of a rod cluster that, due to the rod-rod correlations, forms a corrugated surface.  It would appear that these polymers have a small tendency to follow these corrugations which results in the observed $C_{rp}$ correlations, which just die off at distances where the rods that form these corrugations are no longer correlated.

\begin{figure}[htpb]
	\subfloat[\label{fig:cpp_corr_att}]{
		\includegraphics[width=0.4\textwidth]{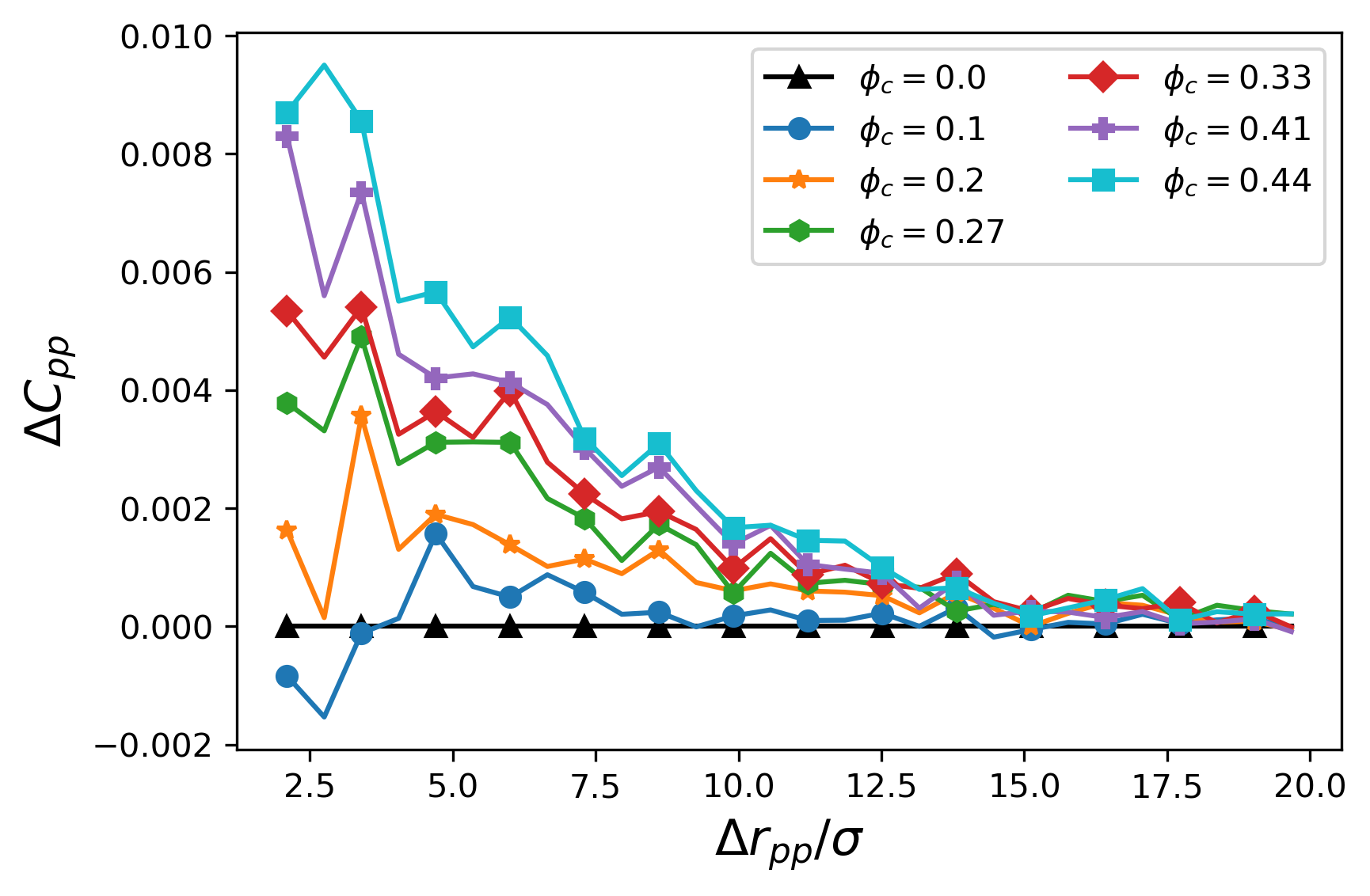}
	}

	\subfloat[\label{fig:cpp_corr_scaled}]{
		\includegraphics[width=0.4\textwidth]{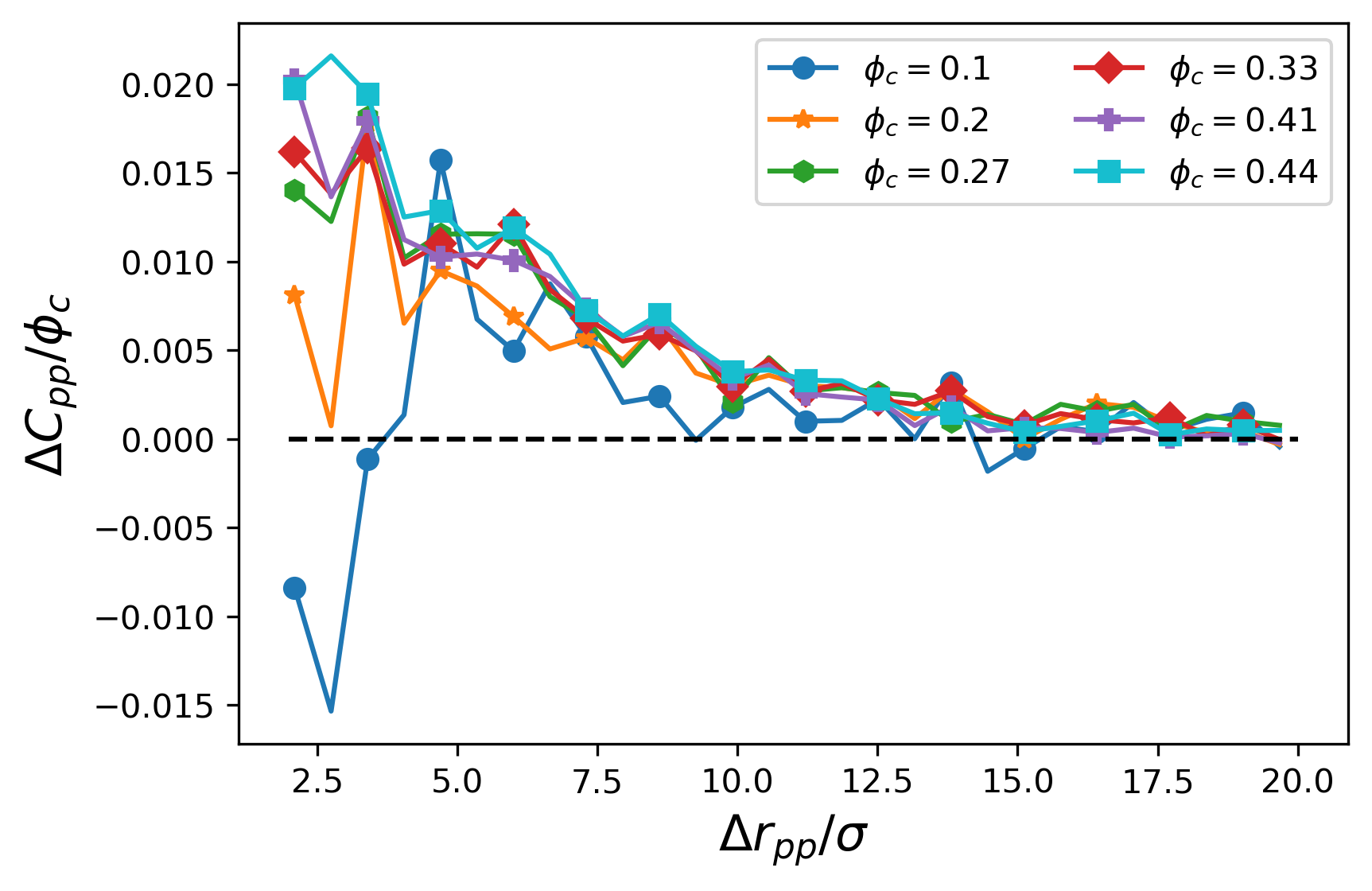}
	}

	\subfloat[\label{fig:cpp_corr_rep}]{
	\includegraphics[width=0.4\textwidth]{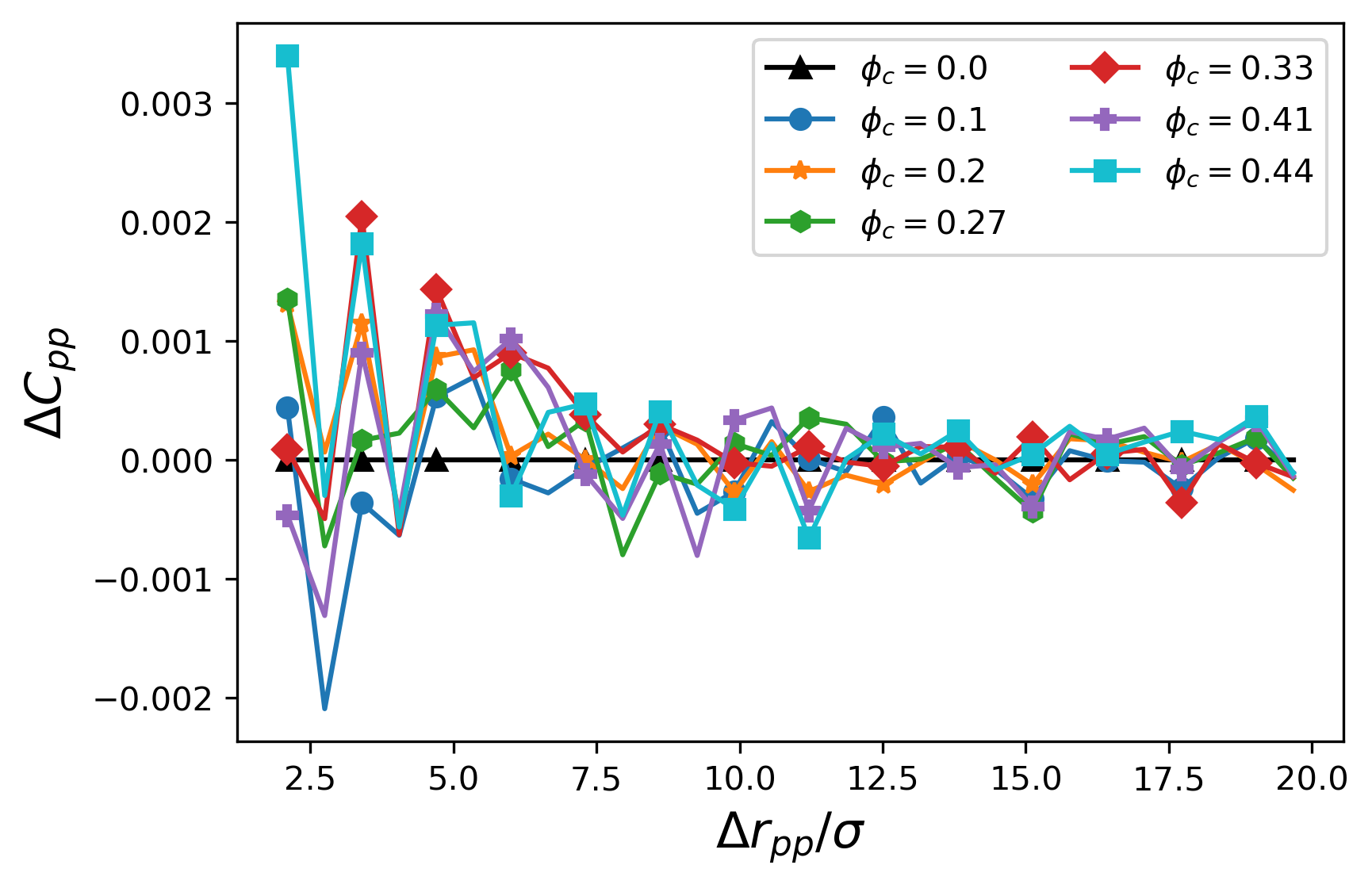}
	}
	\caption{\footnotesize{The relative orientational correlation parameter $\Delta C_{pp} = C_{pp} - C_{pp_{melt}}$ as a function of the polymer-polymer pairwise distance $\Delta r_{pp}$. (a) shows the relative orientational correlation $\Delta C_{pp}$ for the original system while (b) shows the scaled relative orientational correlation $\Delta C_{pp}/\phi_c$ for the same system. (c) shows $\Delta C_{pp}$ for the all-repulsive system (not scaled).}}
	\label{fig:cpp_corr}
\end{figure}

Lastly in this section, we look at the orientation of polymers with respect to each other by introducing a polymer-polymer segmental correlation function like the ones defined for rod-rod and rod-polymer
\begin{equation}\label{eqn:cpp}
	C_{pp}\left(|\Delta \bm{r}|\right) = \left<|\bm{\hat{e}}^p_{i}(\bm{r}) \cdot \bm{\hat{e}}^p_{j}(\bm{r}+\Delta \bm{r})|\right>
\end{equation}
where $e_i$ and $e_j$ are the end-to-end vectors of the $i$th and $j$th polymer segments and $\Delta r = |\Delta \bm{r}|$ is the distance between the centre of mass of the segments. Here, we show the results for segments of length 16 beads (half of a polymer length). In Fig. \ref{fig:cpp_corr}(a), the relative orientational correlation, $\Delta C_{pp} = C_{pp} - C_{pp_{melt}}$, is shown for the attractive system.  While this is typically quite small implying there is only a slight tendency for polymers to align, it is clearly nonzero. At lower concentrations, the behaviour is very close to the melt. However, as more rods are added to the system, $\Delta C_{pp}$ increases and the polymers are more orientationally correlated. This is because the polymers that are aligned with a certain rod are aligned with each other. This becomes even more obvious when $\Delta C_{pp}$ is scaled by the concentration $\phi_c$ as illustrated in Fig. \ref{fig:cpp_corr}(b). As can be seen, the scaled correlation functions collapse implying that the increase in the correlation is a direct result of increase in the number of rods.

In contrast, for the all-repulsive system shown in Fig. \ref{fig:cpp_corr}(c), the increase in the number of rods does not affect $\Delta C_{pp}$ by much and its value only fluctuates around zero. This agrees with what we have seen so far: the rods in the all-repulsive system aggregate into a big cluster with no polymers in between, i.e. direct contact. Therefore, due to lack of contact with the polymers, they cannot alter the conformation of the polymers. This being said, we see some positive values at smaller distances. This is attributed to the polymers at the surface of the large cluster.

An interesting comparison can be made to work by Gorkunov and Osipov \cite{Gorkunov2011} who looked at adding nanoparticles into a liquid crystal matrix.  They found that adding isotropic nanoparticles into the liquid crystal dilutes the liquid crystal and lowers order of the system and consequently, the nematic-isotropic transition temperature. On the other hand, they found that anisotropic particles mimic their nematic host, aligning with the liquid crystal, and as a result improving the nematic ordering of the system. We see a similar effect here:  
In the absence of polymer-nanorod attractive forces, the polymer chains form random walk blobs (see Fig.\ref{fig:dist_rg_rep}) and effectively, play the role of the isotropic nanoparticles in a pool of nanorods. However, when the attractive interactions are present, the polymers attempt to increase their contact surface with the rods. As a result, they take on more elongated conformations (see Fig.\ref{fig:dist_rg_att}) and like anisotropic nanoparticles, they reinforce the nematic ordering of the nanorods. This is reflected in the difference between $\langle q\rangle$ for attractive and repulsive cases in Fig. \ref{fig:order_parameter}. 

There is one caveat in this comparison. Single molecule nanoparticles only have degrees of freedom associated with translation and rotation, but polymer chains have significantly more degrees of freedom (they can change shape). Although aligning with the rods can be favourable, the polymer pays the cost in loss of conformational entropy. At lower concentrations, the population of elongated chains is low enough that the ordering effects can easily compensate for it. However, by adding considerable number of rods, the conformational entropy decreases deeply and drives the system towards a less ordered configuration. As a result, we see a change of trend as the concentration goes up and $\langle q\rangle$ of the attractive system becomes less than that of the all-repulsive system for high $\phi_c$. 	

\section{\label{sec:level4} Discussion and Conclusions}
The dispersion and orientation patterns of nanorods in polymer melts with either attractive or repulsive rod-polymer interactions has been examined as a function of the rod concentration.  We see three competing effects: i) entropy of the rods; ii) energetic interactions between the rods and polymers and; iii) entropy of the polymers.  The all-repulsive interaction system completely phase separates at all rod concentrations whereas the system with attractive polymer-rod interactions does not.  This strongly implicates the role of entropy, in the form of the depletion effect (related to the free volume per particle), as the main driver of phase separation in the system. 

The attractive polymer-rod interactions set up a competition between entropic and enthalpic effects (as this is a constant pressure system it is more appropriate to discuss in terms of enthalpy than energy).
Since the strength of the enthalpic effects are proportional to the number of rods in the system, the dispersion patterns show direct correlation with the rod concentration. At lower concentrations, entropic processes are dominant and ordered clusters of nanorods are created. However, due to the presence of the attractive forces, the rods do not completely phase separate and polymers interpenetrate between the rods of a cluster. At higher concentrations, the energetic effects become significant and the dispersion of the rods improves with the overall cluster size diminishing with concentration.  The polymer interpenetration between the rods is typically referred to as "polymer bridging" and, at least at very strong polymer-rod interactions, is often argued to create an effective rod-rod attraction leading to the formation of rod clusters.  As mentioned above, this does not appear to be the case here as the phase separation seems entirely entropically driven at the strength of polymer-rod interaction we have studied here. 

The orientational ordering of the rods also appears to be strongly affected by a competition between entropy of the rods and entropy of the polymers.  In all systems, the global orientational order of the rods decreases as the concentration of rods increases.  However, In the system with repulsive rod-polymer interactions at low concentrations there is no long-range correlations in the rod orientations implying that the observed system averaged orientational order is primarily a function of having a limited number of oriented clusters and as we add more rods to the system the number of uncorrelated clusters increases hence lowering the average orientional order.  However, at the highest concentrations of rods we do start to see weak long-range correlations in the rod orientations implying true orientional order may start to set in at these concentrations. 
The polymers in the repulsive interaction system show no orientational order and are somewhat compacted at the boundary between the rod phase and polymer melt phase.  Such compaction is consistent with the depletion effect being responsible for the full phase separation seen in that system.   as the concentraion of rods increases, attractive rod-polymer interactions, the long range orientional order of the rods decreases as the concentration of rods increases.  In contrast, the system with attractive rod-polymer interactions shows true long range orientational order at low concentrations of rods.  As the concentration of rods increases these clusters break up and the orientional order decreases to the point where, at the highest concentrations, there is no long range correlations between rod orientations.  In the system with attractive interactions the polymers close to and interpenetrating rod clusters are streched out and oriented along the rods.  As a result, increased rod ordering, while entropically favorable for the rods, would result in a considerable loss of entropy for the polymers in the system which prevents the long range orientional order at high rod concentrations.  This competition between the entropy of the rods and that of the polymers seems to be a under appreciated facet of these systems.

\begin{acknowledgments}
This work was supported by the Natural Science and Engineering Council of Canada (NSERC) Discovery grant (VB and CD) and a Create grant (NA) on Advanced Polymer Composite Materials and Technologies.  Computational resources were provided by the Shared Hierarchical Academic Research Computing Network (SHARCNET) and the Digital Research Alliance of Canada.
\end{acknowledgments}
\appendix
\section{Equilibration}\label{app1}
\begin{figure*}[!htbp]
	\includegraphics[width=\textwidth]{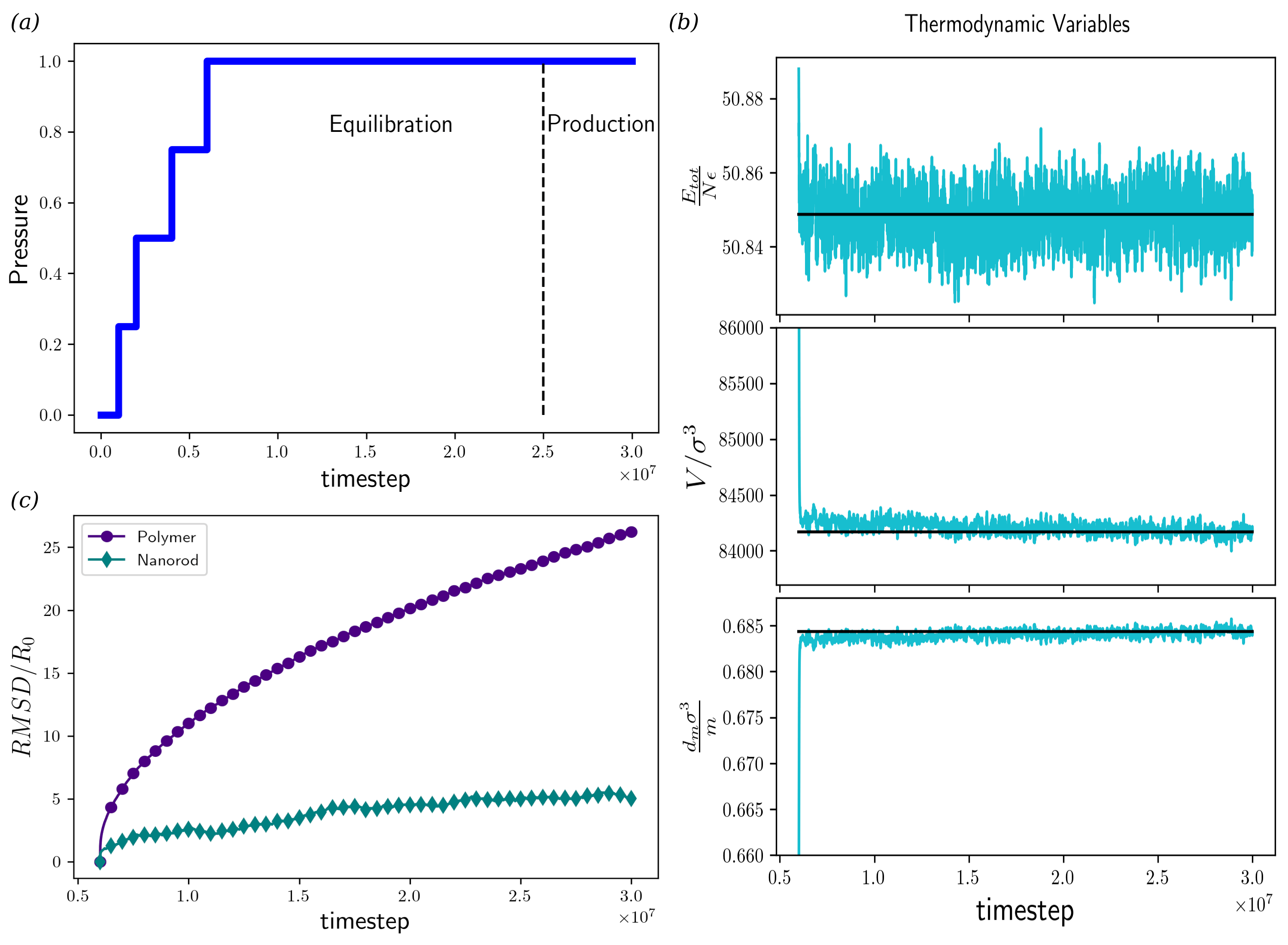}
	
	\caption{\footnotesize{Equilibration procedure is schematically shown in (a). Thermodynamic parameters (b) and RMSD (c) are shown for the system at $\phi_c = 0.44$ and all-repulsive interactions. The thermodynamic properties plateau for the last $5\times 10^6$ steps which shows that the system is properly equilibrated. Both the polymers and the rods move a considerable distance during the equilibration process which shows that the system is not trapped in a kinetically favourable state.}}
	
	\label{fig:thermo_var_rep}
\end{figure*}
The equilibration of the all-repulsive system, like the original simulations, consists of 5 stages:
\begin{enumerate}
	\item $10^6$ steps of NVT at $T=1.0$ and low packing density
	\item $10^6$ steps of NPT at $T=1.0$ and $P=0.25$
	\item $2\times10^6$ steps of NPT at $T=1.0$ and $P=0.5$
	\item $2\times10^6$ steps of NPT at $T=1.0$ and $P=.75$
	\item $14\times10^6$ (or $24\times10^6$ for $\phi_c=0.44$) steps of NPT at $T=1.0$ and $P=1.0$
\end{enumerate}
The final pressure is chosen to achieve a system with a melt-like packing density of $0.3 \leq d_p \leq 0.5$. The analysis is done based on the last $5*10^6$ steps (production). The reported results are averaged over at least 11 uncorrelated configurations for each realization. Fig. \ref{fig:thermo_var_rep} presents the thermodynamic and the particle Root-Mean-Squared-Displacement (RMSD) for a all-repulsive system at the highest concentration $\phi_c = 0.44$. As can be seen, the thermodynamic parameters are reasonably constant and stable for the last $5\times 10^6$ timesteps which shows that the system has at least reached a steady state. The RMSD for both the polymers and the rods takes values of a several times the average radius of gyration of polymers in pure melt $R_0$ implying that the system is not stuck in a local minima and is truly equilibrated.

\begin{figure}[!htpb]
	\subfloat[\label{fig:com_rmsd_att}]{
		\includegraphics[width=0.4\textwidth]{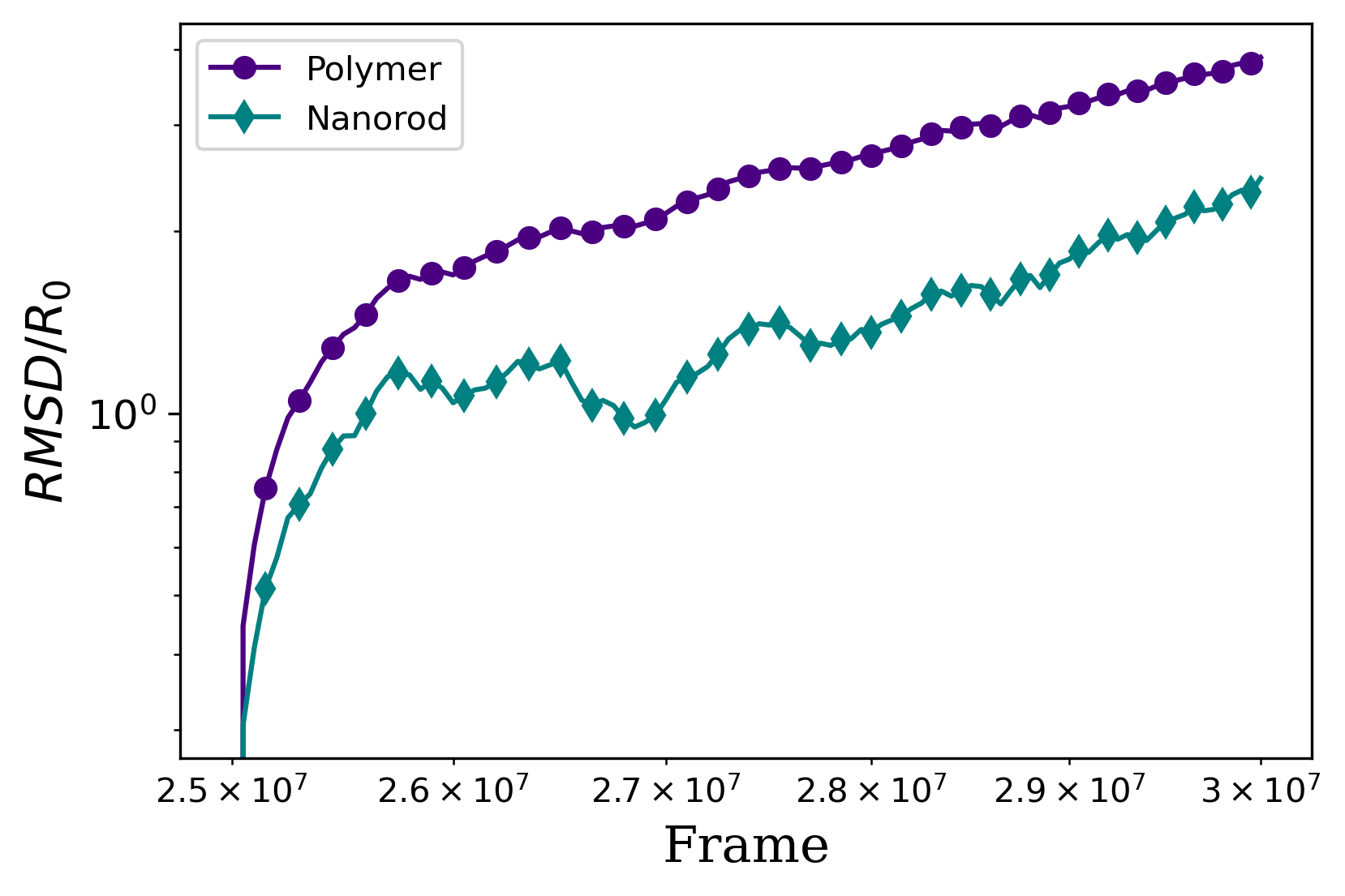}
	}
	
	\subfloat[\label{fig:com_rmsd_rep}]{
		\includegraphics[width=0.4\textwidth]{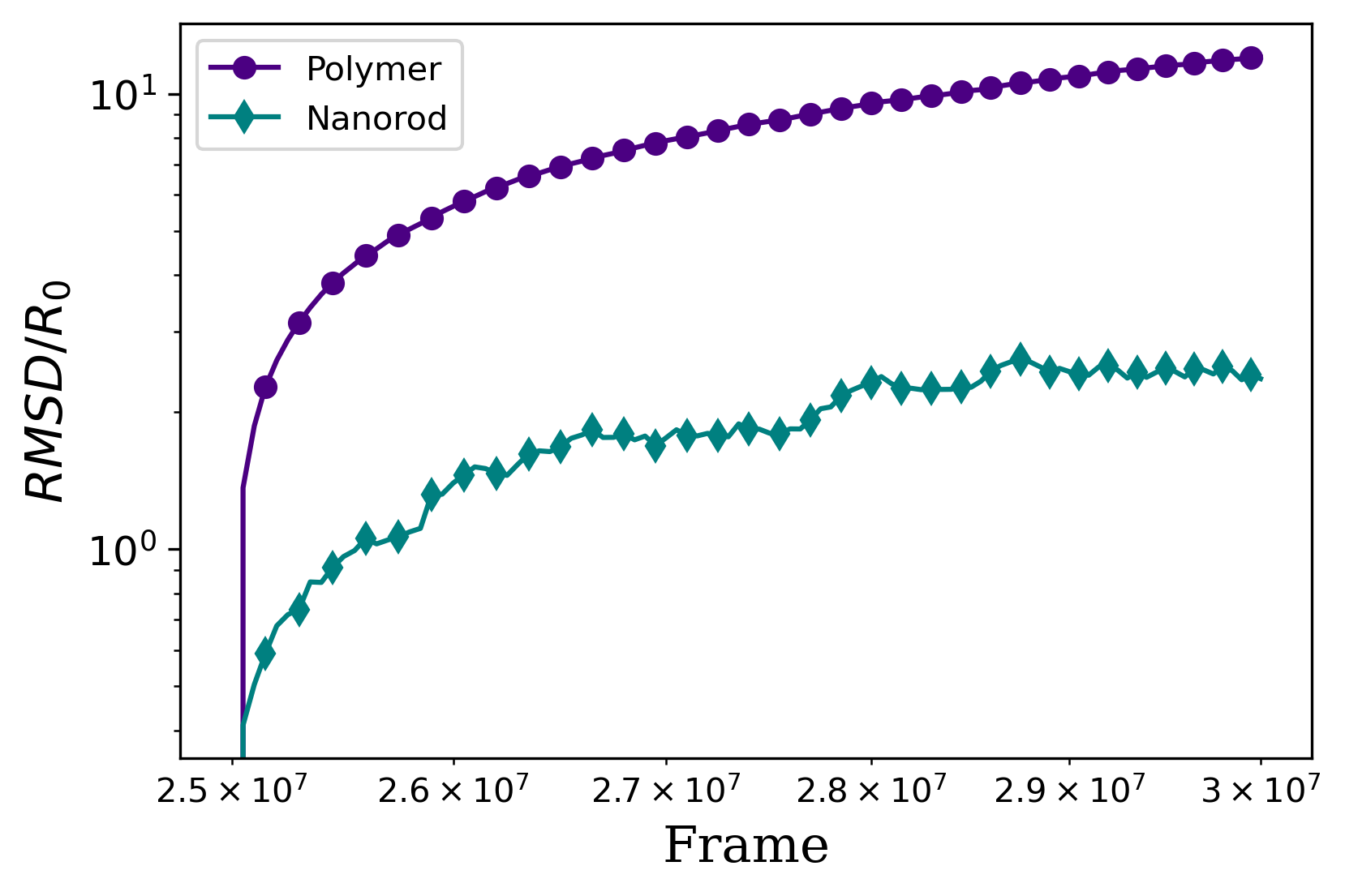}
	}	
	
	\caption{\footnotesize{The Root-Mean-Squared-Displacement of the centre of mass for the attractive (a) and repulsive (b) cases are shown.}}
	\label{fig:com_rmsd}
\end{figure}

The RMSD's shown in Fig. \ref{fig:thermo_var} and \ref{fig:thermo_var_rep} are particle RMSD and take into account the rotational motion of the molecules. We also track the RMSD of centre of mass (COM) of the molecules to further ensure that the nanorods and the polymers move at least a couple of $R_0$ before production. We show the RMSD of COM for the polymers and the nanorods in Fig. \ref{fig:com_rmsd} (a) for the original system and in Fig. \ref{fig:com_rmsd} (b) for the all-repulsive system.

\bibliography{./myrefs1}

\begin{thebibliography}{86}%
\makeatletter
\providecommand \@ifxundefined [1]{%
 \@ifx{#1\undefined}
}%
\providecommand \@ifnum [1]{%
 \ifnum #1\expandafter \@firstoftwo
 \else \expandafter \@secondoftwo
 \fi
}%
\providecommand \@ifx [1]{%
 \ifx #1\expandafter \@firstoftwo
 \else \expandafter \@secondoftwo
 \fi
}%
\providecommand \natexlab [1]{#1}%
\providecommand \enquote  [1]{``#1''}%
\providecommand \bibnamefont  [1]{#1}%
\providecommand \bibfnamefont [1]{#1}%
\providecommand \citenamefont [1]{#1}%
\providecommand \href@noop [0]{\@secondoftwo}%
\providecommand \href [0]{\begingroup \@sanitize@url \@href}%
\providecommand \@href[1]{\@@startlink{#1}\@@href}%
\providecommand \@@href[1]{\endgroup#1\@@endlink}%
\providecommand \@sanitize@url [0]{\catcode `\\12\catcode `\$12\catcode
  `\&12\catcode `\#12\catcode `\^12\catcode `\_12\catcode `\%12\relax}%
\providecommand \@@startlink[1]{}%
\providecommand \@@endlink[0]{}%
\providecommand \url  [0]{\begingroup\@sanitize@url \@url }%
\providecommand \@url [1]{\endgroup\@href {#1}{\urlprefix }}%
\providecommand \urlprefix  [0]{URL }%
\providecommand \Eprint [0]{\href }%
\providecommand \doibase [0]{https://doi.org/}%
\providecommand \selectlanguage [0]{\@gobble}%
\providecommand \bibinfo  [0]{\@secondoftwo}%
\providecommand \bibfield  [0]{\@secondoftwo}%
\providecommand \translation [1]{[#1]}%
\providecommand \BibitemOpen [0]{}%
\providecommand \bibitemStop [0]{}%
\providecommand \bibitemNoStop [0]{.\EOS\space}%
\providecommand \EOS [0]{\spacefactor3000\relax}%
\providecommand \BibitemShut  [1]{\csname bibitem#1\endcsname}%
\let\auto@bib@innerbib\@empty
\bibitem [{\citenamefont {Iijima}(1991)}]{Iijima1991}%
  \BibitemOpen
  \bibfield  {author} {\bibinfo {author} {\bibfnamefont {S.}~\bibnamefont
  {Iijima}},\ }\href {https://doi.org/10.1038/354056a0} {\bibfield  {journal}
  {\bibinfo  {journal} {Nature}\ }\textbf {\bibinfo {volume} {354}},\ \bibinfo
  {pages} {56} (\bibinfo {year} {1991})}\BibitemShut {NoStop}%
\bibitem [{\citenamefont {Ajayan}\ \emph {et~al.}(1994)\citenamefont {Ajayan},
  \citenamefont {Stephan}, \citenamefont {Colliex},\ and\ \citenamefont
  {Trauth}}]{Ajayan1994}%
  \BibitemOpen
  \bibfield  {author} {\bibinfo {author} {\bibfnamefont {P.~M.}\ \bibnamefont
  {Ajayan}}, \bibinfo {author} {\bibfnamefont {O.}~\bibnamefont {Stephan}},
  \bibinfo {author} {\bibfnamefont {C.}~\bibnamefont {Colliex}},\ and\ \bibinfo
  {author} {\bibfnamefont {D.}~\bibnamefont {Trauth}},\ }\href
  {https://doi.org/10.1126/science.265.5176.1212} {\bibfield  {journal}
  {\bibinfo  {journal} {Science}\ }\textbf {\bibinfo {volume} {265}},\ \bibinfo
  {pages} {1212} (\bibinfo {year} {1994})}\BibitemShut {NoStop}%
\bibitem [{\citenamefont {Arash}, \citenamefont {Wang},\ and\ \citenamefont
  {Varadan}(2014)}]{Arash2014}%
  \BibitemOpen
  \bibfield  {author} {\bibinfo {author} {\bibfnamefont {B.}~\bibnamefont
  {Arash}}, \bibinfo {author} {\bibfnamefont {Q.}~\bibnamefont {Wang}},\ and\
  \bibinfo {author} {\bibfnamefont {V.~K.}\ \bibnamefont {Varadan}},\ }\href
  {https://doi.org/10.1038/srep06479} {\bibfield  {journal} {\bibinfo
  {journal} {Scientific Reports}\ }\textbf {\bibinfo {volume} {4}},\ \bibinfo
  {pages} {6479} (\bibinfo {year} {2014})}\BibitemShut {NoStop}%
\bibitem [{\citenamefont {tak Lau}, \citenamefont {Gu},\ and\ \citenamefont
  {Hui}(2006)}]{Lau2006}%
  \BibitemOpen
  \bibfield  {author} {\bibinfo {author} {\bibfnamefont {K.}~\bibnamefont {tak
  Lau}}, \bibinfo {author} {\bibfnamefont {C.}~\bibnamefont {Gu}},\ and\
  \bibinfo {author} {\bibfnamefont {D.}~\bibnamefont {Hui}},\ }\href
  {https://doi.org/10.1016/j.compositesb.2006.02.020} {\bibfield  {journal}
  {\bibinfo  {journal} {Composites Part B: Engineering}\ }\textbf {\bibinfo
  {volume} {37}},\ \bibinfo {pages} {425} (\bibinfo {year} {2006})}\BibitemShut
  {NoStop}%
\bibitem [{\citenamefont {Zhu}, \citenamefont {Pan},\ and\ \citenamefont
  {Roy}(2007)}]{Zhu2007}%
  \BibitemOpen
  \bibfield  {author} {\bibinfo {author} {\bibfnamefont {R.}~\bibnamefont
  {Zhu}}, \bibinfo {author} {\bibfnamefont {E.}~\bibnamefont {Pan}},\ and\
  \bibinfo {author} {\bibfnamefont {A.~K.}\ \bibnamefont {Roy}},\ }\href
  {https://doi.org/10.1016/j.msea.2006.10.054} {\bibfield  {journal} {\bibinfo
  {journal} {Materials Science and Engineering A}\ }\textbf {\bibinfo {volume}
  {447}},\ \bibinfo {pages} {51} (\bibinfo {year} {2007})}\BibitemShut
  {NoStop}%
\bibitem [{\citenamefont {Pal}\ and\ \citenamefont {Kumar}(2016)}]{Pal2016}%
  \BibitemOpen
  \bibfield  {author} {\bibinfo {author} {\bibfnamefont {G.}~\bibnamefont
  {Pal}}\ and\ \bibinfo {author} {\bibfnamefont {S.}~\bibnamefont {Kumar}},\
  }\href {https://doi.org/10.1016/j.paerosci.2015.12.001} {\bibfield  {journal}
  {\bibinfo  {journal} {Progress in Aerospace Sciences}\ }\textbf {\bibinfo
  {volume} {80}},\ \bibinfo {pages} {33} (\bibinfo {year} {2016})}\BibitemShut
  {NoStop}%
\bibitem [{\citenamefont {Sahoo}\ \emph {et~al.}(2010)\citenamefont {Sahoo},
  \citenamefont {Rana}, \citenamefont {Cho}, \citenamefont {Li},\ and\
  \citenamefont {Chan}}]{Sahoo2010}%
  \BibitemOpen
  \bibfield  {author} {\bibinfo {author} {\bibfnamefont {N.~G.}\ \bibnamefont
  {Sahoo}}, \bibinfo {author} {\bibfnamefont {S.}~\bibnamefont {Rana}},
  \bibinfo {author} {\bibfnamefont {J.~W.}\ \bibnamefont {Cho}}, \bibinfo
  {author} {\bibfnamefont {L.}~\bibnamefont {Li}},\ and\ \bibinfo {author}
  {\bibfnamefont {S.~H.}\ \bibnamefont {Chan}},\ }\href
  {https://doi.org/10.1016/j.progpolymsci.2010.03.002} {\bibfield  {journal}
  {\bibinfo  {journal} {Progress in Polymer Science (Oxford)}\ }\textbf
  {\bibinfo {volume} {35}},\ \bibinfo {pages} {837} (\bibinfo {year}
  {2010})}\BibitemShut {NoStop}%
\bibitem [{\citenamefont {Moniruzzaman}\ and\ \citenamefont
  {Winey}(2006)}]{Moniruzzaman2006}%
  \BibitemOpen
  \bibfield  {author} {\bibinfo {author} {\bibfnamefont {M.}~\bibnamefont
  {Moniruzzaman}}\ and\ \bibinfo {author} {\bibfnamefont {K.~I.}\ \bibnamefont
  {Winey}},\ }\href {https://doi.org/10.1021/ma060733p} {\bibfield  {journal}
  {\bibinfo  {journal} {Macromolecules}\ }\textbf {\bibinfo {volume} {39}},\
  \bibinfo {pages} {5194} (\bibinfo {year} {2006})}\BibitemShut {NoStop}%
\bibitem [{\citenamefont {Lozano}\ and\ \citenamefont
  {Barrera}(2001)}]{Lozano2001}%
  \BibitemOpen
  \bibfield  {author} {\bibinfo {author} {\bibfnamefont {K.}~\bibnamefont
  {Lozano}}\ and\ \bibinfo {author} {\bibfnamefont {E.~V.}\ \bibnamefont
  {Barrera}},\ }\href
  {https://doi.org/10.1002/1097-4628(20010103)79:1<125::AID-APP150>3.0.CO;2-D}
  {\bibfield  {journal} {\bibinfo  {journal} {Journal of Applied Polymer
  Science}\ }\textbf {\bibinfo {volume} {79}},\ \bibinfo {pages} {125}
  (\bibinfo {year} {2001})}\BibitemShut {NoStop}%
\bibitem [{\citenamefont {Mordkovich}(2003)}]{Mordkovich2003}%
  \BibitemOpen
  \bibfield  {author} {\bibinfo {author} {\bibfnamefont {V.~Z.}\ \bibnamefont
  {Mordkovich}},\ }\href {https://doi.org/10.1023/A:1026082323244} {\bibfield
  {journal} {\bibinfo  {journal} {Theoretical Foundations of Chemical
  Engineering}\ }\textbf {\bibinfo {volume} {37}},\ \bibinfo {pages} {429}
  (\bibinfo {year} {2003})}\BibitemShut {NoStop}%
\bibitem [{\citenamefont {Al-Saleh}\ and\ \citenamefont
  {Sundararaj}(2011)}]{Al-Saleh2011}%
  \BibitemOpen
  \bibfield  {author} {\bibinfo {author} {\bibfnamefont {M.~H.}\ \bibnamefont
  {Al-Saleh}}\ and\ \bibinfo {author} {\bibfnamefont {U.}~\bibnamefont
  {Sundararaj}},\ }\href {https://doi.org/10.1016/j.compositesa.2011.08.005}
  {\bibfield  {journal} {\bibinfo  {journal} {Composites Part A: Applied
  Science and Manufacturing}\ }\textbf {\bibinfo {volume} {42}},\ \bibinfo
  {pages} {2126} (\bibinfo {year} {2011})}\BibitemShut {NoStop}%
\bibitem [{\citenamefont {Kumar}\ \emph {et~al.}(2002)\citenamefont {Kumar},
  \citenamefont {Doshi}, \citenamefont {Srinivasarao}, \citenamefont {Park},\
  and\ \citenamefont {Schiraldi}}]{Kumar2002}%
  \BibitemOpen
  \bibfield  {author} {\bibinfo {author} {\bibfnamefont {S.}~\bibnamefont
  {Kumar}}, \bibinfo {author} {\bibfnamefont {H.}~\bibnamefont {Doshi}},
  \bibinfo {author} {\bibfnamefont {M.}~\bibnamefont {Srinivasarao}}, \bibinfo
  {author} {\bibfnamefont {J.~O.}\ \bibnamefont {Park}},\ and\ \bibinfo
  {author} {\bibfnamefont {D.~A.}\ \bibnamefont {Schiraldi}},\ }\href
  {https://doi.org/10.1016/S0032-3861(01)00744-3} {\bibfield  {journal}
  {\bibinfo  {journal} {Polymer}\ }\textbf {\bibinfo {volume} {43}},\ \bibinfo
  {pages} {1701} (\bibinfo {year} {2002})}\BibitemShut {NoStop}%
\bibitem [{\citenamefont {Coleman}, \citenamefont {Khan},\ and\ \citenamefont
  {Gun'ko}(2006)}]{Coleman2006a}%
  \BibitemOpen
  \bibfield  {author} {\bibinfo {author} {\bibfnamefont {J.~N.}\ \bibnamefont
  {Coleman}}, \bibinfo {author} {\bibfnamefont {U.}~\bibnamefont {Khan}},\ and\
  \bibinfo {author} {\bibfnamefont {Y.~K.}\ \bibnamefont {Gun'ko}},\ }\href
  {https://doi.org/10.1002/adma.200501851} {\bibfield  {journal} {\bibinfo
  {journal} {Advanced Materials}\ }\textbf {\bibinfo {volume} {18}},\ \bibinfo
  {pages} {689} (\bibinfo {year} {2006})}\BibitemShut {NoStop}%
\bibitem [{\citenamefont {Paul}\ and\ \citenamefont {Dai}(2018)}]{Paul2018}%
  \BibitemOpen
  \bibfield  {author} {\bibinfo {author} {\bibfnamefont {R.}~\bibnamefont
  {Paul}}\ and\ \bibinfo {author} {\bibfnamefont {L.}~\bibnamefont {Dai}},\
  }\href {https://doi.org/10.1080/09276440.2018.1439632} {\bibfield  {journal}
  {\bibinfo  {journal} {Composite Interfaces}\ }\textbf {\bibinfo {volume}
  {25}},\ \bibinfo {pages} {539} (\bibinfo {year} {2018})}\BibitemShut
  {NoStop}%
\bibitem [{\citenamefont {Toepperwein}\ \emph {et~al.}(2011)\citenamefont
  {Toepperwein}, \citenamefont {Karayiannis}, \citenamefont {Riggleman},
  \citenamefont {Kr{\"{o}}ger},\ and\ \citenamefont {{De
  Pablo}}}]{Toepperwein2011}%
  \BibitemOpen
  \bibfield  {author} {\bibinfo {author} {\bibfnamefont {G.~N.}\ \bibnamefont
  {Toepperwein}}, \bibinfo {author} {\bibfnamefont {N.~C.}\ \bibnamefont
  {Karayiannis}}, \bibinfo {author} {\bibfnamefont {R.~A.}\ \bibnamefont
  {Riggleman}}, \bibinfo {author} {\bibfnamefont {M.}~\bibnamefont
  {Kr{\"{o}}ger}},\ and\ \bibinfo {author} {\bibfnamefont {J.~J.}\ \bibnamefont
  {{De Pablo}}},\ }\href {https://doi.org/10.1021/ma102741r} {\bibfield
  {journal} {\bibinfo  {journal} {Macromolecules}\ }\textbf {\bibinfo {volume}
  {44}},\ \bibinfo {pages} {1034} (\bibinfo {year} {2011})}\BibitemShut
  {NoStop}%
\bibitem [{\citenamefont {Sapkota}\ \emph {et~al.}(2017)\citenamefont
  {Sapkota}, \citenamefont {Gooneie}, \citenamefont {Shirole},\ and\
  \citenamefont {{Martinez Garcia}}}]{Sapkota2017}%
  \BibitemOpen
  \bibfield  {author} {\bibinfo {author} {\bibfnamefont {J.}~\bibnamefont
  {Sapkota}}, \bibinfo {author} {\bibfnamefont {A.}~\bibnamefont {Gooneie}},
  \bibinfo {author} {\bibfnamefont {A.}~\bibnamefont {Shirole}},\ and\ \bibinfo
  {author} {\bibfnamefont {J.~C.}\ \bibnamefont {{Martinez Garcia}}},\ }\href
  {https://doi.org/10.1002/app.45279} {\bibfield  {journal} {\bibinfo
  {journal} {Journal of Applied Polymer Science}\ }\textbf {\bibinfo {volume}
  {134}},\ \bibinfo {pages} {45279} (\bibinfo {year} {2017})}\BibitemShut
  {NoStop}%
\bibitem [{\citenamefont {Xie}, \citenamefont {Mai},\ and\ \citenamefont
  {Zhou}(2005)}]{Xie2005}%
  \BibitemOpen
  \bibfield  {author} {\bibinfo {author} {\bibfnamefont {X.~L.}\ \bibnamefont
  {Xie}}, \bibinfo {author} {\bibfnamefont {Y.~W.}\ \bibnamefont {Mai}},\ and\
  \bibinfo {author} {\bibfnamefont {X.~P.}\ \bibnamefont {Zhou}},\ }\href
  {https://doi.org/10.1016/j.mser.2005.04.002} {\bibfield  {journal} {\bibinfo
  {journal} {Materials Science and Engineering R: Reports}\ }\textbf {\bibinfo
  {volume} {49}},\ \bibinfo {pages} {89} (\bibinfo {year} {2005})}\BibitemShut
  {NoStop}%
\bibitem [{\citenamefont {Coleman}\ \emph {et~al.}(2006)\citenamefont
  {Coleman}, \citenamefont {Khan}, \citenamefont {Blau},\ and\ \citenamefont
  {Gun'ko}}]{Coleman2006}%
  \BibitemOpen
  \bibfield  {author} {\bibinfo {author} {\bibfnamefont {J.~N.}\ \bibnamefont
  {Coleman}}, \bibinfo {author} {\bibfnamefont {U.}~\bibnamefont {Khan}},
  \bibinfo {author} {\bibfnamefont {W.~J.}\ \bibnamefont {Blau}},\ and\
  \bibinfo {author} {\bibfnamefont {Y.~K.}\ \bibnamefont {Gun'ko}},\ }\href
  {https://doi.org/10.1016/j.carbon.2006.02.038} {\bibfield  {journal}
  {\bibinfo  {journal} {Carbon}\ }\textbf {\bibinfo {volume} {44}},\ \bibinfo
  {pages} {1624} (\bibinfo {year} {2006})}\BibitemShut {NoStop}%
\bibitem [{\citenamefont {Hirsch}(2002)}]{Hirsch2002}%
  \BibitemOpen
  \bibfield  {author} {\bibinfo {author} {\bibfnamefont {A.}~\bibnamefont
  {Hirsch}},\ }\href
  {https://doi.org/10.1002/1521-3773(20020603)41:11<1853::AID-ANIE1853>3.0.CO;2-N}
  {\bibfield  {journal} {\bibinfo  {journal} {Angewandte Chemie - International
  Edition}\ }\textbf {\bibinfo {volume} {41}},\ \bibinfo {pages} {1853}
  (\bibinfo {year} {2002})}\BibitemShut {NoStop}%
\bibitem [{\citenamefont {Severini}\ \emph {et~al.}(2002)\citenamefont
  {Severini}, \citenamefont {Formaro}, \citenamefont {Pegoraro},\ and\
  \citenamefont {Posca}}]{Severini2002}%
  \BibitemOpen
  \bibfield  {author} {\bibinfo {author} {\bibfnamefont {F.}~\bibnamefont
  {Severini}}, \bibinfo {author} {\bibfnamefont {L.}~\bibnamefont {Formaro}},
  \bibinfo {author} {\bibfnamefont {M.}~\bibnamefont {Pegoraro}},\ and\
  \bibinfo {author} {\bibfnamefont {L.}~\bibnamefont {Posca}},\ }\href
  {https://doi.org/10.1016/S0008-6223(01)00180-4} {\bibfield  {journal}
  {\bibinfo  {journal} {Carbon}\ }\textbf {\bibinfo {volume} {40}},\ \bibinfo
  {pages} {735} (\bibinfo {year} {2002})}\BibitemShut {NoStop}%
\bibitem [{\citenamefont {Ramasubramaniam}, \citenamefont {Chen},\ and\
  \citenamefont {Liu}(2003)}]{Ramasubramaniam2003}%
  \BibitemOpen
  \bibfield  {author} {\bibinfo {author} {\bibfnamefont {R.}~\bibnamefont
  {Ramasubramaniam}}, \bibinfo {author} {\bibfnamefont {J.}~\bibnamefont
  {Chen}},\ and\ \bibinfo {author} {\bibfnamefont {H.}~\bibnamefont {Liu}},\
  }\href {https://doi.org/10.1063/1.1616976} {\bibfield  {journal} {\bibinfo
  {journal} {Applied Physics Letters}\ }\textbf {\bibinfo {volume} {83}},\
  \bibinfo {pages} {2928} (\bibinfo {year} {2003})}\BibitemShut {NoStop}%
\bibitem [{\citenamefont {Zhu}\ \emph {et~al.}(2003)\citenamefont {Zhu},
  \citenamefont {Kim}, \citenamefont {Peng}, \citenamefont {Margrave},
  \citenamefont {Khabashesku},\ and\ \citenamefont {Barrera}}]{Zhu2003}%
  \BibitemOpen
  \bibfield  {author} {\bibinfo {author} {\bibfnamefont {J.}~\bibnamefont
  {Zhu}}, \bibinfo {author} {\bibfnamefont {J.~D.}\ \bibnamefont {Kim}},
  \bibinfo {author} {\bibfnamefont {H.}~\bibnamefont {Peng}}, \bibinfo {author}
  {\bibfnamefont {J.~L.}\ \bibnamefont {Margrave}}, \bibinfo {author}
  {\bibfnamefont {V.~N.}\ \bibnamefont {Khabashesku}},\ and\ \bibinfo {author}
  {\bibfnamefont {E.~V.}\ \bibnamefont {Barrera}},\ }\href
  {https://doi.org/10.1021/nl0342489} {\bibfield  {journal} {\bibinfo
  {journal} {Nano Letters}\ }\textbf {\bibinfo {volume} {3}},\ \bibinfo {pages}
  {1107} (\bibinfo {year} {2003})}\BibitemShut {NoStop}%
\bibitem [{\citenamefont {Grossiord}\ \emph {et~al.}(2006)\citenamefont
  {Grossiord}, \citenamefont {Loos}, \citenamefont {Regev},\ and\ \citenamefont
  {Koning}}]{Grossiord2006}%
  \BibitemOpen
  \bibfield  {author} {\bibinfo {author} {\bibfnamefont {N.}~\bibnamefont
  {Grossiord}}, \bibinfo {author} {\bibfnamefont {J.}~\bibnamefont {Loos}},
  \bibinfo {author} {\bibfnamefont {O.}~\bibnamefont {Regev}},\ and\ \bibinfo
  {author} {\bibfnamefont {C.~E.}\ \bibnamefont {Koning}},\ }\href
  {https://doi.org/10.1021/cm051881h} {\bibfield  {journal} {\bibinfo
  {journal} {Chemistry of Materials}\ }\textbf {\bibinfo {volume} {18}},\
  \bibinfo {pages} {1089} (\bibinfo {year} {2006})}\BibitemShut {NoStop}%
\bibitem [{\citenamefont {Vaisman}, \citenamefont {Wagner},\ and\ \citenamefont
  {Marom}(2006)}]{Vaisman2006}%
  \BibitemOpen
  \bibfield  {author} {\bibinfo {author} {\bibfnamefont {L.}~\bibnamefont
  {Vaisman}}, \bibinfo {author} {\bibfnamefont {H.~D.}\ \bibnamefont
  {Wagner}},\ and\ \bibinfo {author} {\bibfnamefont {G.}~\bibnamefont
  {Marom}},\ }\href {https://doi.org/10.1016/j.cis.2006.11.007} {\bibfield
  {journal} {\bibinfo  {journal} {Advances in Colloid and Interface Science}\
  }\textbf {\bibinfo {volume} {128-130}},\ \bibinfo {pages} {37} (\bibinfo
  {year} {2006})}\BibitemShut {NoStop}%
\bibitem [{\citenamefont {Ma}\ \emph {et~al.}(2010)\citenamefont {Ma},
  \citenamefont {Siddiqui}, \citenamefont {Marom},\ and\ \citenamefont
  {Kim}}]{Ma2010}%
  \BibitemOpen
  \bibfield  {author} {\bibinfo {author} {\bibfnamefont {P.~C.}\ \bibnamefont
  {Ma}}, \bibinfo {author} {\bibfnamefont {N.~A.}\ \bibnamefont {Siddiqui}},
  \bibinfo {author} {\bibfnamefont {G.}~\bibnamefont {Marom}},\ and\ \bibinfo
  {author} {\bibfnamefont {J.~K.}\ \bibnamefont {Kim}},\ }\href
  {https://doi.org/10.1016/j.compositesa.2010.07.003} {\bibfield  {journal}
  {\bibinfo  {journal} {Composites Part A: Applied Science and Manufacturing}\
  }\textbf {\bibinfo {volume} {41}},\ \bibinfo {pages} {1345} (\bibinfo {year}
  {2010})}\BibitemShut {NoStop}%
\bibitem [{\citenamefont {Kalra}, \citenamefont {Escobedo},\ and\ \citenamefont
  {Joo}(2010)}]{Kalra2010}%
  \BibitemOpen
  \bibfield  {author} {\bibinfo {author} {\bibfnamefont {V.}~\bibnamefont
  {Kalra}}, \bibinfo {author} {\bibfnamefont {F.}~\bibnamefont {Escobedo}},\
  and\ \bibinfo {author} {\bibfnamefont {Y.~L.}\ \bibnamefont {Joo}},\ }\href
  {https://doi.org/10.1063/1.3277671} {\bibfield  {journal} {\bibinfo
  {journal} {Journal of Chemical Physics}\ }\textbf {\bibinfo {volume} {132}}
  (\bibinfo {year} {2010}),\ 10.1063/1.3277671}\BibitemShut {NoStop}%
\bibitem [{\citenamefont {Lekkerkerker}\ and\ \citenamefont
  {Stroobants}(1994)}]{Lekkerkerker1994}%
  \BibitemOpen
  \bibfield  {author} {\bibinfo {author} {\bibfnamefont {H.~N.}\ \bibnamefont
  {Lekkerkerker}}\ and\ \bibinfo {author} {\bibfnamefont {A.}~\bibnamefont
  {Stroobants}},\ }\href {https://doi.org/10.1007/BF02458781} {\bibfield
  {journal} {\bibinfo  {journal} {Il Nuovo Cimento D}\ }\textbf {\bibinfo
  {volume} {16}},\ \bibinfo {pages} {949} (\bibinfo {year} {1994})}\BibitemShut
  {NoStop}%
\bibitem [{\citenamefont {Bolhuis}\ \emph {et~al.}(1997)\citenamefont
  {Bolhuis}, \citenamefont {Stroobants}, \citenamefont {Frenkel},\ and\
  \citenamefont {Lekkerkerker}}]{Bolhuis1997}%
  \BibitemOpen
  \bibfield  {author} {\bibinfo {author} {\bibfnamefont {P.~G.}\ \bibnamefont
  {Bolhuis}}, \bibinfo {author} {\bibfnamefont {A.}~\bibnamefont {Stroobants}},
  \bibinfo {author} {\bibfnamefont {D.}~\bibnamefont {Frenkel}},\ and\ \bibinfo
  {author} {\bibfnamefont {H.~N.}\ \bibnamefont {Lekkerkerker}},\ }\href
  {https://doi.org/10.1063/1.474508} {\bibfield  {journal} {\bibinfo  {journal}
  {Journal of Chemical Physics}\ }\textbf {\bibinfo {volume} {107}},\ \bibinfo
  {pages} {1551} (\bibinfo {year} {1997})}\BibitemShut {NoStop}%
\bibitem [{\citenamefont {Surve}, \citenamefont {Pryamitsyn},\ and\
  \citenamefont {Ganesan}(2007)}]{Surve2007}%
  \BibitemOpen
  \bibfield  {author} {\bibinfo {author} {\bibfnamefont {M.}~\bibnamefont
  {Surve}}, \bibinfo {author} {\bibfnamefont {V.}~\bibnamefont {Pryamitsyn}},\
  and\ \bibinfo {author} {\bibfnamefont {V.}~\bibnamefont {Ganesan}},\ }\href
  {https://doi.org/10.1021/ma061603j} {\bibfield  {journal} {\bibinfo
  {journal} {Macromolecules}\ }\textbf {\bibinfo {volume} {40}},\ \bibinfo
  {pages} {344} (\bibinfo {year} {2007})}\BibitemShut {NoStop}%
\bibitem [{\citenamefont {Hall}\ and\ \citenamefont
  {Schweizer}(2010)}]{Hall2010}%
  \BibitemOpen
  \bibfield  {author} {\bibinfo {author} {\bibfnamefont {L.~M.}\ \bibnamefont
  {Hall}}\ and\ \bibinfo {author} {\bibfnamefont {K.~S.}\ \bibnamefont
  {Schweizer}},\ }\href {https://doi.org/10.1039/b919160g} {\bibfield
  {journal} {\bibinfo  {journal} {Soft Matter}\ }\textbf {\bibinfo {volume}
  {6}},\ \bibinfo {pages} {1015} (\bibinfo {year} {2010})}\BibitemShut
  {NoStop}%
\bibitem [{\citenamefont {Starr}, \citenamefont {Schr{\o}der},\ and\
  \citenamefont {Glotzer}(2002)}]{Starr2002}%
  \BibitemOpen
  \bibfield  {author} {\bibinfo {author} {\bibfnamefont {F.~W.}\ \bibnamefont
  {Starr}}, \bibinfo {author} {\bibfnamefont {T.~B.}\ \bibnamefont
  {Schr{\o}der}},\ and\ \bibinfo {author} {\bibfnamefont {S.~C.}\ \bibnamefont
  {Glotzer}},\ }\href {https://doi.org/10.1021/ma010626p} {\bibfield  {journal}
  {\bibinfo  {journal} {Macromolecules}\ }\textbf {\bibinfo {volume} {35}},\
  \bibinfo {pages} {4481} (\bibinfo {year} {2002})}\BibitemShut {NoStop}%
\bibitem [{\citenamefont {Savenko}\ and\ \citenamefont
  {Dijkstra}(2006)}]{Savenko2006}%
  \BibitemOpen
  \bibfield  {author} {\bibinfo {author} {\bibfnamefont {S.~V.}\ \bibnamefont
  {Savenko}}\ and\ \bibinfo {author} {\bibfnamefont {M.}~\bibnamefont
  {Dijkstra}},\ }\href {https://doi.org/10.1063/1.2202853} {\bibfield
  {journal} {\bibinfo  {journal} {Journal of Chemical Physics}\ }\textbf
  {\bibinfo {volume} {124}},\ \bibinfo {pages} {244909} (\bibinfo {year}
  {2006})}\BibitemShut {NoStop}%
\bibitem [{\citenamefont {Hu}, \citenamefont {Sheng},\ and\ \citenamefont
  {Tsao}(2013)}]{Hu2013a}%
  \BibitemOpen
  \bibfield  {author} {\bibinfo {author} {\bibfnamefont {S.~W.}\ \bibnamefont
  {Hu}}, \bibinfo {author} {\bibfnamefont {Y.~J.}\ \bibnamefont {Sheng}},\ and\
  \bibinfo {author} {\bibfnamefont {H.~K.}\ \bibnamefont {Tsao}},\ }\href
  {https://doi.org/10.1039/c3sm50825k} {\bibfield  {journal} {\bibinfo
  {journal} {Soft Matter}\ }\textbf {\bibinfo {volume} {9}},\ \bibinfo {pages}
  {7261} (\bibinfo {year} {2013})}\BibitemShut {NoStop}%
\bibitem [{\citenamefont {Hore}\ and\ \citenamefont
  {Composto}(2014)}]{Hore2013}%
  \BibitemOpen
  \bibfield  {author} {\bibinfo {author} {\bibfnamefont {M.~J.~A.}\
  \bibnamefont {Hore}}\ and\ \bibinfo {author} {\bibfnamefont {R.~J.}\
  \bibnamefont {Composto}},\ }\href {https://doi.org/10.1021/ma402179w}
  {\bibfield  {journal} {\bibinfo  {journal} {Macromolecules}\ }\textbf
  {\bibinfo {volume} {47}},\ \bibinfo {pages} {875} (\bibinfo {year}
  {2014})}\BibitemShut {NoStop}%
\bibitem [{\citenamefont {Gao}\ \emph {et~al.}(2014)\citenamefont {Gao},
  \citenamefont {Liu}, \citenamefont {Shen}, \citenamefont {Zhang},\ and\
  \citenamefont {Cao}}]{Gao2014}%
  \BibitemOpen
  \bibfield  {author} {\bibinfo {author} {\bibfnamefont {Y.}~\bibnamefont
  {Gao}}, \bibinfo {author} {\bibfnamefont {J.}~\bibnamefont {Liu}}, \bibinfo
  {author} {\bibfnamefont {J.}~\bibnamefont {Shen}}, \bibinfo {author}
  {\bibfnamefont {L.}~\bibnamefont {Zhang}},\ and\ \bibinfo {author}
  {\bibfnamefont {D.}~\bibnamefont {Cao}},\ }\href
  {https://doi.org/10.1016/j.polymer.2014.01.042} {\bibfield  {journal}
  {\bibinfo  {journal} {Polymer}\ }\textbf {\bibinfo {volume} {55}},\ \bibinfo
  {pages} {1273} (\bibinfo {year} {2014})}\BibitemShut {NoStop}%
\bibitem [{\citenamefont {Sankar}\ and\ \citenamefont
  {Tripathy}(2015)}]{Sankar2015}%
  \BibitemOpen
  \bibfield  {author} {\bibinfo {author} {\bibfnamefont {U.~K.}\ \bibnamefont
  {Sankar}}\ and\ \bibinfo {author} {\bibfnamefont {M.}~\bibnamefont
  {Tripathy}},\ }\href {https://doi.org/10.1021/ma501292d} {\bibfield
  {journal} {\bibinfo  {journal} {Macromolecules}\ }\textbf {\bibinfo {volume}
  {48}},\ \bibinfo {pages} {432} (\bibinfo {year} {2015})}\BibitemShut
  {NoStop}%
\bibitem [{\citenamefont {Milchev}\ \emph {et~al.}(2020)\citenamefont
  {Milchev}, \citenamefont {Egorov}, \citenamefont {Midya}, \citenamefont
  {Binder},\ and\ \citenamefont {Nikoubashman}}]{Milchev2020}%
  \BibitemOpen
  \bibfield  {author} {\bibinfo {author} {\bibfnamefont {A.}~\bibnamefont
  {Milchev}}, \bibinfo {author} {\bibfnamefont {S.~A.}\ \bibnamefont {Egorov}},
  \bibinfo {author} {\bibfnamefont {J.}~\bibnamefont {Midya}}, \bibinfo
  {author} {\bibfnamefont {K.}~\bibnamefont {Binder}},\ and\ \bibinfo {author}
  {\bibfnamefont {A.}~\bibnamefont {Nikoubashman}},\ }\href
  {https://doi.org/10.1021/acsmacrolett.0c00668} {\bibfield  {journal}
  {\bibinfo  {journal} {ACS Macro Letters}\ }\textbf {\bibinfo {volume} {9}},\
  \bibinfo {pages} {1779} (\bibinfo {year} {2020})},\ \Eprint
  {https://arxiv.org/abs/2009.06271} {arXiv:2009.06271} \BibitemShut {NoStop}%
\bibitem [{\citenamefont {Lu}, \citenamefont {Wu},\ and\ \citenamefont
  {Jayaraman}(2021)}]{Lu2021}%
  \BibitemOpen
  \bibfield  {author} {\bibinfo {author} {\bibfnamefont {S.}~\bibnamefont
  {Lu}}, \bibinfo {author} {\bibfnamefont {Z.}~\bibnamefont {Wu}},\ and\
  \bibinfo {author} {\bibfnamefont {A.}~\bibnamefont {Jayaraman}},\ }\href
  {https://doi.org/10.1021/acs.jpcb.1c00097} {\bibfield  {journal} {\bibinfo
  {journal} {Journal of Physical Chemistry B}\ } (\bibinfo {year} {2021}),\
  10.1021/acs.jpcb.1c00097}\BibitemShut {NoStop}%
\bibitem [{\citenamefont {Erigi}, \citenamefont {Dhumal},\ and\ \citenamefont
  {Tripathy}(2021)}]{Erigi2021}%
  \BibitemOpen
  \bibfield  {author} {\bibinfo {author} {\bibfnamefont {U.}~\bibnamefont
  {Erigi}}, \bibinfo {author} {\bibfnamefont {U.}~\bibnamefont {Dhumal}},\ and\
  \bibinfo {author} {\bibfnamefont {M.}~\bibnamefont {Tripathy}},\ }\href
  {https://doi.org/10.1063/5.0038186} {\bibfield  {journal} {\bibinfo
  {journal} {Journal of Chemical Physics}\ }\textbf {\bibinfo {volume} {154}},\
  \bibinfo {pages} {124903} (\bibinfo {year} {2021})}\BibitemShut {NoStop}%
\bibitem [{\citenamefont {Tang}\ and\ \citenamefont {Karoos}(1997)}]{Tang1997}%
  \BibitemOpen
  \bibfield  {author} {\bibinfo {author} {\bibfnamefont {L.~G.}\ \bibnamefont
  {Tang}}\ and\ \bibinfo {author} {\bibfnamefont {J.~L.}\ \bibnamefont
  {Karoos}},\ }\href {https://doi.org/10.1002/pc.10265} {\bibfield  {journal}
  {\bibinfo  {journal} {Polymer Composites}\ }\textbf {\bibinfo {volume}
  {18}},\ \bibinfo {pages} {100} (\bibinfo {year} {1997})}\BibitemShut
  {NoStop}%
\bibitem [{\citenamefont {Zhang}\ \emph
  {et~al.}(2013{\natexlab{a}})\citenamefont {Zhang}, \citenamefont {Liu},
  \citenamefont {Huang},\ and\ \citenamefont {Liu}}]{Zhang2013}%
  \BibitemOpen
  \bibfield  {author} {\bibinfo {author} {\bibfnamefont {R.~L.}\ \bibnamefont
  {Zhang}}, \bibinfo {author} {\bibfnamefont {Y.}~\bibnamefont {Liu}}, \bibinfo
  {author} {\bibfnamefont {Y.~D.}\ \bibnamefont {Huang}},\ and\ \bibinfo
  {author} {\bibfnamefont {L.}~\bibnamefont {Liu}},\ }\href
  {https://doi.org/10.1016/j.apsusc.2013.09.174} {\bibfield  {journal}
  {\bibinfo  {journal} {Applied Surface Science}\ }\textbf {\bibinfo {volume}
  {287}},\ \bibinfo {pages} {423} (\bibinfo {year}
  {2013}{\natexlab{a}})}\BibitemShut {NoStop}%
\bibitem [{\citenamefont {Zhang}\ \emph
  {et~al.}(2013{\natexlab{b}})\citenamefont {Zhang}, \citenamefont {Liu},
  \citenamefont {Hao}, \citenamefont {Jiao}, \citenamefont {Yang},\ and\
  \citenamefont {Wang}}]{Zhang2013a}%
  \BibitemOpen
  \bibfield  {author} {\bibinfo {author} {\bibfnamefont {S.}~\bibnamefont
  {Zhang}}, \bibinfo {author} {\bibfnamefont {W.~B.}\ \bibnamefont {Liu}},
  \bibinfo {author} {\bibfnamefont {L.~F.}\ \bibnamefont {Hao}}, \bibinfo
  {author} {\bibfnamefont {W.~C.}\ \bibnamefont {Jiao}}, \bibinfo {author}
  {\bibfnamefont {F.}~\bibnamefont {Yang}},\ and\ \bibinfo {author}
  {\bibfnamefont {R.~G.}\ \bibnamefont {Wang}},\ }\href
  {https://doi.org/10.1016/j.compscitech.2013.08.035} {\bibfield  {journal}
  {\bibinfo  {journal} {Composites Science and Technology}\ }\textbf {\bibinfo
  {volume} {88}},\ \bibinfo {pages} {120} (\bibinfo {year}
  {2013}{\natexlab{b}})}\BibitemShut {NoStop}%
\bibitem [{\citenamefont {Karger-Kocsis}, \citenamefont {Mahmood},\ and\
  \citenamefont {Pegoretti}(2015)}]{Karger-Kocsis2015}%
  \BibitemOpen
  \bibfield  {author} {\bibinfo {author} {\bibfnamefont {J.}~\bibnamefont
  {Karger-Kocsis}}, \bibinfo {author} {\bibfnamefont {H.}~\bibnamefont
  {Mahmood}},\ and\ \bibinfo {author} {\bibfnamefont {A.}~\bibnamefont
  {Pegoretti}},\ }\href {https://doi.org/10.1016/j.pmatsci.2015.02.003}
  {\bibfield  {journal} {\bibinfo  {journal} {Progress in Materials Science}\
  }\textbf {\bibinfo {volume} {73}},\ \bibinfo {pages} {1} (\bibinfo {year}
  {2015})}\BibitemShut {NoStop}%
\bibitem [{\citenamefont {Park}, \citenamefont {Kalra},\ and\ \citenamefont
  {Joo}(2014)}]{Park2014}%
  \BibitemOpen
  \bibfield  {author} {\bibinfo {author} {\bibfnamefont {J.~H.}\ \bibnamefont
  {Park}}, \bibinfo {author} {\bibfnamefont {V.}~\bibnamefont {Kalra}},\ and\
  \bibinfo {author} {\bibfnamefont {Y.~L.}\ \bibnamefont {Joo}},\ }\href
  {https://doi.org/10.1063/1.4868986} {\bibfield  {journal} {\bibinfo
  {journal} {Journal of Chemical Physics}\ }\textbf {\bibinfo {volume} {140}}
  (\bibinfo {year} {2014}),\ 10.1063/1.4868986}\BibitemShut {NoStop}%
\bibitem [{\citenamefont {Aztatzi-Pluma}\ \emph {et~al.}(2016)\citenamefont
  {Aztatzi-Pluma}, \citenamefont {Castrej{\'{o}}n-Gonz{\'{a}}lez},
  \citenamefont {Almendarez-Camarillo}, \citenamefont {Alvarado},\ and\
  \citenamefont {Dur{\'{a}}n-Morales}}]{Aztatzi-Pluma2016}%
  \BibitemOpen
  \bibfield  {author} {\bibinfo {author} {\bibfnamefont {D.}~\bibnamefont
  {Aztatzi-Pluma}}, \bibinfo {author} {\bibfnamefont {E.~O.}\ \bibnamefont
  {Castrej{\'{o}}n-Gonz{\'{a}}lez}}, \bibinfo {author} {\bibfnamefont
  {A.}~\bibnamefont {Almendarez-Camarillo}}, \bibinfo {author} {\bibfnamefont
  {J.~F.}\ \bibnamefont {Alvarado}},\ and\ \bibinfo {author} {\bibfnamefont
  {Y.}~\bibnamefont {Dur{\'{a}}n-Morales}},\ }\href
  {https://doi.org/10.1021/acs.jpcc.5b08136} {\bibfield  {journal} {\bibinfo
  {journal} {Journal of Physical Chemistry C}\ }\textbf {\bibinfo {volume}
  {120}},\ \bibinfo {pages} {2371} (\bibinfo {year} {2016})}\BibitemShut
  {NoStop}%
\bibitem [{\citenamefont {Liao}\ and\ \citenamefont {Li}(2001)}]{Liao2001}%
  \BibitemOpen
  \bibfield  {author} {\bibinfo {author} {\bibfnamefont {K.}~\bibnamefont
  {Liao}}\ and\ \bibinfo {author} {\bibfnamefont {S.}~\bibnamefont {Li}},\
  }\href {https://doi.org/10.1063/1.1428116} {\bibfield  {journal} {\bibinfo
  {journal} {Applied Physics Letters}\ }\textbf {\bibinfo {volume} {79}},\
  \bibinfo {pages} {4225} (\bibinfo {year} {2001})}\BibitemShut {NoStop}%
\bibitem [{\citenamefont {Xu}\ \emph {et~al.}(2002)\citenamefont {Xu},
  \citenamefont {Thwe}, \citenamefont {Shearwood},\ and\ \citenamefont
  {Liao}}]{Xu2002}%
  \BibitemOpen
  \bibfield  {author} {\bibinfo {author} {\bibfnamefont {X.}~\bibnamefont
  {Xu}}, \bibinfo {author} {\bibfnamefont {M.~M.}\ \bibnamefont {Thwe}},
  \bibinfo {author} {\bibfnamefont {C.}~\bibnamefont {Shearwood}},\ and\
  \bibinfo {author} {\bibfnamefont {K.}~\bibnamefont {Liao}},\ }\href
  {https://doi.org/10.1063/1.1511532} {\bibfield  {journal} {\bibinfo
  {journal} {Applied Physics Letters}\ }\textbf {\bibinfo {volume} {81}},\
  \bibinfo {pages} {2833} (\bibinfo {year} {2002})}\BibitemShut {NoStop}%
\bibitem [{\citenamefont {Wong}\ \emph {et~al.}(2003)\citenamefont {Wong},
  \citenamefont {Paramsothy}, \citenamefont {Xu}, \citenamefont {Ren},
  \citenamefont {Li},\ and\ \citenamefont {Liao}}]{Wong2003}%
  \BibitemOpen
  \bibfield  {author} {\bibinfo {author} {\bibfnamefont {M.}~\bibnamefont
  {Wong}}, \bibinfo {author} {\bibfnamefont {M.}~\bibnamefont {Paramsothy}},
  \bibinfo {author} {\bibfnamefont {X.~J.}\ \bibnamefont {Xu}}, \bibinfo
  {author} {\bibfnamefont {Y.}~\bibnamefont {Ren}}, \bibinfo {author}
  {\bibfnamefont {S.}~\bibnamefont {Li}},\ and\ \bibinfo {author}
  {\bibfnamefont {K.}~\bibnamefont {Liao}},\ }\href
  {https://doi.org/10.1016/j.polymer.2003.10.011} {\bibfield  {journal}
  {\bibinfo  {journal} {Polymer}\ }\textbf {\bibinfo {volume} {44}},\ \bibinfo
  {pages} {7757} (\bibinfo {year} {2003})}\BibitemShut {NoStop}%
\bibitem [{\citenamefont {Huang}\ \emph {et~al.}(2015)\citenamefont {Huang},
  \citenamefont {Jiang}, \citenamefont {Hor}, \citenamefont {Gupta},
  \citenamefont {Zhang}, \citenamefont {Stebe}, \citenamefont {Feng},
  \citenamefont {Turner},\ and\ \citenamefont {Lee}}]{Huang2015}%
  \BibitemOpen
  \bibfield  {author} {\bibinfo {author} {\bibfnamefont {Y.~R.}\ \bibnamefont
  {Huang}}, \bibinfo {author} {\bibfnamefont {Y.}~\bibnamefont {Jiang}},
  \bibinfo {author} {\bibfnamefont {J.~L.}\ \bibnamefont {Hor}}, \bibinfo
  {author} {\bibfnamefont {R.}~\bibnamefont {Gupta}}, \bibinfo {author}
  {\bibfnamefont {L.}~\bibnamefont {Zhang}}, \bibinfo {author} {\bibfnamefont
  {K.~J.}\ \bibnamefont {Stebe}}, \bibinfo {author} {\bibfnamefont
  {G.}~\bibnamefont {Feng}}, \bibinfo {author} {\bibfnamefont {K.~T.}\
  \bibnamefont {Turner}},\ and\ \bibinfo {author} {\bibfnamefont
  {D.}~\bibnamefont {Lee}},\ }\href {https://doi.org/10.1039/c4nr05464d}
  {\bibfield  {journal} {\bibinfo  {journal} {Nanoscale}\ }\textbf {\bibinfo
  {volume} {7}},\ \bibinfo {pages} {798} (\bibinfo {year} {2015})}\BibitemShut
  {NoStop}%
\bibitem [{\citenamefont {Manohar}, \citenamefont {Stebe},\ and\ \citenamefont
  {Lee}(2017)}]{Manohar2017}%
  \BibitemOpen
  \bibfield  {author} {\bibinfo {author} {\bibfnamefont {N.}~\bibnamefont
  {Manohar}}, \bibinfo {author} {\bibfnamefont {K.~J.}\ \bibnamefont {Stebe}},\
  and\ \bibinfo {author} {\bibfnamefont {D.}~\bibnamefont {Lee}},\ }\href
  {https://doi.org/10.1021/ACSMACROLETT.7B00392/ASSET/IMAGES/LARGE/MZ-2017-00392N_0005.JPEG}
  {\bibfield  {journal} {\bibinfo  {journal} {ACS Macro Letters}\ }\textbf
  {\bibinfo {volume} {6}},\ \bibinfo {pages} {1104} (\bibinfo {year}
  {2017})}\BibitemShut {NoStop}%
\bibitem [{\citenamefont {Venkatesh}, \citenamefont {Han},\ and\ \citenamefont
  {Lee}(2019)}]{Venkatesh2019}%
  \BibitemOpen
  \bibfield  {author} {\bibinfo {author} {\bibfnamefont {R.~B.}\ \bibnamefont
  {Venkatesh}}, \bibinfo {author} {\bibfnamefont {S.~H.}\ \bibnamefont {Han}},\
  and\ \bibinfo {author} {\bibfnamefont {D.}~\bibnamefont {Lee}},\ }\href
  {https://doi.org/10.1039/c9nh00130a} {\bibfield  {journal} {\bibinfo
  {journal} {Nanoscale Horizons}\ }\textbf {\bibinfo {volume} {4}},\ \bibinfo
  {pages} {933} (\bibinfo {year} {2019})}\BibitemShut {NoStop}%
\bibitem [{\citenamefont {Jiang}\ \emph {et~al.}(2018)\citenamefont {Jiang},
  \citenamefont {Hor}, \citenamefont {Lee},\ and\ \citenamefont
  {Turner}}]{Jiang2018}%
  \BibitemOpen
  \bibfield  {author} {\bibinfo {author} {\bibfnamefont {Y.}~\bibnamefont
  {Jiang}}, \bibinfo {author} {\bibfnamefont {J.~L.}\ \bibnamefont {Hor}},
  \bibinfo {author} {\bibfnamefont {D.}~\bibnamefont {Lee}},\ and\ \bibinfo
  {author} {\bibfnamefont {K.~T.}\ \bibnamefont {Turner}},\ }\href
  {https://doi.org/10.1021/acsami.8b15027} {\bibfield  {journal} {\bibinfo
  {journal} {ACS Applied Materials and Interfaces}\ }\textbf {\bibinfo {volume}
  {10}},\ \bibinfo {pages} {44011} (\bibinfo {year} {2018})}\BibitemShut
  {NoStop}%
\bibitem [{\citenamefont {Swain}\ \emph {et~al.}(2022)\citenamefont {Swain},
  \citenamefont {{Das A}}, \citenamefont {Chandran},\ and\ \citenamefont
  {Basu}}]{Swain2022}%
  \BibitemOpen
  \bibfield  {author} {\bibinfo {author} {\bibfnamefont {A.}~\bibnamefont
  {Swain}}, \bibinfo {author} {\bibfnamefont {N.}~\bibnamefont {{Das A}}},
  \bibinfo {author} {\bibfnamefont {S.}~\bibnamefont {Chandran}},\ and\
  \bibinfo {author} {\bibfnamefont {J.~K.}\ \bibnamefont {Basu}},\ }\href
  {https://doi.org/10.1039/d1sm01681d} {\bibfield  {journal} {\bibinfo
  {journal} {Soft Matter}\ }\textbf {\bibinfo {volume} {18}},\ \bibinfo {pages}
  {1005} (\bibinfo {year} {2022})}\BibitemShut {NoStop}%
\bibitem [{\citenamefont {Humphrey}, \citenamefont {Dalke},\ and\ \citenamefont
  {Schulten}(1996)}]{Humphrey1996}%
  \BibitemOpen
  \bibfield  {author} {\bibinfo {author} {\bibfnamefont {W.}~\bibnamefont
  {Humphrey}}, \bibinfo {author} {\bibfnamefont {A.}~\bibnamefont {Dalke}},\
  and\ \bibinfo {author} {\bibfnamefont {K.}~\bibnamefont {Schulten}},\ }\href
  {https://doi.org/10.1016/0263-7855(96)00018-5} {\bibfield  {journal}
  {\bibinfo  {journal} {Journal of Molecular Graphics}\ }\textbf {\bibinfo
  {volume} {14}},\ \bibinfo {pages} {33} (\bibinfo {year} {1996})}\BibitemShut
  {NoStop}%
\bibitem [{\citenamefont {Kremer}\ and\ \citenamefont
  {Grest}(1990)}]{Kremer1990}%
  \BibitemOpen
  \bibfield  {author} {\bibinfo {author} {\bibfnamefont {K.}~\bibnamefont
  {Kremer}}\ and\ \bibinfo {author} {\bibfnamefont {G.~S.}\ \bibnamefont
  {Grest}},\ }\href {https://doi.org/10.1063/1.458541} {\bibfield  {journal}
  {\bibinfo  {journal} {The Journal of Chemical Physics}\ }\textbf {\bibinfo
  {volume} {92}},\ \bibinfo {pages} {5057} (\bibinfo {year}
  {1990})}\BibitemShut {NoStop}%
\bibitem [{\citenamefont {Toepperwein}, \citenamefont {Riggleman},\ and\
  \citenamefont {{De Pablo}}(2012)}]{Toepperwein2012}%
  \BibitemOpen
  \bibfield  {author} {\bibinfo {author} {\bibfnamefont {G.~N.}\ \bibnamefont
  {Toepperwein}}, \bibinfo {author} {\bibfnamefont {R.~A.}\ \bibnamefont
  {Riggleman}},\ and\ \bibinfo {author} {\bibfnamefont {J.~J.}\ \bibnamefont
  {{De Pablo}}},\ }\href {https://doi.org/10.1021/ma2017277} {\bibfield
  {journal} {\bibinfo  {journal} {Macromolecules}\ }\textbf {\bibinfo {volume}
  {45}},\ \bibinfo {pages} {543} (\bibinfo {year} {2012})}\BibitemShut
  {NoStop}%
\bibitem [{\citenamefont {Park}\ and\ \citenamefont {Joo}(2014)}]{Park2014a}%
  \BibitemOpen
  \bibfield  {author} {\bibinfo {author} {\bibfnamefont {J.~H.}\ \bibnamefont
  {Park}}\ and\ \bibinfo {author} {\bibfnamefont {Y.~L.}\ \bibnamefont {Joo}},\
  }\href {https://doi.org/10.1039/c4sm00096j} {\bibfield  {journal} {\bibinfo
  {journal} {Soft Matter}\ }\textbf {\bibinfo {volume} {10}},\ \bibinfo {pages}
  {3494} (\bibinfo {year} {2014})}\BibitemShut {NoStop}%
\bibitem [{\citenamefont {Gao}\ \emph {et~al.}(2015)\citenamefont {Gao},
  \citenamefont {Cao}, \citenamefont {Liu}, \citenamefont {Shen}, \citenamefont
  {Wu},\ and\ \citenamefont {Zhang}}]{Gao2015}%
  \BibitemOpen
  \bibfield  {author} {\bibinfo {author} {\bibfnamefont {Y.}~\bibnamefont
  {Gao}}, \bibinfo {author} {\bibfnamefont {D.}~\bibnamefont {Cao}}, \bibinfo
  {author} {\bibfnamefont {J.}~\bibnamefont {Liu}}, \bibinfo {author}
  {\bibfnamefont {J.}~\bibnamefont {Shen}}, \bibinfo {author} {\bibfnamefont
  {Y.}~\bibnamefont {Wu}},\ and\ \bibinfo {author} {\bibfnamefont
  {L.}~\bibnamefont {Zhang}},\ }\href {https://doi.org/10.1039/c5cp01953b}
  {\bibfield  {journal} {\bibinfo  {journal} {Physical Chemistry Chemical
  Physics}\ }\textbf {\bibinfo {volume} {17}},\ \bibinfo {pages} {22959}
  (\bibinfo {year} {2015})}\BibitemShut {NoStop}%
\bibitem [{\citenamefont {Shen}\ \emph {et~al.}(2018)\citenamefont {Shen},
  \citenamefont {Li}, \citenamefont {Zhang}, \citenamefont {Lin}, \citenamefont
  {Li}, \citenamefont {Shen}, \citenamefont {Ganesan},\ and\ \citenamefont
  {Liu}}]{Shen2018}%
  \BibitemOpen
  \bibfield  {author} {\bibinfo {author} {\bibfnamefont {J.}~\bibnamefont
  {Shen}}, \bibinfo {author} {\bibfnamefont {X.}~\bibnamefont {Li}}, \bibinfo
  {author} {\bibfnamefont {L.}~\bibnamefont {Zhang}}, \bibinfo {author}
  {\bibfnamefont {X.}~\bibnamefont {Lin}}, \bibinfo {author} {\bibfnamefont
  {H.}~\bibnamefont {Li}}, \bibinfo {author} {\bibfnamefont {X.}~\bibnamefont
  {Shen}}, \bibinfo {author} {\bibfnamefont {V.}~\bibnamefont {Ganesan}},\ and\
  \bibinfo {author} {\bibfnamefont {J.}~\bibnamefont {Liu}},\ }\href
  {https://doi.org/10.1021/acs.macromol.8b00183} {\bibfield  {journal}
  {\bibinfo  {journal} {Macromolecules}\ }\textbf {\bibinfo {volume} {51}},\
  \bibinfo {pages} {2641} (\bibinfo {year} {2018})}\BibitemShut {NoStop}%
\bibitem [{\citenamefont {Paiva}\ \emph {et~al.}(2019)\citenamefont {Paiva},
  \citenamefont {Boromand}, \citenamefont {Maia}, \citenamefont {Secchi},
  \citenamefont {Calado},\ and\ \citenamefont {Khani}}]{Paiva2019}%
  \BibitemOpen
  \bibfield  {author} {\bibinfo {author} {\bibfnamefont {F.}~\bibnamefont
  {Paiva}}, \bibinfo {author} {\bibfnamefont {A.}~\bibnamefont {Boromand}},
  \bibinfo {author} {\bibfnamefont {J.}~\bibnamefont {Maia}}, \bibinfo {author}
  {\bibfnamefont {A.}~\bibnamefont {Secchi}}, \bibinfo {author} {\bibfnamefont
  {V.}~\bibnamefont {Calado}},\ and\ \bibinfo {author} {\bibfnamefont
  {S.}~\bibnamefont {Khani}},\ }\href {https://doi.org/10.1063/1.5100134}
  {\bibfield  {journal} {\bibinfo  {journal} {Journal of Chemical Physics}\
  }\textbf {\bibinfo {volume} {151}},\ \bibinfo {pages} {114905} (\bibinfo
  {year} {2019})}\BibitemShut {NoStop}%
\bibitem [{\citenamefont {Karatrantos}\ \emph {et~al.}(2011)\citenamefont
  {Karatrantos}, \citenamefont {Composto}, \citenamefont {Winey},\ and\
  \citenamefont {Clarke}}]{Karatrantos2011}%
  \BibitemOpen
  \bibfield  {author} {\bibinfo {author} {\bibfnamefont {A.}~\bibnamefont
  {Karatrantos}}, \bibinfo {author} {\bibfnamefont {R.~J.}\ \bibnamefont
  {Composto}}, \bibinfo {author} {\bibfnamefont {K.~I.}\ \bibnamefont
  {Winey}},\ and\ \bibinfo {author} {\bibfnamefont {N.}~\bibnamefont
  {Clarke}},\ }\href {https://doi.org/10.1021/ma201359s} {\bibfield  {journal}
  {\bibinfo  {journal} {Macromolecules}\ }\textbf {\bibinfo {volume} {44}},\
  \bibinfo {pages} {9830} (\bibinfo {year} {2011})}\BibitemShut {NoStop}%
\bibitem [{\citenamefont {Karatrantos}, \citenamefont {Clarke},\ and\
  \citenamefont {Kr{\"{o}}ger}(2016)}]{Karatrantos2016}%
  \BibitemOpen
  \bibfield  {author} {\bibinfo {author} {\bibfnamefont {A.}~\bibnamefont
  {Karatrantos}}, \bibinfo {author} {\bibfnamefont {N.}~\bibnamefont
  {Clarke}},\ and\ \bibinfo {author} {\bibfnamefont {M.}~\bibnamefont
  {Kr{\"{o}}ger}},\ }\href {https://doi.org/10.1080/15583724.2015.1090450}
  {\bibfield  {journal} {\bibinfo  {journal} {Polymer Reviews}\ }\textbf
  {\bibinfo {volume} {56}},\ \bibinfo {pages} {385} (\bibinfo {year}
  {2016})}\BibitemShut {NoStop}%
\bibitem [{\citenamefont {Tuinier}, \citenamefont {Taniguchi},\ and\
  \citenamefont {Wensink}(2007)}]{Tuinier2007}%
  \BibitemOpen
  \bibfield  {author} {\bibinfo {author} {\bibfnamefont {R.}~\bibnamefont
  {Tuinier}}, \bibinfo {author} {\bibfnamefont {T.}~\bibnamefont {Taniguchi}},\
  and\ \bibinfo {author} {\bibfnamefont {H.~H.}\ \bibnamefont {Wensink}},\
  }\href {https://doi.org/10.1140/epje/i2007-10197-0} {\bibfield  {journal}
  {\bibinfo  {journal} {European Physical Journal E}\ }\textbf {\bibinfo
  {volume} {23}},\ \bibinfo {pages} {355} (\bibinfo {year} {2007})}\BibitemShut
  {NoStop}%
\bibitem [{\citenamefont {Wang}\ \emph {et~al.}(2021)\citenamefont {Wang},
  \citenamefont {O'Connor}, \citenamefont {Grest}, \citenamefont {Zheng},
  \citenamefont {Rubinstein},\ and\ \citenamefont {Ge}}]{Wang2021}%
  \BibitemOpen
  \bibfield  {author} {\bibinfo {author} {\bibfnamefont {J.}~\bibnamefont
  {Wang}}, \bibinfo {author} {\bibfnamefont {T.~C.}\ \bibnamefont {O'Connor}},
  \bibinfo {author} {\bibfnamefont {G.~S.}\ \bibnamefont {Grest}}, \bibinfo
  {author} {\bibfnamefont {Y.}~\bibnamefont {Zheng}}, \bibinfo {author}
  {\bibfnamefont {M.}~\bibnamefont {Rubinstein}},\ and\ \bibinfo {author}
  {\bibfnamefont {T.}~\bibnamefont {Ge}},\ }\href
  {https://doi.org/10.1021/acs.macromol.1c00989} {\bibfield  {journal}
  {\bibinfo  {journal} {Macromolecules}\ }\textbf {\bibinfo {volume} {54}},\
  \bibinfo {pages} {7051} (\bibinfo {year} {2021})}\BibitemShut {NoStop}%
\bibitem [{\citenamefont {Lu}\ and\ \citenamefont {Jayaraman}(2021)}]{Lu2021a}%
  \BibitemOpen
  \bibfield  {author} {\bibinfo {author} {\bibfnamefont {S.}~\bibnamefont
  {Lu}}\ and\ \bibinfo {author} {\bibfnamefont {A.}~\bibnamefont {Jayaraman}},\
  }\href {https://doi.org/10.1021/acsmacrolett.1c00503} {\bibfield  {journal}
  {\bibinfo  {journal} {ACS Macro Letters}\ }\textbf {\bibinfo {volume} {10}},\
  \bibinfo {pages} {1416} (\bibinfo {year} {2021})}\BibitemShut {NoStop}%
\bibitem [{\citenamefont {Robbins}\ and\ \citenamefont
  {M{\"{u}}ser}(2000)}]{Robbins2000}%
  \BibitemOpen
  \bibfield  {author} {\bibinfo {author} {\bibfnamefont {M.}~\bibnamefont
  {Robbins}}\ and\ \bibinfo {author} {\bibfnamefont {M.}~\bibnamefont
  {M{\"{u}}ser}},\ }in\ \href {https://doi.org/10.1201/9780849377877.ch20}
  {\emph {\bibinfo {booktitle} {Modern Tribology Handbook: Volume One:
  Principles of Tribology}}}\ (\bibinfo  {publisher} {CRC Press},\ \bibinfo
  {year} {2000})\ pp.\ \bibinfo {pages} {717--765},\ \Eprint
  {https://arxiv.org/abs/0001056} {arXiv:0001056 [cond-mat]} \BibitemShut
  {NoStop}%
\bibitem [{\citenamefont {Berman}, \citenamefont {Erdemir},\ and\ \citenamefont
  {Sumant}(2018)}]{Berman2018}%
  \BibitemOpen
  \bibfield  {author} {\bibinfo {author} {\bibfnamefont {D.}~\bibnamefont
  {Berman}}, \bibinfo {author} {\bibfnamefont {A.}~\bibnamefont {Erdemir}},\
  and\ \bibinfo {author} {\bibfnamefont {A.~V.}\ \bibnamefont {Sumant}},\
  }\href {https://doi.org/10.1021/acsnano.7b09046} {\enquote {\bibinfo {title}
  {{Approaches for Achieving Superlubricity in Two-Dimensional Materials}},}\ }
  (\bibinfo {year} {2018})\BibitemShut {NoStop}%
\bibitem [{\citenamefont {Thompson}\ \emph {et~al.}(2022)\citenamefont
  {Thompson}, \citenamefont {Aktulga}, \citenamefont {Berger}, \citenamefont
  {Bolintineanu}, \citenamefont {Brown}, \citenamefont {Crozier}, \citenamefont
  {{in 't Veld}}, \citenamefont {Kohlmeyer}, \citenamefont {Moore},
  \citenamefont {Nguyen}, \citenamefont {Shan}, \citenamefont {Stevens},
  \citenamefont {Tranchida}, \citenamefont {Trott},\ and\ \citenamefont
  {Plimpton}}]{Thompson2022}%
  \BibitemOpen
  \bibfield  {author} {\bibinfo {author} {\bibfnamefont {A.~P.}\ \bibnamefont
  {Thompson}}, \bibinfo {author} {\bibfnamefont {H.~M.}\ \bibnamefont
  {Aktulga}}, \bibinfo {author} {\bibfnamefont {R.}~\bibnamefont {Berger}},
  \bibinfo {author} {\bibfnamefont {D.~S.}\ \bibnamefont {Bolintineanu}},
  \bibinfo {author} {\bibfnamefont {W.~M.}\ \bibnamefont {Brown}}, \bibinfo
  {author} {\bibfnamefont {P.~S.}\ \bibnamefont {Crozier}}, \bibinfo {author}
  {\bibfnamefont {P.~J.}\ \bibnamefont {{in 't Veld}}}, \bibinfo {author}
  {\bibfnamefont {A.}~\bibnamefont {Kohlmeyer}}, \bibinfo {author}
  {\bibfnamefont {S.~G.}\ \bibnamefont {Moore}}, \bibinfo {author}
  {\bibfnamefont {T.~D.}\ \bibnamefont {Nguyen}}, \bibinfo {author}
  {\bibfnamefont {R.}~\bibnamefont {Shan}}, \bibinfo {author} {\bibfnamefont
  {M.~J.}\ \bibnamefont {Stevens}}, \bibinfo {author} {\bibfnamefont
  {J.}~\bibnamefont {Tranchida}}, \bibinfo {author} {\bibfnamefont
  {C.}~\bibnamefont {Trott}},\ and\ \bibinfo {author} {\bibfnamefont {S.~J.}\
  \bibnamefont {Plimpton}},\ }\href {https://doi.org/10.1016/j.cpc.2021.108171}
  {\bibfield  {journal} {\bibinfo  {journal} {Computer Physics Communications}\
  }\textbf {\bibinfo {volume} {271}},\ \bibinfo {pages} {108171} (\bibinfo
  {year} {2022})}\BibitemShut {NoStop}%
\bibitem [{\citenamefont {Jewett}\ \emph {et~al.}(2021)\citenamefont {Jewett},
  \citenamefont {Stelter}, \citenamefont {Lambert}, \citenamefont {Saladi},
  \citenamefont {Roscioni}, \citenamefont {Ricci}, \citenamefont {Autin},
  \citenamefont {Maritan}, \citenamefont {Bashusqeh}, \citenamefont {Keyes},
  \citenamefont {Dame}, \citenamefont {Shea}, \citenamefont {Jensen},\ and\
  \citenamefont {Goodsell}}]{Jewett2021}%
  \BibitemOpen
  \bibfield  {author} {\bibinfo {author} {\bibfnamefont {A.~I.}\ \bibnamefont
  {Jewett}}, \bibinfo {author} {\bibfnamefont {D.}~\bibnamefont {Stelter}},
  \bibinfo {author} {\bibfnamefont {J.}~\bibnamefont {Lambert}}, \bibinfo
  {author} {\bibfnamefont {S.~M.}\ \bibnamefont {Saladi}}, \bibinfo {author}
  {\bibfnamefont {O.~M.}\ \bibnamefont {Roscioni}}, \bibinfo {author}
  {\bibfnamefont {M.}~\bibnamefont {Ricci}}, \bibinfo {author} {\bibfnamefont
  {L.}~\bibnamefont {Autin}}, \bibinfo {author} {\bibfnamefont
  {M.}~\bibnamefont {Maritan}}, \bibinfo {author} {\bibfnamefont {S.~M.}\
  \bibnamefont {Bashusqeh}}, \bibinfo {author} {\bibfnamefont {T.}~\bibnamefont
  {Keyes}}, \bibinfo {author} {\bibfnamefont {R.~T.}\ \bibnamefont {Dame}},
  \bibinfo {author} {\bibfnamefont {J.~E.}\ \bibnamefont {Shea}}, \bibinfo
  {author} {\bibfnamefont {G.~J.}\ \bibnamefont {Jensen}},\ and\ \bibinfo
  {author} {\bibfnamefont {D.~S.}\ \bibnamefont {Goodsell}},\ }\href
  {https://doi.org/10.1016/j.jmb.2021.166841} {\bibfield  {journal} {\bibinfo
  {journal} {Journal of Molecular Biology}\ ,\ \bibinfo {pages} {166841}}
  (\bibinfo {year} {2021})}\BibitemShut {NoStop}%
\bibitem [{\citenamefont {Auhl}\ \emph {et~al.}(2003)\citenamefont {Auhl},
  \citenamefont {Everaers}, \citenamefont {Grest}, \citenamefont {Kremer},\
  and\ \citenamefont {Plimpton}}]{Auhl2003}%
  \BibitemOpen
  \bibfield  {author} {\bibinfo {author} {\bibfnamefont {R.}~\bibnamefont
  {Auhl}}, \bibinfo {author} {\bibfnamefont {R.}~\bibnamefont {Everaers}},
  \bibinfo {author} {\bibfnamefont {G.~S.}\ \bibnamefont {Grest}}, \bibinfo
  {author} {\bibfnamefont {K.}~\bibnamefont {Kremer}},\ and\ \bibinfo {author}
  {\bibfnamefont {S.~J.}\ \bibnamefont {Plimpton}},\ }\href
  {https://doi.org/10.1063/1.1628670} {\bibfield  {journal} {\bibinfo
  {journal} {Journal of Chemical Physics}\ }\textbf {\bibinfo {volume} {119}},\
  \bibinfo {pages} {12718} (\bibinfo {year} {2003})},\ \Eprint
  {https://arxiv.org/abs/0306026} {arXiv:0306026 [cond-mat]} \BibitemShut
  {NoStop}%
\bibitem [{\citenamefont {Hall}\ and\ \citenamefont
  {Schweizer}(2008)}]{Hall2008}%
  \BibitemOpen
  \bibfield  {author} {\bibinfo {author} {\bibfnamefont {L.~M.}\ \bibnamefont
  {Hall}}\ and\ \bibinfo {author} {\bibfnamefont {K.~S.}\ \bibnamefont
  {Schweizer}},\ }\href {https://doi.org/10.1063/1.2938379} {\bibfield
  {journal} {\bibinfo  {journal} {Journal of Chemical Physics}\ }\textbf
  {\bibinfo {volume} {128}},\ \bibinfo {pages} {234901} (\bibinfo {year}
  {2008})}\BibitemShut {NoStop}%
\bibitem [{\citenamefont {Mozafar}\ and\ \citenamefont
  {Denniston}(2022)}]{Mozafar2022}%
  \BibitemOpen
  \bibfield  {author} {\bibinfo {author} {\bibfnamefont {O.}~\bibnamefont
  {Mozafar}}\ and\ \bibinfo {author} {\bibfnamefont {C.}~\bibnamefont
  {Denniston}},\ }\href {https://doi.org/10.1103/PhysRevE.105.064109}
  {\bibfield  {journal} {\bibinfo  {journal} {Physical Review E}\ }\textbf
  {\bibinfo {volume} {105}},\ \bibinfo {pages} {64109} (\bibinfo {year}
  {2022})}\BibitemShut {NoStop}%
\bibitem [{\citenamefont {Gartner}\ and\ \citenamefont
  {Jayaraman}(2019)}]{Gartner2019}%
  \BibitemOpen
  \bibfield  {author} {\bibinfo {author} {\bibfnamefont {T.~E.}\ \bibnamefont
  {Gartner}}\ and\ \bibinfo {author} {\bibfnamefont {A.}~\bibnamefont
  {Jayaraman}},\ }\href {https://doi.org/10.1021/acs.macromol.8b01836}
  {\enquote {\bibinfo {title} {{Modeling and Simulations of Polymers: A
  Roadmap}},}\ } (\bibinfo {year} {2019})\BibitemShut {NoStop}%
\bibitem [{\citenamefont {Du}, \citenamefont {Fischer},\ and\ \citenamefont
  {Winey}(2003)}]{Du2003}%
  \BibitemOpen
  \bibfield  {author} {\bibinfo {author} {\bibfnamefont {F.}~\bibnamefont
  {Du}}, \bibinfo {author} {\bibfnamefont {J.~E.}\ \bibnamefont {Fischer}},\
  and\ \bibinfo {author} {\bibfnamefont {K.~I.}\ \bibnamefont {Winey}},\ }\href
  {https://doi.org/10.1002/polb.10701} {\bibfield  {journal} {\bibinfo
  {journal} {Journal of Polymer Science, Part B: Polymer Physics}\ }\textbf
  {\bibinfo {volume} {41}},\ \bibinfo {pages} {3333} (\bibinfo {year}
  {2003})}\BibitemShut {NoStop}%
\bibitem [{\citenamefont {Zhao}\ \emph {et~al.}(2016)\citenamefont {Zhao},
  \citenamefont {Byshkin}, \citenamefont {Cong}, \citenamefont {Kawakatsu},
  \citenamefont {Guadagno}, \citenamefont {{De Nicola}}, \citenamefont {Yu},
  \citenamefont {Milano},\ and\ \citenamefont {Dong}}]{Zhao2016}%
  \BibitemOpen
  \bibfield  {author} {\bibinfo {author} {\bibfnamefont {Y.}~\bibnamefont
  {Zhao}}, \bibinfo {author} {\bibfnamefont {M.}~\bibnamefont {Byshkin}},
  \bibinfo {author} {\bibfnamefont {Y.}~\bibnamefont {Cong}}, \bibinfo {author}
  {\bibfnamefont {T.}~\bibnamefont {Kawakatsu}}, \bibinfo {author}
  {\bibfnamefont {L.}~\bibnamefont {Guadagno}}, \bibinfo {author}
  {\bibfnamefont {A.}~\bibnamefont {{De Nicola}}}, \bibinfo {author}
  {\bibfnamefont {N.}~\bibnamefont {Yu}}, \bibinfo {author} {\bibfnamefont
  {G.}~\bibnamefont {Milano}},\ and\ \bibinfo {author} {\bibfnamefont
  {B.}~\bibnamefont {Dong}},\ }\href {https://doi.org/10.1039/c6nr03304k}
  {\bibfield  {journal} {\bibinfo  {journal} {Nanoscale}\ }\textbf {\bibinfo
  {volume} {8}},\ \bibinfo {pages} {15538} (\bibinfo {year}
  {2016})}\BibitemShut {NoStop}%
\bibitem [{\citenamefont {Mahnke}, \citenamefont {Kaupu{\v{z}}s},\ and\
  \citenamefont {Lubashevsky}(2009)}]{Mahnke2009}%
  \BibitemOpen
  \bibfield  {author} {\bibinfo {author} {\bibfnamefont {R.}~\bibnamefont
  {Mahnke}}, \bibinfo {author} {\bibfnamefont {J.}~\bibnamefont
  {Kaupu{\v{z}}s}},\ and\ \bibinfo {author} {\bibfnamefont {I.}~\bibnamefont
  {Lubashevsky}},\ }\href {https://doi.org/10.1002/9783527626090} {\emph
  {\bibinfo {title} {Physics of Stochastic Processes: How Randomness Acts in
  Time}}}\ (\bibinfo  {publisher} {Wiley},\ \bibinfo {year} {2009})\ pp.\
  \bibinfo {pages} {1--430}\BibitemShut {NoStop}%
\bibitem [{\citenamefont {Asakura}\ and\ \citenamefont
  {Oosawa}(1954)}]{Asakura1954}%
  \BibitemOpen
  \bibfield  {author} {\bibinfo {author} {\bibfnamefont {S.}~\bibnamefont
  {Asakura}}\ and\ \bibinfo {author} {\bibfnamefont {F.}~\bibnamefont
  {Oosawa}},\ }\href {https://doi.org/10.1063/1.1740347} {\bibfield  {journal}
  {\bibinfo  {journal} {The Journal of Chemical Physics}\ }\textbf {\bibinfo
  {volume} {22}},\ \bibinfo {pages} {1255} (\bibinfo {year}
  {1954})}\BibitemShut {NoStop}%
\bibitem [{\citenamefont {Vrij}(1977)}]{Vrij1977}%
  \BibitemOpen
  \bibfield  {author} {\bibinfo {author} {\bibfnamefont {A.}~\bibnamefont
  {Vrij}},\ }in\ \href {https://doi.org/10.1016/b978-0-08-021570-9.50016-4}
  {\emph {\bibinfo {booktitle} {Colloid and Surface Science}}}\ (\bibinfo
  {publisher} {Pergamon},\ \bibinfo {year} {1977})\ pp.\ \bibinfo {pages}
  {471--483}\BibitemShut {NoStop}%
\bibitem [{\citenamefont {Biben}, \citenamefont {Bladon},\ and\ \citenamefont
  {Frenkel}(1996)}]{Biben1996}%
  \BibitemOpen
  \bibfield  {author} {\bibinfo {author} {\bibfnamefont {T.}~\bibnamefont
  {Biben}}, \bibinfo {author} {\bibfnamefont {P.}~\bibnamefont {Bladon}},\ and\
  \bibinfo {author} {\bibfnamefont {D.}~\bibnamefont {Frenkel}},\ }\href
  {https://doi.org/10.1088/0953-8984/8/50/008} {\bibfield  {journal} {\bibinfo
  {journal} {Journal of Physics Condensed Matter}\ }\textbf {\bibinfo {volume}
  {8}},\ \bibinfo {pages} {10799} (\bibinfo {year} {1996})}\BibitemShut
  {NoStop}%
\bibitem [{\citenamefont {Striolo}\ \emph {et~al.}(2004)\citenamefont
  {Striolo}, \citenamefont {Colina}, \citenamefont {Gubbins}, \citenamefont
  {Elvassore},\ and\ \citenamefont {Lue}}]{Striolo2004}%
  \BibitemOpen
  \bibfield  {author} {\bibinfo {author} {\bibfnamefont {A.}~\bibnamefont
  {Striolo}}, \bibinfo {author} {\bibfnamefont {C.~M.}\ \bibnamefont {Colina}},
  \bibinfo {author} {\bibfnamefont {K.~E.}\ \bibnamefont {Gubbins}}, \bibinfo
  {author} {\bibfnamefont {N.}~\bibnamefont {Elvassore}},\ and\ \bibinfo
  {author} {\bibfnamefont {L.}~\bibnamefont {Lue}},\ }\href
  {https://doi.org/10.1080/0892702042000197649} {\bibfield  {journal} {\bibinfo
   {journal} {Molecular Simulation}\ }\textbf {\bibinfo {volume} {30}},\
  \bibinfo {pages} {437} (\bibinfo {year} {2004})}\BibitemShut {NoStop}%
\bibitem [{\citenamefont {Vroege}\ and\ \citenamefont
  {Lekkerkerker}(1992)}]{Vroege1992}%
  \BibitemOpen
  \bibfield  {author} {\bibinfo {author} {\bibfnamefont {G.~J.}\ \bibnamefont
  {Vroege}}\ and\ \bibinfo {author} {\bibfnamefont {H.~N.}\ \bibnamefont
  {Lekkerkerker}},\ }\href {https://doi.org/10.1088/0034-4885/55/8/003}
  {\bibfield  {journal} {\bibinfo  {journal} {Reports on Progress in Physics}\
  }\textbf {\bibinfo {volume} {55}},\ \bibinfo {pages} {1241} (\bibinfo {year}
  {1992})}\BibitemShut {NoStop}%
\bibitem [{\citenamefont {Changizrezaei}\ and\ \citenamefont
  {Denniston}(2017)}]{Changizrezaei2017}%
  \BibitemOpen
  \bibfield  {author} {\bibinfo {author} {\bibfnamefont {S.}~\bibnamefont
  {Changizrezaei}}\ and\ \bibinfo {author} {\bibfnamefont {C.}~\bibnamefont
  {Denniston}},\ }\href {https://doi.org/10.1103/PhysRevE.96.032702} {\bibfield
   {journal} {\bibinfo  {journal} {Physical Review E}\ }\textbf {\bibinfo
  {volume} {96}},\ \bibinfo {pages} {32702} (\bibinfo {year}
  {2017})}\BibitemShut {NoStop}%
\bibitem [{\citenamefont {Mottram}\ and\ \citenamefont
  {Newton}(2014)}]{Mottram2014}%
  \BibitemOpen
  \bibfield  {author} {\bibinfo {author} {\bibfnamefont {N.~J.}\ \bibnamefont
  {Mottram}}\ and\ \bibinfo {author} {\bibfnamefont {C.~J.~P.}\ \bibnamefont
  {Newton}},\ }\href {http://arxiv.org/abs/1409.3542} {\enquote {\bibinfo
  {title} {{Introduction to Q-tensor theory}},}\ } (\bibinfo {year} {2014}),\
  \Eprint {https://arxiv.org/abs/1409.3542} {arXiv:1409.3542} \BibitemShut
  {NoStop}%
\bibitem [{\citenamefont {Majumdar}(2010)}]{Majumdar2010}%
  \BibitemOpen
  \bibfield  {author} {\bibinfo {author} {\bibfnamefont {A.}~\bibnamefont
  {Majumdar}},\ }\href {https://doi.org/10.1017/S0956792509990210} {\bibfield
  {journal} {\bibinfo  {journal} {European Journal of Applied Mathematics}\
  }\textbf {\bibinfo {volume} {21}},\ \bibinfo {pages} {181} (\bibinfo {year}
  {2010})}\BibitemShut {NoStop}%
\bibitem [{\citenamefont {Aoyagi}, \citenamefont {Takimoto},\ and\
  \citenamefont {Doi}(2001)}]{Aoyagi2001}%
  \BibitemOpen
  \bibfield  {author} {\bibinfo {author} {\bibfnamefont {T.}~\bibnamefont
  {Aoyagi}}, \bibinfo {author} {\bibfnamefont {J.~I.}\ \bibnamefont
  {Takimoto}},\ and\ \bibinfo {author} {\bibfnamefont {M.}~\bibnamefont
  {Doi}},\ }\href {https://doi.org/10.1063/1.1377015} {\bibfield  {journal}
  {\bibinfo  {journal} {Journal of Chemical Physics}\ }\textbf {\bibinfo
  {volume} {115}},\ \bibinfo {pages} {552} (\bibinfo {year}
  {2001})}\BibitemShut {NoStop}%
\bibitem [{\citenamefont {Gorkunov}\ and\ \citenamefont
  {Osipov}(2011)}]{Gorkunov2011}%
  \BibitemOpen
  \bibfield  {author} {\bibinfo {author} {\bibfnamefont {M.~V.}\ \bibnamefont
  {Gorkunov}}\ and\ \bibinfo {author} {\bibfnamefont {M.~A.}\ \bibnamefont
  {Osipov}},\ }\href {https://doi.org/10.1039/c0sm01398f} {\bibfield  {journal}
  {\bibinfo  {journal} {Soft Matter}\ }\textbf {\bibinfo {volume} {7}},\
  \bibinfo {pages} {4348} (\bibinfo {year} {2011})}\BibitemShut {NoStop}%
\end{thebibliography}%
\bibliographystyle{aipnum4-2}

\end{document}